\documentstyle[epsf]{article}
\newcommand{\bookfig}[5]{
\begin{figure}\centering\fbox{\epsfysize=#5cm \epsfbox{#1}} 
\caption[#2]{#4}\label{#3}
\end{figure}
}
\font\tenbb=msbm10
\font\sevenbb=msbm7
\font\fivebb=msbm5
\newfam\bbfam
\textfont\bbfam=\tenbb \scriptfont\bbfam=\sevenbb
\scriptscriptfont\bbfam=\fivebb
\def\bb{\fam\bbfam}

\def\Cb{{\bb C}}
\def\Nb{{\bb N}}
\def\Rb{{\bb R}}

\begin{document}

%%XXX \special{src: 37 CK1.TEX} %Inserted by TeXtelmExtel

\title{Hopf Algebras, Renormalization and Noncommutative Geometry}
\author{Alain CONNES\\
IHES 
\and 
Dirk KREIMER\\ Mainz Univ.}
\date{ August 1998, IHES/M/98/60}

%%XXX \special{src: 43 CK1.TEX} %Inserted by TeXtelmExtel

\maketitle

%%XXX \special{src: 47 CK1.TEX} %Inserted by TeXtelmExtel

\begin{abstract} 
We explore the relation between the Hopf algebra associated to the renormalization 
of QFT and the Hopf algebra associated to the NCG computations of tranverse index 
theory for foliations.
\end{abstract}

%%XXX \special{src: 55 CK1.TEX} %Inserted by TeXtelmExtel

\section*{Introduction} 

%%XXX \special{src: 59 CK1.TEX} %Inserted by TeXtelmExtel

In \cite{K} it was shown that the combinatorics of the subtraction procedure 
inherent to perturbative renormalization gives rise to a Hopf algebra ${\cal H}_R$
which provides a conceptual framework to understand the intricacies of the forest 
formula of Zimmermann.

%%XXX \special{src: 66 CK1.TEX} %Inserted by TeXtelmExtel

In \cite{CM}, it was shown that the delicate computational problem which arises from 
the transverse hypoelliptic theory of foliations, as formulated in noncommutative 
geometry, can only be settled thanks to a Hopf algebra ${\cal H}_T$
associated to each integer codimension. This Hopf algebra reduces transverse 
geometry, to a universal geometry of affine nature.

%%XXX \special{src: 74 CK1.TEX} %Inserted by TeXtelmExtel

The aim of this paper is to establish a close relation between the above Hopf 
algebras ${\cal H}_R$ and ${\cal H}_T$. We shall first recall the first results of 
\cite{CM} which describe in the simplest case of codimension $1$ 
the presentation of the Hopf algebra ${\cal H}_T$.

%%XXX \special{src: 81 CK1.TEX} %Inserted by TeXtelmExtel

We shall then explain the origin of the Hopf algebra ${\cal H}_R$ from the 
renormalization of the divergences of QFT and show following \cite{K} how ${\cal 
H}_R$ is used in concrete problems of renormalization theory.
In the appendix we  include the case of 
overlapping divergences.

%%XXX \special{src: 89 CK1.TEX} %Inserted by TeXtelmExtel

We then give the presentation of the simplest model of ${\cal H}_R$ namely the Hopf 
algebra of rooted trees, and show that it is uniquely characterized as the solution 
of a universal problem in Hochschild cohomology. We then determine the formal Lie 
algebra $G$ such that ${\cal H}_R$ is obtained as the dual of the envelopping 
algebra of this Lie algebra. It turns out to be a refinement of the Lie algebra of 
formal vector fields in one dimension. We then show that many of the results of 
\cite{CM} actually extend to this refinement of formal vector fields.
These results indicate that parallel to the ordinary differential calculus which 
underlies the transverse structure of foliations, the recipes of renormalization 
theory should be considered as a refined form of calculus and should be 
understandable on a conceptual ground.

%%%%%%%%%%
Concretely, in the first section we show in Theorem (8)
that the algebraic rules of the
Hopf algebra ${\cal H}_T$ are the expression of the group law
of composition of diffeomorphisms of ${\bf R}$
in terms of the coordinates $\delta_n$ given by the Taylor expansion
of $-\log(\psi^\prime(x))$ at $x=0$. In particular 
this shows that the antipode in ${\cal H}_T$
is, modulo a change of variables, the same as the operation of
inversion
of a formal power series for the composition law.

In the second section we begin by the simplest and most explicit examples
of divergent integrals of the kind that are met in Quantum Field Theory
computations.

We describe in this toy case the explicit counterterm construction
and the immediate problem which arises
from divergent subintegrations and explain how the Hopf algebra ${\cal H}_R$
finds the combinatorial solution of the subtraction problem from its
antipode.

We next explain why the same holds in QFT  (the treatment of overlapping divergences
is postponed to the appendix).

In the third section we exhibit the precise relation and analogy between 
${\cal H}_R$ and ${\cal H}_T$ to the point that the antipode in ${\cal H}_R$
appears as a direct analogue of the antipode in ${\cal H}_T$
which we understood above as the inversion of formal power series.
The key nuance between the two Hopf algebras is that where ${\cal H}_T$
uses integers to label the Taylor expansion, the Hopf algebra
${\cal H}_R$ uses rooted trees for labels. 
 
%%XXX \special{src: 103 CK1.TEX} %Inserted by TeXtelmExtel

\section{The Hopf algebra  ${\cal H}_T$}

%%XXX \special{src: 107 CK1.TEX} %Inserted by TeXtelmExtel

The computation of the local index formula for transversally hypoelliptic operators 
(\cite{CM}) is governed by a very specific Hopf algebra ${\cal H}_T$, which only 
depends upon the codimension $n$ of the foliation. The structure of this Hopf 
algebra, its relation with the Lie algebra of formal vector fields as well as the 
computation of its cyclic cohomology have been done in \cite{CM}.

%%XXX \special{src: 115 CK1.TEX} %Inserted by TeXtelmExtel

In order to pursue the analogy between this development and the discovery by D.~K. 
(\cite{K}) of the Hopf algebra underlying renormalization, we shall recall here in 
all details the presentation and first properties of the Hopf algebra ${\cal H}_T$ 
in the simplest case of codimension one. Useless to say this does not dispense one 
from consulting \cite{CM}, in particular in connection with the specific 
representation of ${\cal H}_T$ on crossed product algebras and the corresponding 
analysis.

%%XXX \special{src: 125 CK1.TEX} %Inserted by TeXtelmExtel

We first define a bialgebra by generators and relations. As an algebra we 
view ${\cal H}_T$ as the envelopping algebra of the Lie algebra which is the 
linear span of $Y$, $X$, $\delta_n$, $n \geq 1$ with the relations,
\begin{equation}
[Y,X] = X , [Y,\delta_n] = n \, \delta_n , [\delta_n , \delta_m] = 0 \ \forall \, n 
, m \geq 1 , [X,\delta_n] = \delta_{n+1} \ \forall \, n \geq 1 \, . \label{eq1}
\end{equation}
The coproduct $\Delta$ in ${\cal H}_T$ is defined by
\begin{equation}
\Delta \, Y = Y \otimes 1 + 1 \otimes Y \ , \ \Delta \, X = X \otimes 1 + 1 \otimes 
X + \delta_1 \otimes Y \ , \ \Delta \, \delta_1 = \delta_1 \otimes 1 + 1 \otimes 
\delta_1 \label{eq2}
\end{equation}
with $\Delta \, \delta_n$ defined by induction using (\ref{eq1}) and the equality,
\begin{equation}
\Delta (h_1 \, h_2) = \Delta h_1 \, \Delta h_2 \qquad \forall \, h_j \in {\cal H}_T 
\, . \label{eq3}
\end{equation}

%%XXX \special{src: 146 CK1.TEX} %Inserted by TeXtelmExtel

\noindent {\bf Lemma 1.} {\it The above presentation defines a Hopf algebra ${\cal 
H}_T$.}

%%XXX \special{src: 151 CK1.TEX} %Inserted by TeXtelmExtel

\smallskip

%%XXX \special{src: 155 CK1.TEX} %Inserted by TeXtelmExtel

\noindent {\bf Proof.} One checks that the Lie algebra relations (\ref{eq1}) are 
fulfilled by the elements $\Delta(Y)$, $\Delta(X)$, $\Delta(\delta_1)$, so 
that, by the universal property of the envelopping algebra, $\Delta$ extends to an 
algebra homomorphism,
\begin{equation}
\Delta : {\cal H}_T \rightarrow {\cal H}_T \otimes {\cal H}_T \label{eq4}
\end{equation}
and using the uniqueness of the extension, one also checks the coassociativity.

%%XXX \special{src: 166 CK1.TEX} %Inserted by TeXtelmExtel

One needs to show the existence of the antipode $S$. It is characterized abstractly 
as the inverse of the element $L(a) = a$ in the algebra of linear maps $L$ from 
${\cal H}_T$ to ${\cal H}_T$ endowed with the product
\begin{equation}
(L_1 * L_2) (a) = \sum \, L_1 (a_{(1)}) \, L_2 (a_{(2)}) \qquad \Delta \, a = 
\sum \, a_{(1)} \otimes a_{(2)} \ , \ a \in {\cal H}_T \, . \label{eq5}
\end{equation}
A simple computation shows that $S$ is the unique antiautomorphism of ${\cal H}_T 
(n)$ such that,
\begin{equation}
S(Y)=-Y  \qquad  S(\delta_1)=- \delta_1  \qquad S(X)= -X +\, \delta_1 Y \, . 
\label{eq6}
\end{equation}
Note that the square of $S$ is not the identity.~$\Box$

%%XXX \special{src: 183 CK1.TEX} %Inserted by TeXtelmExtel

\smallskip

%%XXX \special{src: 187 CK1.TEX} %Inserted by TeXtelmExtel

In order to understand the Hopf algebra ${\cal H}_T$, we first analyse the 
commutative subalgebra generated by the $\delta_n$.

%%XXX \special{src: 192 CK1.TEX} %Inserted by TeXtelmExtel

For each $n$ we let ${\cal H}_n$ be the subalgebra generated by $\delta_1 , \ldots 
, \delta_n$,
\begin{equation}
{\cal H}_n = \{ P (\delta_1 , \ldots , \delta_n) \ ; \ P \ \hbox{polynomial in} \ n 
\ \hbox{variables} \} \, . \label{eq7}
\end{equation}
We let ${\cal H}_{n,0}$ be the ideal,
\begin{equation}
{\cal H}_{n,0} = \{ P ; P(0) = 0 \} \, . \label{eq8}
\end{equation}
By induction on $n$ one proves the following

%%XXX \special{src: 206 CK1.TEX} %Inserted by TeXtelmExtel

\smallskip

%%XXX \special{src: 210 CK1.TEX} %Inserted by TeXtelmExtel

\noindent {\bf Lemma 2.} {\it For each $n$ there exists $R_{n-1} \in 
{\cal H}_{n-1,0} \otimes {\cal H}_{n-1,0}$ such that $\Delta \, \delta_n = \delta_n 
\otimes 1 + 1 \otimes \delta_n + R_{n-1}$.}

%%XXX \special{src: 216 CK1.TEX} %Inserted by TeXtelmExtel

\smallskip

%%XXX \special{src: 220 CK1.TEX} %Inserted by TeXtelmExtel

\noindent {\bf Proof.} One has $ \Delta \, \delta_1 = \delta_1 \otimes 1 + 1 \otimes 
\delta_1 $, and a simple computation shows that,
\begin{equation}
\Delta \, \delta_2 = \delta_2 \otimes 1 + 1 \otimes \delta_2 + \delta_1 \otimes 
\delta_1 \, ,\label{eq9}
\end{equation}
and,
\begin{equation}
\Delta \, \delta_3 = \delta_3 \otimes 1 + 1 \otimes \delta_3 + \delta_1^2 \otimes 
\delta_1  + \delta_2 \otimes \delta_1  + 3 \delta_1 \otimes \delta_2 \, . 
\label{eq10}
\end{equation}
In general, one determines $R_n$ by induction, using
\begin{equation}
R_n = [X \otimes 1 + 1 \otimes X , R_{n-1}] + n \, \delta_1 \otimes \delta_n + 
[\delta_1 \otimes Y , R_{n-1}] \, . \label{eq11}
\end{equation}
Since $[X, {\cal H}_{n-1,0}] \subset {\cal H}_{n,0}$ and $[Y, {\cal H}_{n-1,0}] 
\subset {\cal H}_{n-1,0} \subset{\cal H}_{n,0}$, one gets that $R_n \in 
{\cal H}_{n,0} \otimes {\cal H}_{n,0}$.~$\Box$

%%XXX \special{src: 243 CK1.TEX} %Inserted by TeXtelmExtel

\smallskip

%%XXX \special{src: 247 CK1.TEX} %Inserted by TeXtelmExtel

\noindent The equality (\ref{eq10}) shows that ${\cal H}_n$ is not cocommutative for 
$n \geq 3$. However, since it is commutative, we shall determine the corresponding 
Lie algebra, using the Milnor-Moore theorem.

%%XXX \special{src: 253 CK1.TEX} %Inserted by TeXtelmExtel

\noindent Let ${\cal A}_n^1$ be the Lie algebra of jets of order $(n+1)$ of vector 
fields on the line, 
\[
f(x) \, \partial / \partial x \quad , \quad f(0) = f' (0) = 0 
\]
modulo $x^{n+2} \, \partial / \partial x$.

%%XXX \special{src: 262 CK1.TEX} %Inserted by TeXtelmExtel

\smallskip

%%XXX \special{src: 266 CK1.TEX} %Inserted by TeXtelmExtel

\noindent {\bf Proposition 3.} {\it The Hopf algebra ${\cal H}_n$ is the 
dual of the envelopping agebra ${\cal U} ({\cal A}_n^1)$, ${\cal H}_n = {\cal U} 
({\cal A}_n^1)^*$.}

%%XXX \special{src: 272 CK1.TEX} %Inserted by TeXtelmExtel

\smallskip

%%XXX \special{src: 276 CK1.TEX} %Inserted by TeXtelmExtel

\noindent {\bf Proof.}  For each $k \leq n$ we introduce a linear form $Z_{k,n}$ on 
${\cal H}_n$
\begin{equation}
\langle Z_{k,n} , P \rangle = \left( {\partial \over \partial \, \delta_k} \, P 
\right) 
(0) \, . \label{eq12}
\end{equation}
One has by construction,
\begin{equation}
\langle Z_{k,n} , PQ \rangle = \langle Z_{k,n} , P \rangle \, Q (0) + P(0) \, 
\langle Z_{k,n} , Q \rangle \, . \label{eq13}
\end{equation}
Note that $\varepsilon$, $\langle \varepsilon , P \rangle = P(0)$ is the counit of
${\cal H}_n$,
\begin{equation}
\langle L \otimes \varepsilon , \Delta \, P \rangle = \langle \varepsilon \otimes L 
, 
\Delta \, P \rangle = \langle L,P \rangle \qquad \forall \, P \in {\cal H}_n \, . 
\label{eq14}
\end{equation}
(Check both sides on a monomial $P = \delta_1^{a_1} \ldots \delta_n^{a_n}$.)

%%XXX \special{src: 300 CK1.TEX} %Inserted by TeXtelmExtel

\noindent Thus in the dual agebra ${\cal H}_n^*$ one can write (\ref{eq13}) as
\begin{equation}
\Delta \, Z_{k,n} = Z_{k,n} \otimes 1 + 1 \otimes Z_{k,n} \, . \label{eq15}
\end{equation}
Moreover the $Z_{k,n}$ form a basis of the linear space of solutions of 
(\ref{eq15}) and we just need to determine the Lie algebra structure determined by 
the 
bracket.

%%XXX \special{src: 311 CK1.TEX} %Inserted by TeXtelmExtel

\noindent Let,
\begin{equation}
Z'_{k,n} = (k+1) ! \, Z_{k,n} \, . \label{eq16}
\end{equation}

%%XXX \special{src: 318 CK1.TEX} %Inserted by TeXtelmExtel

Let us show that $[Z'_{k,n} , Z'_{\ell,n}] = 0$ if $ k + \ell > n $, and that,
\begin{equation}
[Z'_{k,n} , Z'_{\ell,n}] = (\ell - k) \, Z'_{k+\ell , n} \, , \label{eq17}
\end{equation}
if $k + \ell \leq n$. Let $P = \delta_1^{a_1} \ldots \delta_n^{a_n}$ be a 
monomial. We need to compute $\langle \Delta \, P , Z_{k,n} \otimes Z_{\ell , n} - 
Z_{\ell , n} \otimes Z_{k,n} \rangle$. One has
\[
\Delta \, P = (\delta_1 \otimes 1 + 1 \otimes \delta_1)^{a_1} \, (\delta_2 \otimes 1 
+ 1 \otimes \delta_2 + R_1)^{a_2} \ldots (\delta_n \otimes 1 + 1 \otimes \delta_n + 
R_{n-1})^{a_n} \, .
\]
We look for the terms in $\delta_k \otimes \delta_{\ell}$ or $\delta_{\ell} \otimes 
\delta_k$ and take the difference. The latter is non zero only if all $a_j = 0$ 
except $a_q = 1$. Moreover since $R_m$ is homogeneous of degree $m+1$ 
one gets $q=k+\ell$ and in particular $[Z'_{k,n} , Z'_{\ell ,n}] = 0$ if 
$k + \ell > n$. One then computes by induction using (\ref{eq11}) the bilinear 
part of $R_m$. One has $R_1^{(1)} = \delta_1 \otimes \delta_1$, and from 
(\ref{eq11})
\begin{equation}
R_n^{(1)} = [(X \otimes 1 + 1 \otimes X) , R_{n-1}^{(1)}] + n \, \delta_1 \otimes 
\delta_n \, . \label{eq18}
\end{equation}
This gives
\begin{equation}
R_{n-1}^{(1)} = \delta_{n-1} \otimes \delta_1 + C_n^1 \, \delta_{n-2} \otimes 
\delta_2 + \ldots + C_n^{n-2} \, \delta_1 \otimes \delta_{n-1} \, . \label{eq19}
\end{equation}
Thus the coefficient of $\delta_k \otimes \delta_{\ell}$ is $C_{k+\ell}^{\ell - 1}$ 
and we get
\begin{equation}
[Z_{k,n} , Z_{\ell , n}] = (C_{k+\ell}^{\ell - 1} - C_{k+\ell}^{k - 1}) 
\, Z_{k+\ell , n} \, . \label{eq20}
\end{equation}
One has ${(k+1) ! \, (\ell + 1) ! \over (k+\ell + 1) !} \, (C_{k+\ell}^{\ell - 1} - 
C_{k+\ell}^{k - 1}) = {\ell (\ell + 1) - k (k+1) \over k + \ell + 1} = \ell - k$ 
thus 
one gets (\ref{eq17}). The elements $Z_{k,n} = {x^{k+1} \over (k+1) !} \, \partial / 
\partial x$ of the Lie algebra ${\cal A}_n^1$ are related by (\ref{eq16}) to 
$Z'_{k,n} = x^{k+1} \, \partial / \partial x$ which satisfy the Lie algebra 
relations 
(\ref{eq17}). The result then follows from the Milnor-Moore theorem.~$\Box$

%%XXX \special{src: 363 CK1.TEX} %Inserted by TeXtelmExtel

\smallskip

%%XXX \special{src: 367 CK1.TEX} %Inserted by TeXtelmExtel

\noindent The ${\cal A}_n^1$ form a projective system of Lie algebras, with limit 
the 
Lie algebra ${\cal A}^1$ of formal vector fields which vanish at order 2 at 0. Thus 
the inductive limit ${\cal H}^1$ of the Hopf algebras ${\cal H}_n$ is,
\begin{equation}
{\cal H}^1 = {\cal U} ({\cal A}^1)^* \, . \label{eq21}
\end{equation}
The Lie algebra ${\cal A}^1$ is a {\it graded} Lie algebra, with one parameter group 
of automorphisms,
\begin{equation}
\alpha_t \, (Z_n) = e^{nt} \, Z_n \label{eq22}
\end{equation}
which extends to ${\cal U} ({\cal A}^1)$ and transposes to ${\cal U} ({\cal A}^1)^*$ 
as
\begin{equation}
\langle [Y,P] , a \rangle = \left\langle P , {\partial \over \partial \, t} \, 
\alpha_t \, (a)_{t=0} \right\rangle \qquad \forall \, P \in {\cal H}^1 \ , \ a \in 
{\cal U} ({\cal A}^1) \, . \label{eq23}
\end{equation}
Indeed $(\alpha_t)^t$ is a one parameter group of automorphisms of ${\cal H}^1$ 
such that
\begin{equation}
\alpha_t^t \, (\delta_n) = e^{nt} \, \delta_n \, . \label{eq24}
\end{equation}

%%XXX \special{src: 394 CK1.TEX} %Inserted by TeXtelmExtel

\noindent Now, using the Poincar\'e-Birkhoff-Witt theorem, we take the basis of 
${\cal U} ({\cal A}^1)$ given by the monomials,
\begin{equation}
Z_n^{a_n} \, Z_{n-1}^{a_{n-1}} \ldots Z_2^{a_2} \, Z_1^{a_1} \ , \ a_j \geq 0 \, . 
\label{eq25}
\end{equation}
To each $L \in {\cal U} ({\cal A}^1)^*$ one associates the formal power series
\begin{equation}
\sum \, {L (Z_n^{a_n} \ldots Z_1^{a_1}) \over a_n ! \ldots a_1 !} \, 
x_1^{a_1} \ldots x_n^{a_n} \, , \label{eq26}
\end{equation}
in the commuting variables $x_j$, $j \in \Nb$.

%%XXX \special{src: 409 CK1.TEX} %Inserted by TeXtelmExtel

\noindent It follows from \cite{Dix}~2.7.5 that we obtain in this way an 
isomorphism of the algebra of polynomials $P (\delta_1 , \ldots , \delta_n)$ on 
the algebra of polynomials in the $x_j$'s. To determine the formula for 
$\delta_n$ in terms of the $x_j$'s, we just need to compute
\begin{equation}
\langle \delta_n , Z_n^{a_n} \ldots Z_1^{a_1} \rangle \, . \label{eq27}
\end{equation}
Note that, by homogeneity, (\ref{eq27}) vanishes unless $\sum \, j \, a_j = n$.

%%XXX \special{src: 420 CK1.TEX} %Inserted by TeXtelmExtel

\noindent For $n=1$, we get
\begin{equation}
\rho \, (\delta_1) = x_1 \label{eq28}
\end{equation}
where $\rho$ is the above isomorphism.

%%XXX \special{src: 428 CK1.TEX} %Inserted by TeXtelmExtel

\noindent We determine $\rho \, (\delta_n)$ by induction, using the derivation
\begin{equation}
D(P) = \sum \, \delta_{n+1} \, {\partial \over \partial \, \delta_n} \, (P) 
\label{eq29}
\end{equation}
(which corresponds to $P \rightarrow [X,P]$).

%%XXX \special{src: 437 CK1.TEX} %Inserted by TeXtelmExtel

\noindent One has by construction,
\begin{equation}
\langle \delta_n , a \rangle = \langle \delta_{n-1} , D^t (a) \rangle \qquad \forall 
\, a \in {\cal U} ({\cal A}^1) \label{eq30}
\end{equation}
where $D^t$ is the transpose of $D$.

%%XXX \special{src: 446 CK1.TEX} %Inserted by TeXtelmExtel

\noindent By definition of $Z_n$ as a linear form (\ref{eq12}) one has,
\begin{equation}
D^t \, Z_n = Z_{n-1} \ , \ n \geq 2 \ , \ D^t \, Z_1 = 0 \, . \label{eq31}
\end{equation}
Moreover the compatibility of $D^t$ with the coproduct of ${\cal H}^1$ is given by
\begin{equation}
D^t (ab) = D^t (a) \, b + a \, D^t (b) + (\delta_1 \, a) \, \partial_t \, b 
\qquad \forall \, a,b \in {\cal U} ({\cal A}^1) \label{eq32}
\end{equation}
where $a \rightarrow \delta_1 \, a$ is the natural action of the algebra ${\cal 
H}^1$ on its dual
\begin{equation}
\langle P , \delta_1 \, a \rangle = \langle P \, \delta_1 , a \rangle \qquad \forall 
\, P \in {\cal H}^1 \, , \ a \in {\cal U} ({\cal A}^1) \, . \label{eq33}
\end{equation}

%%XXX \special{src: 464 CK1.TEX} %Inserted by TeXtelmExtel

\smallskip

%%XXX \special{src: 468 CK1.TEX} %Inserted by TeXtelmExtel

\noindent {\bf Lemma 4.} {\it When restricted to ${\cal U} ({\cal A}^2)$, 
$D^t$ is the unique derivation, with values in ${\cal U} ({\cal A}^1)$ 
satisfying (\ref{eq32}), moreover
\[
D^t (Z_n^{a_n} \ldots Z_2^{a_2} \, Z_1^{a_1}) = D^t (Z_n^{a_n} \ldots 
Z_2^{a_2}) \, Z_1^{a_1} + Z_n^{a_n} \ldots Z_2^{a_2} \, {a_1 (a_1 - 1) 
\over 2} \, Z_1^{a_1 - 1} \, .
\]
}

%%XXX \special{src: 480 CK1.TEX} %Inserted by TeXtelmExtel

\noindent {\bf Proof.} The equality $\Delta \, \delta_1 = \delta_1 \otimes 1 + 1 
\otimes \delta_1$ shows that $a \rightarrow \delta_1 \, a$ is a derivation of ${\cal 
U} ({\cal A}^1)$. One has $\delta_1 \, Z_n = 0$ for $n \ne 1$ so that $\delta_1 = 0$ 
on ${\cal U} ({\cal A}^2)$ and the first statement follows from (\ref{eq31}) and 
(\ref{eq32}). The second statement follows from,
\begin{equation}
D^t (Z_1^m) = {m (m-1) \over 2} \, Z_1^{m-1} \label{eq34}
\end{equation}
which one proves by induction on $m$ using (\ref{eq32}).~$\Box$

%%XXX \special{src: 492 CK1.TEX} %Inserted by TeXtelmExtel

\smallskip

%%XXX \special{src: 496 CK1.TEX} %Inserted by TeXtelmExtel

\noindent Motivated by the first part of the lemma, we enlarge the Lie algebra 
${\cal A}^1$ by adjoining an element $Z_{-1}$ such that,
\begin{equation}
[Z_{-1} , Z_n] = Z_{n-1} \qquad \forall \, n \geq 2 \, , \label{eq35}
\end{equation}
we then define $Z_0$ by
\begin{equation}
[Z_{-1} , Z_1] = Z_0 \ , \ [Z_0 , Z_k] = k \, Z_k \, . \label{eq36}
\end{equation}
The obtained Lie algebra $\cal A$, is the Lie algebra of formal vector 
fields with $Z_0 = x \, {\partial \over \partial \, x}$, $Z_{-1} = {\partial \over 
\partial \, x}$ and as above $Z_n = {x^{n+1} \over (n+1)!} \, {\partial \over 
\partial \, x}$.

%%XXX \special{src: 512 CK1.TEX} %Inserted by TeXtelmExtel

\noindent Let $\cal L$ be the left ideal in ${\cal U} ({\cal A})$ generated by 
$Z_{-1}$, $Z_0$,

%%XXX \special{src: 517 CK1.TEX} %Inserted by TeXtelmExtel

\smallskip

%%XXX \special{src: 521 CK1.TEX} %Inserted by TeXtelmExtel

\noindent {\bf Proposition 5.} {\it The linear map $D^t : {\cal U} ({\cal A}^1) 
\rightarrow {\cal U} ({\cal A}^1)$ is uniquely determined by the equality $D^t (a) = 
[Z_{-1} , a]$ {\rm mod} $\cal L$.}

%%XXX \special{src: 527 CK1.TEX} %Inserted by TeXtelmExtel

\smallskip

%%XXX \special{src: 531 CK1.TEX} %Inserted by TeXtelmExtel

\noindent {\bf Proof.} Let us compare $D^t$ with the bracket with $Z_{-1}$. By Lemma 
4, they agree on ${\cal U} ({\cal A}^2)$. Let us compute $[Z_{-1} , Z_1^m ]$. 
One has
\begin{equation}
[Z_{-1} , Z_1^m ] = {m (m-1) \over 2} \, Z_1^{m-1} + m \, Z_1^{m-1} 
\, Z_0 \, . \label{eq37}
\end{equation}
For each monomial $Z_n^{a_n} \ldots Z_1^{a_1}$ one has $D^t (a) - [Z_{-1} , a] \in 
{\cal L}$. Thus this holds for any $a \in {\cal U} ({\cal A}^1)$. Moreover, using 
the 
basis of ${\cal U} ({\cal A})$ given by the  
\[
Z_n^{a_n} \ldots Z_1^{a_1} \, Z_0^{a_0} \, Z_{-1}^{a_{-1}}
\]
we see that ${\cal U} ({\cal A})$ is the direct sum ${\cal L} \oplus {\cal U} ({\cal 
A}^1)$.~$\Box$

%%XXX \special{src: 550 CK1.TEX} %Inserted by TeXtelmExtel

\smallskip

%%XXX \special{src: 554 CK1.TEX} %Inserted by TeXtelmExtel

\noindent We now define a linear form $L_0$ on ${\cal U} ({\cal A})$ by
\begin{equation}
L_0 (Z_n^{a_n} \ldots Z_1^{a_1} \, Z_0^{a_0} \, Z_{-1}^{a_{-1}}) = 0 \ 
\hbox{unless} \ a_0 = 1 \, , \ a_j = 0 \quad \forall \, j \, , \label{eq38}
\end{equation}
and $L_0 (Z_0) = 1$.

%%XXX \special{src: 563 CK1.TEX} %Inserted by TeXtelmExtel

\smallskip

%%XXX \special{src: 567 CK1.TEX} %Inserted by TeXtelmExtel

\noindent {\bf Lemma 6.} {\it For any $n \geq 1$ one has}
\[
\langle \delta_n , a \rangle = L_0 ([ \underbrace{\ldots}_{n \, {\rm times}} 
[Z_{-1} , a ] \ldots ]) \qquad \forall \, a \in {\cal U} ({\cal A}^1) \, .
\]

%%XXX \special{src: 575 CK1.TEX} %Inserted by TeXtelmExtel

\smallskip

%%XXX \special{src: 579 CK1.TEX} %Inserted by TeXtelmExtel

\noindent {\bf Proof.} Let us first check it for $n=1$. We let $a = 
Z_n^{a_n} \ldots Z_1^{a_1}$. Then the degree of $a$ is $\sum \, j \, a_j$ 
and $L_0 ([Z_{-1} , a ]) \ne 0$ requires $\sum \, j \, a_j = 1$ so that 
the only possibility is $a_1 = 1$, $a_j = 0 \quad \forall \, j$. In this case 
one gets $L_0 ([Z_{-1} , Z_1]) = L_0 (Z_0) = 1$. Thus by (\ref{eq28}) we get the 
equality of Lemma 6 for $n=1$.

%%XXX \special{src: 588 CK1.TEX} %Inserted by TeXtelmExtel

\smallskip

%%XXX \special{src: 592 CK1.TEX} %Inserted by TeXtelmExtel

\noindent For the general case note first that $\cal L$ is stable under 
right multiplication by $Z_{-1}$ and hence by the derivation $[Z_{-1} , 
\cdot]$. Thus one has
\begin{equation}
(D^t)^n \, (a) = [Z_{-1} , \ldots [Z_{-1} , a] \ldots ] \ \hbox{mod} \ 
{\cal L} \qquad \forall \, a \in {\cal U} ({\cal A}^1) \, . \label{eq39}
\end{equation}
Now for $a \in {\cal L}$ one has $L_0 ([Z_{-1} , a ]) = 0$. Indeed writing
\[
a = ( Z_n^{a_n} \ldots Z_1^{a_1})(Z_0^{a_0} \, Z_{-1}^{a_{-1}}) = bc
\]
with $b \in {\cal U} ({\cal A}^1)$, $c = Z_0^{a_0} \, Z_{-1}^{a_{-1}}$, one has 
$[Z_{-1} , a] = [Z_{-1} , b] \, c + b \, [Z_{-1} , c]$.

%%XXX \special{src: 608 CK1.TEX} %Inserted by TeXtelmExtel

Since $b \in {\cal U} ({\cal A}^1)$ and $[Z_{-1} , c]$ has strictly negative degree 
one has $L_0 (b \, [Z_{-1} , c])$ $= 0$. Let $Z_n^{b_n} \ldots Z_1^{b_1} \, 
Z_0^{b_0}$ be a non zero component of $[Z_{-1} , b]$, then unless all 
$b_i$ are 0 it contributes by 0 to $L_0 ([Z_{-1} , b] \, c)$. But 
$[Z_{-1} , b] \in {\cal U} ({\cal A}^0)_0$ has no constant term. Thus one has
\begin{equation}
L_0 ([Z_{-1} , a]) = 0 \qquad \forall \, a = Z_n^{a_n} \ldots Z_1^{a_1} \, 
Z_0^{a_0} \, Z_{-1}^{a_{-1}} \label{eq40}
\end{equation}
except if all $a_j = 0$, $j \ne 1$ and $a_1 = 1$. $L_0 ([Z_{-1} , Z_1]) = 1$.

%%XXX \special{src: 621 CK1.TEX} %Inserted by TeXtelmExtel

\noindent Using (\ref{eq31}) one has $\langle \delta_n , a \rangle = \langle 
\delta_1 
, (D^t)^{n-1} \, (a) \rangle$ and the lemma follows.~$\Box$

%%XXX \special{src: 627 CK1.TEX} %Inserted by TeXtelmExtel

\smallskip

%%XXX \special{src: 631 CK1.TEX} %Inserted by TeXtelmExtel

\noindent One can now easily compute the first values of $\rho \, 
(\delta_n)$, $\rho \, (\delta_1) = x_1$, $\rho \, (\delta_2) = x_2 + {x_1^2 \over 
2}$, $\rho \, (\delta_3) = x_3 + x_2 \, x_1 + { x_1^3 \over 2}$, $\rho \, 
(\delta_4) = x_4 + x_3 \, x_1 + {2} \, x_2^2 + {2} \, x_2 \, 
x_1^2 + {3 \over 4} \, x_1^4$.

%%XXX \special{src: 639 CK1.TEX} %Inserted by TeXtelmExtel

\noindent The affine structure provided by the $\delta_n$ has the following 
compatibility with left multiplication in ${\cal U} ({\cal A}^1)$.

%%XXX \special{src: 644 CK1.TEX} %Inserted by TeXtelmExtel

\smallskip

%%XXX \special{src: 648 CK1.TEX} %Inserted by TeXtelmExtel

\noindent {\bf Lemma 7.} a) {\it One has $R_{n-1} = \sum \, 
R_{n-1}^k \otimes \delta_k$, $R_{n-1}^k \in {\cal H}_{n-1,0}$.} 

%%XXX \special{src: 653 CK1.TEX} %Inserted by TeXtelmExtel

\noindent b) {\it For fixed $a_0 \in {\cal U} ({\cal A}^1)$ there are $\lambda_n^k 
\in \Cb$ such that}
\[
\langle \delta_n , (a_0 \, a) \rangle = \langle \delta_n , a_0 \rangle \, 
\varepsilon 
(a) + \sum \lambda_n^k \, \langle \delta_k , a \rangle \, .
\]

%%XXX \special{src: 663 CK1.TEX} %Inserted by TeXtelmExtel

\smallskip

%%XXX \special{src: 667 CK1.TEX} %Inserted by TeXtelmExtel

\noindent {\bf Proof.} a) By induction using (\ref{eq7}). b) Follows, using 
$\lambda_n^k = \langle R_{n-1}^k , a_0 \rangle$.~$\Box$

%%XXX \special{src: 672 CK1.TEX} %Inserted by TeXtelmExtel

\smallskip

%%XXX \special{src: 676 CK1.TEX} %Inserted by TeXtelmExtel

\noindent The antipode $S$ in ${\cal U} ({\cal A}^1)$ is the unique antiautomorphism 
such that
\begin{equation}
S \, Z_n = - Z_n \qquad \forall \, n \, . \label{eq41}
\end{equation}
It is non trivial to express in terms of the coordinates $\delta_n$.

%%XXX \special{src: 685 CK1.TEX} %Inserted by TeXtelmExtel

\smallskip

%%XXX \special{src: 689 CK1.TEX} %Inserted by TeXtelmExtel

\noindent In fact if we use the basis $Z_j$ of ${\cal A}^1$ but in reverse 
order to construct the map $\rho$ we obtain a map $\widetilde{\rho}$ whose first 
values are $\widetilde{\rho} \, (\delta_1) = z_1$, $\widetilde{\rho} \, (\delta_2) = 
z_2 + {z_1^2 \over 2}$, $\widetilde{\rho} \, (\delta_3) = z_3 + 3 \, z_1 \, z_2 + {1 
\over 2} \, z_1^3$, $\widetilde{\rho} \, (\delta_4) = z_4 + {2} \, z_2^2 + 6 \, 
z_1 \, z_3 + {9} \, z_1^2 \, z_2 + {3 \over 4} \, z_1^4$.

%%XXX \special{src: 698 CK1.TEX} %Inserted by TeXtelmExtel

\noindent One has 
\[
\langle \delta_n , S \, (Z_m^{a_m} \ldots Z_1^{a_1}) \rangle = 
(-1)^{\sum a_j} \, \langle \delta_n , Z_1^{a_1} \ldots Z_m^{a_m} \rangle
\]
so that
\[
\rho \, (S^t \, \delta_n)= \sum \langle \delta_n , S \, (Z_m^{a_m} \ldots 
Z_1^{a_1}) \rangle \, x_1^{a_1} \ldots x_m^{a_m} =
\]
\[
= \sum (-1)^{\sum a_j} 
\, \langle \delta_n , Z_1^{a_1} \ldots Z_m^{a_m} \rangle \, x_1^{a_1} \ldots 
x_m^{a_m} = \widetilde{\rho} \, (\delta_n)
\]
with $z_j = -x_j$ in the latter expression. 

%%XXX \special{src: 717 CK1.TEX} %Inserted by TeXtelmExtel

\noindent Thus $\rho \, (S^t \, \delta_1) = -x_1$, $\rho \, (S^t \, \delta_2) = 
-x_2 + {x_1^2 \over 2}$, $\rho \, (S^t \, \delta_3) = -x_3 + 3 \, x_1 \, x_2 
- { x_1^3 \over 2}$, $\rho \, (S^t \, \delta_4) = -x_4 + {2} \, x_2^2 + 6 \, x_1 \, 
x_3 - {9} \, x_1^2 \, x_2 + {3\over 4} \, x_1^4$. We thus get
\begin{equation}
S^t \, \delta_1 = -\delta_1 \ , \ S^t \, \delta_2 = -\delta_2 + \delta_1^2 \ , \ S^t 
\, \delta_3 = -\delta_3 + 4 \, \delta_1 \, \delta_2 - 2 \, \delta_1^3 \ , \ \ldots 
\label{eq42}
\end{equation}

%%XXX \special{src: 729 CK1.TEX} %Inserted by TeXtelmExtel

The meaning of all the above computations and their relation to the standard 
calculus of Taylor expansions is clarified by the following theorem (\cite{CM}).

%%XXX \special{src: 734 CK1.TEX} %Inserted by TeXtelmExtel

\smallskip

%%XXX \special{src: 738 CK1.TEX} %Inserted by TeXtelmExtel

\noindent {\bf Theorem 8.} {\it Let $G_2$ be the group of formal diffeomorphisms of 
$\Rb$, of the form  $\psi (x) = \, x + \, o (x)$. For each $n$, let $\gamma_n$ be 
the 
functional on $G_2$ defined by,} 
\[
\gamma_n (\psi^{-1})= (\partial_x^n \log \psi' (x))_{x=0} \, . 
\]
{\it The equality $ \Theta (\delta_n) =\, \gamma_n$ determines a canonical 
isomorphism $\Theta$ of the Hopf algebra ${\cal H}^1$ with the Hopf algebra of 
coordinates on the group $G_2$.}

%%XXX \special{src: 751 CK1.TEX} %Inserted by TeXtelmExtel

\smallskip

%%XXX \special{src: 755 CK1.TEX} %Inserted by TeXtelmExtel

We refer to Theorem 8 of \cite{CM} for the proof, as well as for the more elaborate 
structure of the Hopf algebra ${\cal H}_T$. This theorem certainly shows that the 
antipode, i.e. the map $ \psi \rightarrow \psi^{-1}$ is certainly non trivial to 
compute. Note also that the expression $\sigma = \delta_2 - {1 \over 2} \, 
\delta_1^2$ is uniquely characterized by
\begin{equation}
\rho \, (\sigma) = x_2 \label{eq43}
\end{equation}
which suggests to define higher analogues of the Schwartzian as $\rho^{-1} (x_n)$.

%%XXX \special{src: 767 CK1.TEX} %Inserted by TeXtelmExtel

\section{The physics of renormalization and the Hopf algebra of rooted trees}

%%%%%%%%%%%%%
%%XXX \special{src: 771 CK1.TEX} %Inserted by TeXtelmExtel

In this section we want to motivate the Hopf algebra structure behind 
the process of renormalization in Quantum Field Theories (QFTs)
\cite{K} and show how relations to the Hopf algebra of the previous section
emerge.

%%XXX \special{src: 778 CK1.TEX} %Inserted by TeXtelmExtel

The renormalization procedure appears as the cure for the disease caused by the 
unavoidable presence of UV divergences in QFTs which describe the physics of local 
quantized fields. Such QFTs describe successfully all known particle physics 
phenomenology.

%%%%%%%
The point of departure of the renormalization procedure
is to alter the original Lagrangian
by an infinite series of counterterms labelled by Feynman graphs,
whose sole purpose is to cancel the UV-divergences coming
from  the presence of ill-defined 
integrals in the perturbative expansion of the theory.

Recall that the perturbative expansion of the functional integral
appears as a sum labelled by Feynman graphs $\Gamma$.
To each of these graphs
corresponds an integral $I_\Gamma$ which is in general ill-defined.
To compensate for the resulting infinities
one adds to the original Lagrangian $L_0$
which appears as the argument of the
exponential, an infinite series of counterterms  $\sum_\Gamma L_\Gamma$,
each  term in the series corresponding to a Feynman graph $\Gamma$.
The difficulty in finding the cut-off dependent counterterm
Lagrangian $\sum L_\Gamma$ comes only from the presence of ill-defined
subintegrations (usually dubbed subdivergences) in the integral
$I_\Gamma$. Indeed in the special case of a diagram
without subdivergences the counterterm is simply
(in the MS scheme) just the pole part of $I_\Gamma$.

As soon as subdivergences are present the extraction of $L_\Gamma$
is much more complicated since we want to take into account the previous
subtractions which is necessary to maintain locality
in the theory.

This obviously generates complicated combinatorial problems,
which for the first time, acquire mathematical meaning thanks to the
Hopf algebra ${\cal H}_R$.
 
%%XXX \special{src: 790 CK1.TEX} %Inserted by TeXtelmExtel

\subsection*{A Toy Model}

%%XXX \special{src: 794 CK1.TEX} %Inserted by TeXtelmExtel

It is possible to study the basic properties of the renormalization procedure
with the help of toy models, to which we now turn.

%%XXX \special{src: 799 CK1.TEX} %Inserted by TeXtelmExtel

In the following we will consider integrals of the form
\[
x(c):=\int_0^\infty \frac{1}{y+c}dy
\]
for $c>0$, which are to be regarded as functions of the parameter $c$.

%%XXX \special{src: 807 CK1.TEX} %Inserted by TeXtelmExtel

As it stands such an integral is ill-defined, due to its divergence at the upper 
boundary. Power counting reveals the presence of a logarithmic singularity,
and in this respect the integral  behaves  no better or worse than a logarithmic 
divergent integral in QFT, which one typically  confronts  due to
the presence of UV divergences in loop integrations.

%%XXX \special{src: 815 CK1.TEX} %Inserted by TeXtelmExtel

We will introduce a regularization,
\[
x(c)=\int_0^\infty  y^{-\epsilon}\frac{1}{y+c}dy,
\]
where $\epsilon$ is a small positive parameter.

%%XXX \special{src: 823 CK1.TEX} %Inserted by TeXtelmExtel

We now easily evaluate the above integral
\[
x(c)=B(\epsilon,1-\epsilon) c^{-\epsilon},
\]
where the presence of the pole term $\sim 
\Gamma(\epsilon)=\Gamma(1+\epsilon)/\epsilon$ indicates the {\em UV divergence}
in the integral \footnote{$B(\epsilon,1-\epsilon)=\Gamma(1+\epsilon)
\Gamma(1-\epsilon)/\epsilon$, $\Gamma(1+x)=\exp(-\gamma 
x)\exp(\sum_{j=2}^\infty\zeta(j)x^j/j)$, $\mid\!x\!\mid<1$.}.

%%XXX \special{src: 835 CK1.TEX} %Inserted by TeXtelmExtel

The process of renormalization demands the subtraction of this UV divergence,
and at this level we can straightforwardly proceed by a simple subtraction
\[
x(c)-x(1)=\int_0^\infty  y^{-\epsilon} \frac{(1-c)}{(y+c)(y+1)}dy
=B(\epsilon,1-\epsilon) (c^{-\epsilon}-1)
\]
which is evidently finite if we send $\epsilon\to 0$.

%%XXX \special{src: 845 CK1.TEX} %Inserted by TeXtelmExtel

Here, $-x(1)$ acts as the counterterm for the ill-defined
function $x(c)$, and the difference $x(c)-x(1)$ corresponds
to the renormalized function associated to $x(c)$.

%%XXX \special{src: 851 CK1.TEX} %Inserted by TeXtelmExtel

Physicists have good reason to demand that a counterterm like
$-x(1)$ above is independent of the external parameter $c$,
as to maintain locality in the theory. Before we explain this in more detail
we want to generalize this simple example to the presence of subdivergences.

%%XXX \special{src: 858 CK1.TEX} %Inserted by TeXtelmExtel

We consider
\begin{eqnarray*}
x_2(c) & := & \int_0^\infty\int_0^\infty  y_1^{-\epsilon}y_2^{-\epsilon}
\frac{1}{y_1+c}\frac{1}{y_2+y_1}dy_1 dy_2\\
 & = & \int_0^\infty  y_1^{-\epsilon}\frac{1}{y_1+c}x(y_1) dy_1.
\end{eqnarray*}

%%XXX \special{src: 867 CK1.TEX} %Inserted by TeXtelmExtel

We say that $x_2(c)$ has the function $x(y_1)$ as a subdivergence, but it still
is overall divergent itself.

%%XXX \special{src: 872 CK1.TEX} %Inserted by TeXtelmExtel

Powercounting reveals that there is a divergent sector when the $y_2$ 
integration variable tends to infinity for any fixed $y_1$, and when $y_1,y_2$ tend
to infinity jointly. There are no divergences when $y_2$ is kept fixed and $y_1$ 
tends to infinity, though. All the divergences are of logarithmic nature.

%%XXX \special{src: 879 CK1.TEX} %Inserted by TeXtelmExtel

Having successfully eliminated the divergence in the previous example
by a naive subtraction procedure, it is interesting to see if we can eliminate the 
divergences in $x_2(c)$ by subtracting $x_2(1)$:
\begin{eqnarray*}
x_2(c)-x_2(1) & = & 
\int_0^\infty  y_1^{-\epsilon}y_2^{-\epsilon}
\frac{(1-c)}{(y_1+c)(y_1+1)}\frac{1}{y_2+y_1}dy_1 dy_2\\
 & = & 
B(\epsilon,1-\epsilon)\int_0^\infty   y_1^{-2\epsilon}
\frac{(1-c)}{(y_1+c)(y_1+1)}dy_1\\
 & = & 
B(\epsilon,1-\epsilon)\left[ B(2\epsilon,1-2\epsilon)c^{-2\epsilon}
-B(2\epsilon,1-2\epsilon)\right]\\
 & = & 
-\frac{\log(c)}{\epsilon}+\mbox{finite terms}.
\end{eqnarray*}
Unfortunately, this expression still suffers from a divergence in the $y_2$ 
integration, and we were thus not successful with this naive attempt. 

%%XXX \special{src: 900 CK1.TEX} %Inserted by TeXtelmExtel

Actually, we find that the divergence is $\sim \log(c)$.
The parameter $c$ in our toy model is the remaining scale of 
the Green function. In realistic QFTs, this scale is furnished 
typically by an external momentum $q$, say, and divergences of the form
$\log(q^2)/\epsilon$ are non-local divergences: upon Fourier-transformation, they 
involve the logarithm of a differential operator, for example the logarithm of an 
external $q^2$ would translate as $log(\Box)$. Such terms can not be absorbed
by local counterterms, and are strictly to be avoided if one wants to remain
in the context of a local field theory. In the context of field theory,
locality restricts counterterms to be polynomial in momenta.

%%XXX \special{src: 913 CK1.TEX} %Inserted by TeXtelmExtel

Correspondingly, in the context of our toy model, we thus look for counter\-terms 
which
are at most polynomial in the parameter $c$.

%%XXX \special{src: 919 CK1.TEX} %Inserted by TeXtelmExtel

The failure above was twofold: the naive subtraction 
$-x_2(1)$ not only failed to render $x_2(c)$ finite, but also
this failure could only be absorbed by a non-local counterterm
$\sim \log(c)/\epsilon$. To find  a local counterterm, some more work is needed.

%%XXX \special{src: 926 CK1.TEX} %Inserted by TeXtelmExtel

Following the guidance of field theory we associate
to $x_2(c)$ (corresponding to a bare Green function) a function which has its 
subdivergences subtracted (a transition in field theory achieved by the 
$\overline{R}$ 
operation):
\begin{eqnarray*}
\overline{x_2}(c) & := & x_2(c)-x(c)x(1)\equiv
\int_0^\infty  y_1^{-\epsilon}y_2^{-\epsilon}
\frac{1}{y_1+c}\left(\frac{1}{y_2+y_1}-\frac{1}{y_2+1}\right)dy_1 dy_2\\
 &  = & 
B(\epsilon,1-\epsilon)\left[B(2\epsilon,1-2\epsilon)c^{-2\epsilon}-
 B(\epsilon,1-\epsilon)c^{-\epsilon}\right].
\end{eqnarray*}
Note that the subtraction term $-x(c)x(1)$ involves the
{\em counterterm} $-x(1)$ times the analytic expression,
$x(c)$, which we obtain from $x_2(c)$ when 
we set the subdivergence $x(y_1)$ in $x_2(c)$ to one.

%%XXX \special{src: 946 CK1.TEX} %Inserted by TeXtelmExtel

We  realize that $x_2^R(c)=\lim_{\epsilon\to 
0}[\overline{x_2}(c)-\overline{x_2}(1)]$ 
is a well-defined finite expression, the finite renormalized Green function 
$x_2^R(c)$, 
\begin{eqnarray*}
x_2^R(c) & = & \lim_{\epsilon\to 0}\left\{
B(\epsilon,1-\epsilon)\left[B(2\epsilon,1-2\epsilon)c^{-2\epsilon}-
 B(\epsilon,1-\epsilon)c^{-\epsilon}\right]\right.\\
 & & \left.-B(\epsilon,1-\epsilon)\left[B(2\epsilon,1-2\epsilon)-
 B(\epsilon,1-\epsilon)\right]\right\}\\
 & = &
\frac{1}{2}\log^2(c),
\end{eqnarray*}
and thus identify
\[
-\overline{x_2}(1)=
-B(\epsilon,1-\epsilon)[B(2\epsilon,1-2\epsilon)-
 B(\epsilon,1-\epsilon)]
\]
with the counterterm associated to $x_2(c)$.

%%XXX \special{src: 969 CK1.TEX} %Inserted by TeXtelmExtel

Note that $x_2^R(1)=0$, by construction. The renormalized Green function
$x_2^R(c)$ becomes a power series in $\log(c)$ (without constant term).
Note further that we can write an integral representation for it
which eliminates the necessity to introduce a regularization at all:
$$
x_2^R(c)=\int_0^\infty \int_0^\infty  
\left[\frac{1}{x+c}\left[\frac{1}{y+x}-\frac{1}{y+1}\right]-
\frac{1}{x+1}\left[\frac{1}{y+x}-\frac{1}{y+1}\right]\right]dydx.
$$
This could be directly obtained following the BPHZ approach,
and what we have just seen is the equivalence between on-shell
renormalization (subtraction at the on-shell value $c=1$)
and the BPHZ renormalization in the toy model.

%%XXX \special{src: 985 CK1.TEX} %Inserted by TeXtelmExtel

The above example shows how to find a local counterterm for an ill-defined integral
with ill-defined subintegrations. We first eliminated the ill-defined subintegration
by a counterterm, and then proceeded to construct the counter\-term
for the integral as a whole. In QFT one proceeds in the same manner.
A bare Green-function, given by an ill-defined integral, will suffer
from a plethora of ill-defined subintegrations in general.

%%XXX \special{src: 994 CK1.TEX} %Inserted by TeXtelmExtel

These subintegrations can be disjoint, nested or overlapping
\cite{Collins}. We will see later that the overlapping case resolves into the other 
ones. This result was effectively already obtained in \cite{K,habil,bdk},
and also known to others. An example how to resolve overlapping divergences in the 
case of $\phi^3$ theory in six dimensions will be given in an appendix.

%%XXX \special{src: 1002 CK1.TEX} %Inserted by TeXtelmExtel

Thus, we introduce at this stage a generalization of the above toy model allowing 
only for 
arbitrary nested or disjoint subdivergences. 

%%XXX \special{src: 1008 CK1.TEX} %Inserted by TeXtelmExtel

This motivates to generalize the example to functions $x_t(c)$ of an external 
parameter $c$, indexed by a rooted tree $t$, due to the fact that any
configurations of nested or disjoint subdivergences can be described by a rooted 
tree.
The formal definition of a rooted tree is postponed to the next section, while here 
we continue to gain experience in the treatment of functions having nested and 
disjoint subdivergences.

%%XXX \special{src: 1018 CK1.TEX} %Inserted by TeXtelmExtel

We define for a tree $t$ with $m$ vertices, enumerated such that the
root has number 1,
\[
x_t(c):=\int_0^\infty \frac{1}{y_1+c}\prod_{i=2}^m\frac{1}{y_i+y_{j(i)}} 
y_m^{-\epsilon}dy_m\ldots
y_1^{-\epsilon}dy_1,\;\forall c>0,
\]
where $j(i)$ is the number of the vertex to which the $i$-th vertex
is connected via its incoming edge.

%%XXX \special{src: 1030 CK1.TEX} %Inserted by TeXtelmExtel

We can write this as
\[
x_t(c):=\int  \frac{1}{y+c}\prod_{j=1}^r x_{t_j}(y) y^{-\epsilon}dy,
\]
if the root of $t$ connects to  $r$ trees $t_j$. Fig.(\ref{f1})
defines some simple rooted trees.  

%%XXX \special{src: 1039 CK1.TEX} %Inserted by TeXtelmExtel

Note that each vertex $v_i$ of the rooted tree corresponds
to an integration variable $x_i$, and that an edge connecting
$v_j$ to $v_i$ towards the root indicates that the $x_j$ integration
is nested in the $x_i$ integration. Integration variables which correspond to
vertices which are not connected by an edge correspond to disjoint
integrations.

%%XXX \special{src: 1048 CK1.TEX} %Inserted by TeXtelmExtel

\bookfig{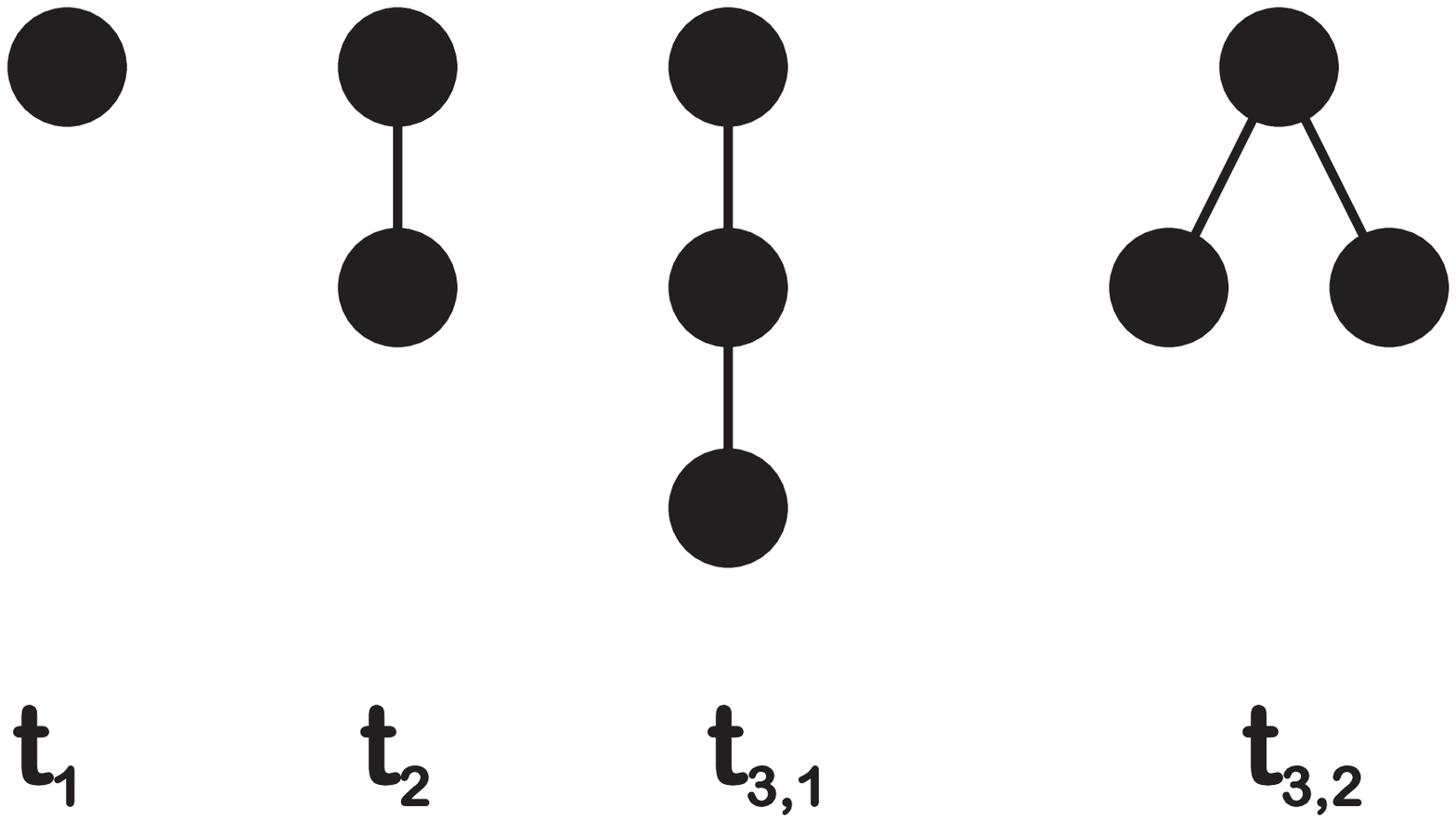}{Toy model}{f1}{A toy model realizing rooted trees.
We define the first couple of rooted trees $t_1,t_2,t_{3_1}, t_{3_2}$.
The root is always drawn as the uppermost vertex. $t_2$ gives rise to
the function $x_2(c)$.
}{4}

%%XXX \special{src: 1056 CK1.TEX} %Inserted by TeXtelmExtel

For the rooted trees defined in Fig.(\ref{f1}) we find the following
analytic expressions:
\begin{eqnarray*}
x_{t_1}(c) & = & \int_0^\infty \frac{y^{-\epsilon}}{y+c}dy,\\
x_{t_2}(c) & = & \int_0^\infty \frac{y^{-\epsilon}\;x_{t_1}(y)}{y+c}dy,\\
x_{t_{3_1}}(c) & = & \int_0^\infty \frac{y^{-\epsilon}\;x_{t_2}(y)}{y+c}dy,\\
x_{t_{3_2}}(c) & = & \int_0^\infty 
\frac{y^{-\epsilon}\;x_{t_1}(y)\;x_{t_1}(y)}{y+c}dy.
\end{eqnarray*}
Note that $x_2(c)\equiv x_{t_2}(c)$.
\subsection*{The Hopf algebra ${\cal H}_{R}$}
The previous remarks motivate  to introduce a Hopf algebra based on rooted trees.
We still postpone all formal definitions to  the next section and simply note that  
a 
{\em rooted tree} $t$ is a connected and simply-connected set of oriented edges and 
vertices such that there is precisely one distinguished vertex with no incoming 
edge. 
This vertex is called the root of $t$. Further, every edge connects two vertices and
the {\em fertility} $f(v)$ of a vertex $v$ is the number of edges outgoing from $v$.

%%XXX \special{src: 1078 CK1.TEX} %Inserted by TeXtelmExtel

We consider the algebra of polynomials over ${\bf Q}$ in rooted trees.

%%XXX \special{src: 1082 CK1.TEX} %Inserted by TeXtelmExtel

Note that for any rooted tree $t$ with root $r$ 
we have $f(r)$ trees $t_1$, $\ldots$,
$t_{f(r)}$ which are the trees attached to $r$.

%%XXX \special{src: 1088 CK1.TEX} %Inserted by TeXtelmExtel

Let $B_-$ be the operator which removes the root
$r$ from a tree $t$:
\begin{equation}
B_-: t\to B_-(t)=t_1 t_2\ldots t_{f(r)}.
\end{equation}
Fig.(\ref{guill-}) gives an example.
\bookfig{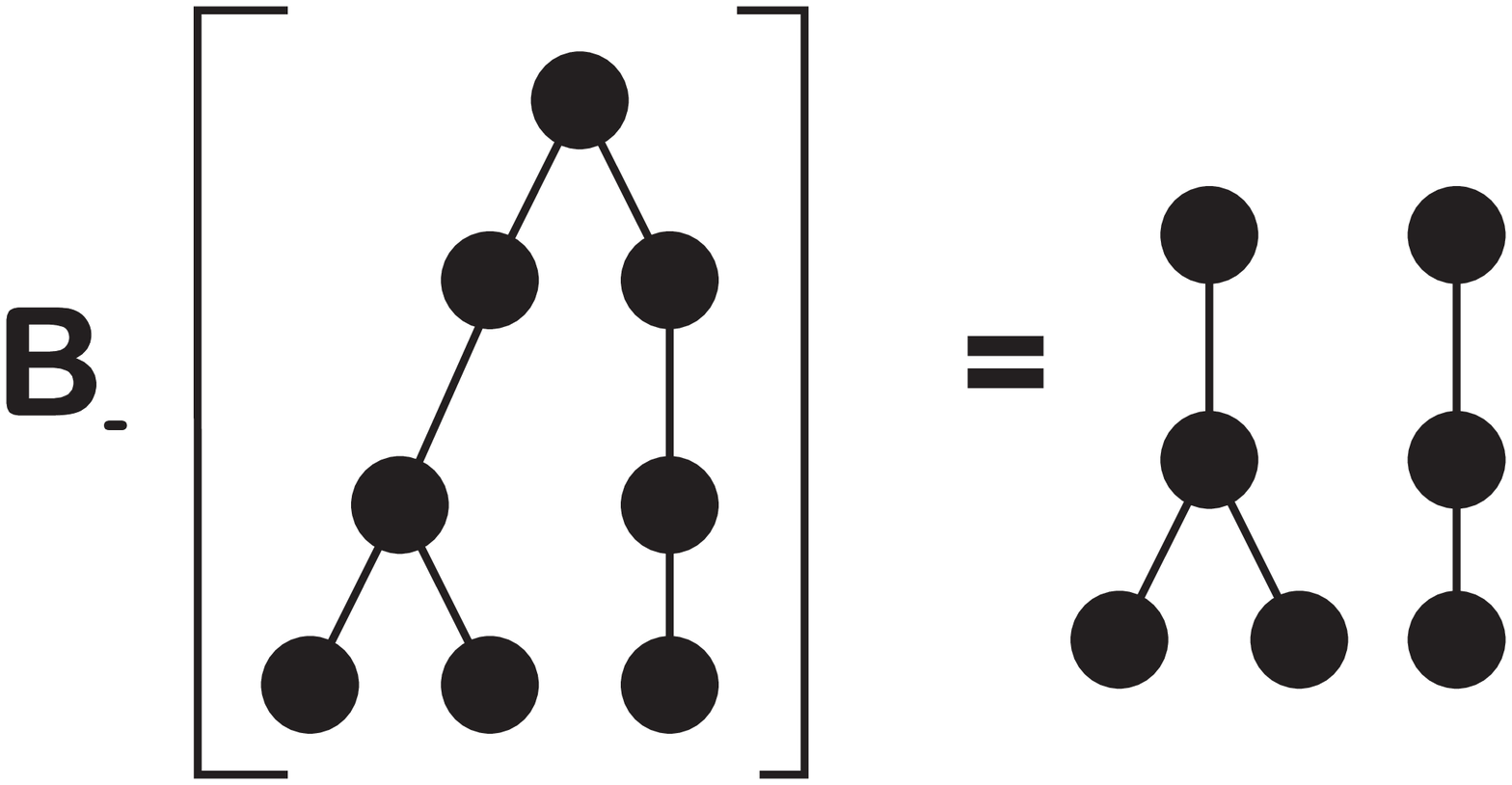}{$B_-$}{guill-}{The action
of $B_-$ on a rooted tree.}{3}

%%XXX \special{src: 1099 CK1.TEX} %Inserted by TeXtelmExtel

Let $B_+$ the operation which maps a monomial of $n$ rooted trees to a new rooted
tree $t$ which has a root $r$ with fertility $f(r)=n$ which connects to the $n$ 
roots 
of $t_1,\ldots,t_n$.
\begin{equation}
B_+: t_1\ldots t_n\to B_+(t_1\ldots t_n)=t.
\end{equation}
This is clearly the inverse to the action of $B_-$.

%%XXX \special{src: 1110 CK1.TEX} %Inserted by TeXtelmExtel

\bookfig{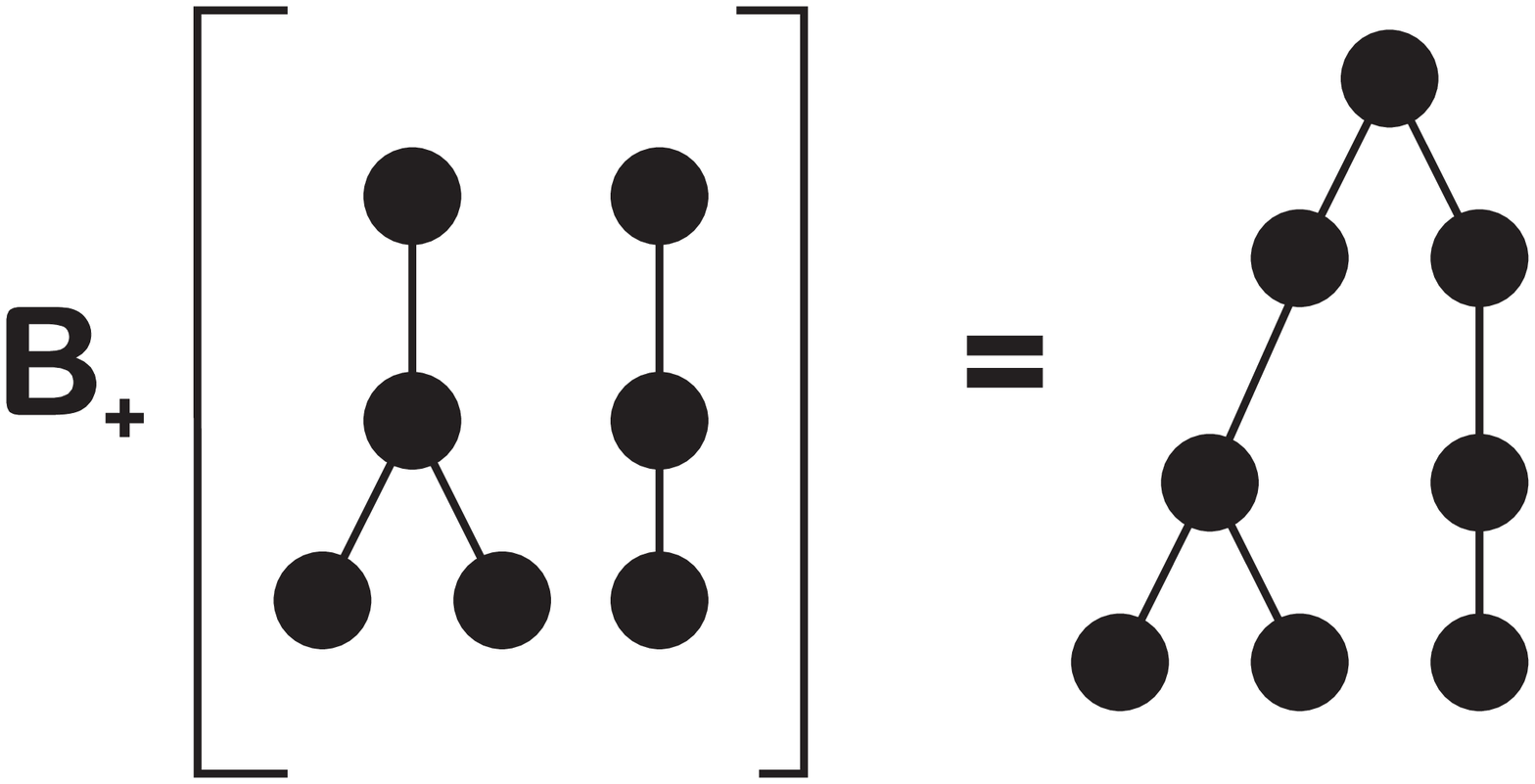}{$B_+$}{guill+}{The action
of $B_+$ on a monomial of  trees.}{3}

%%XXX \special{src: 1115 CK1.TEX} %Inserted by TeXtelmExtel

One has
\begin{equation}
B_+(B_-(t)))=B_-(B_+(t)))=t
\end{equation}
for any rooted tree $t$. Fig.(\ref{guill+}) gives an example.
We further set $B_-(t_1)=1$, $B_+(1)=t_1$.

%%XXX \special{src: 1124 CK1.TEX} %Inserted by TeXtelmExtel

We will introduce a Hopf algebra on such rooted trees by using the possibility
to cut such trees in pieces. We start with the most elementary possibility.
An {\em elementary cut} is a cut of a rooted tree at a single chosen edge, as 
indicated in Fig.(\ref{ecut}). We will formalize all these notions in the next 
section. By such a cutting procedure, we will obtain the possibility to
define a coproduct in a moment, as we can use the resulting pieces on either side 
of 
the coproduct.

%%XXX \special{src: 1135 CK1.TEX} %Inserted by TeXtelmExtel

\bookfig{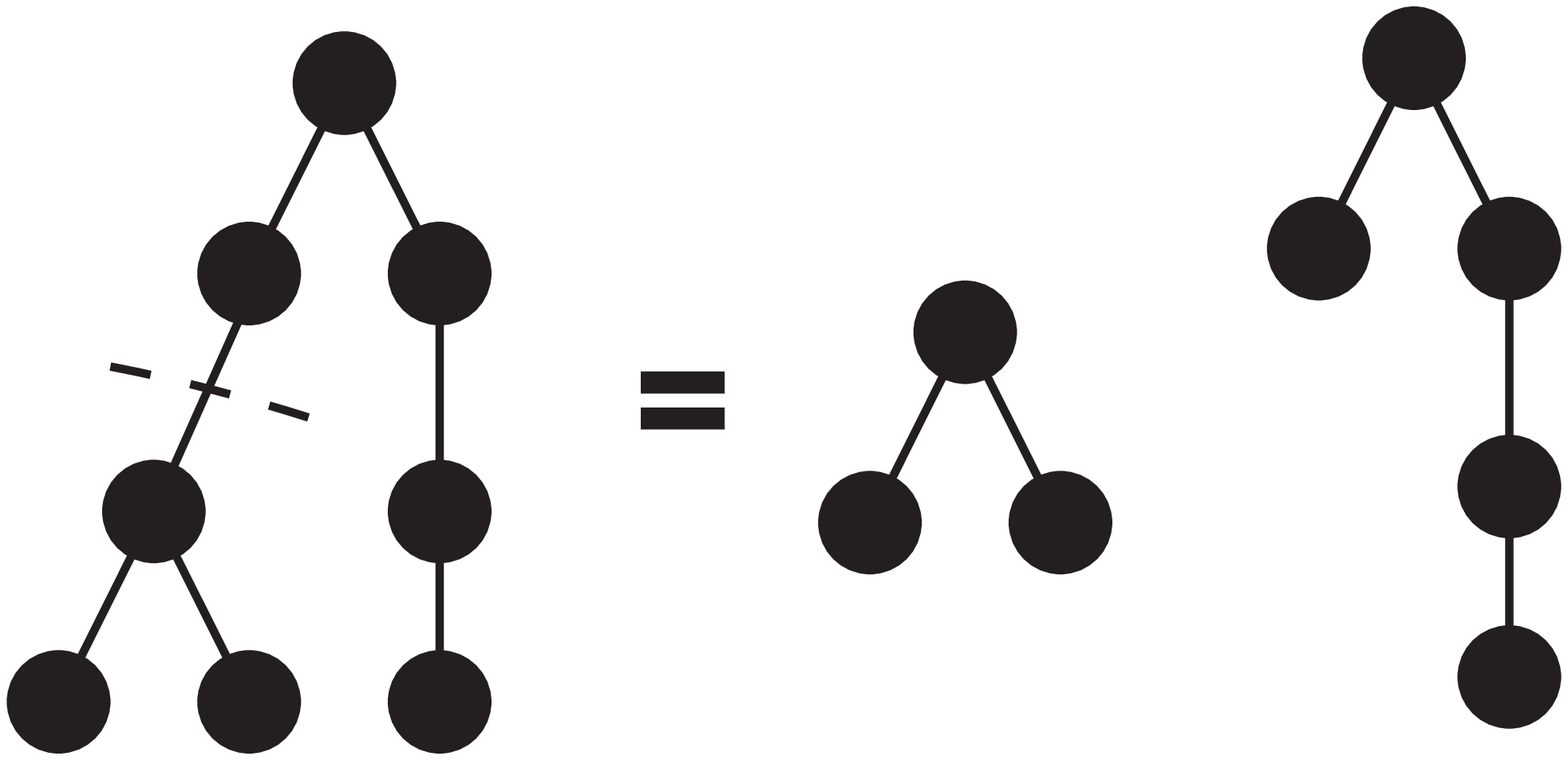}{Elementary cut.}{ecut}{An elementary cut splits a rooted
tree $t$ into two components $t_1,t_2$.}{4}

%%XXX \special{src: 1140 CK1.TEX} %Inserted by TeXtelmExtel

But before doing so we  finally introduce the notion of
an {\em admissible cut}, also called a {\em simple cut}. 
It is any assignment of elementary cuts to a rooted tree $t$ such that any path from
any vertex of the tree to the root has at most one elementary cut. Fig.(\ref{cut}) 
gives an example.

%%XXX \special{src: 1148 CK1.TEX} %Inserted by TeXtelmExtel

\bookfig{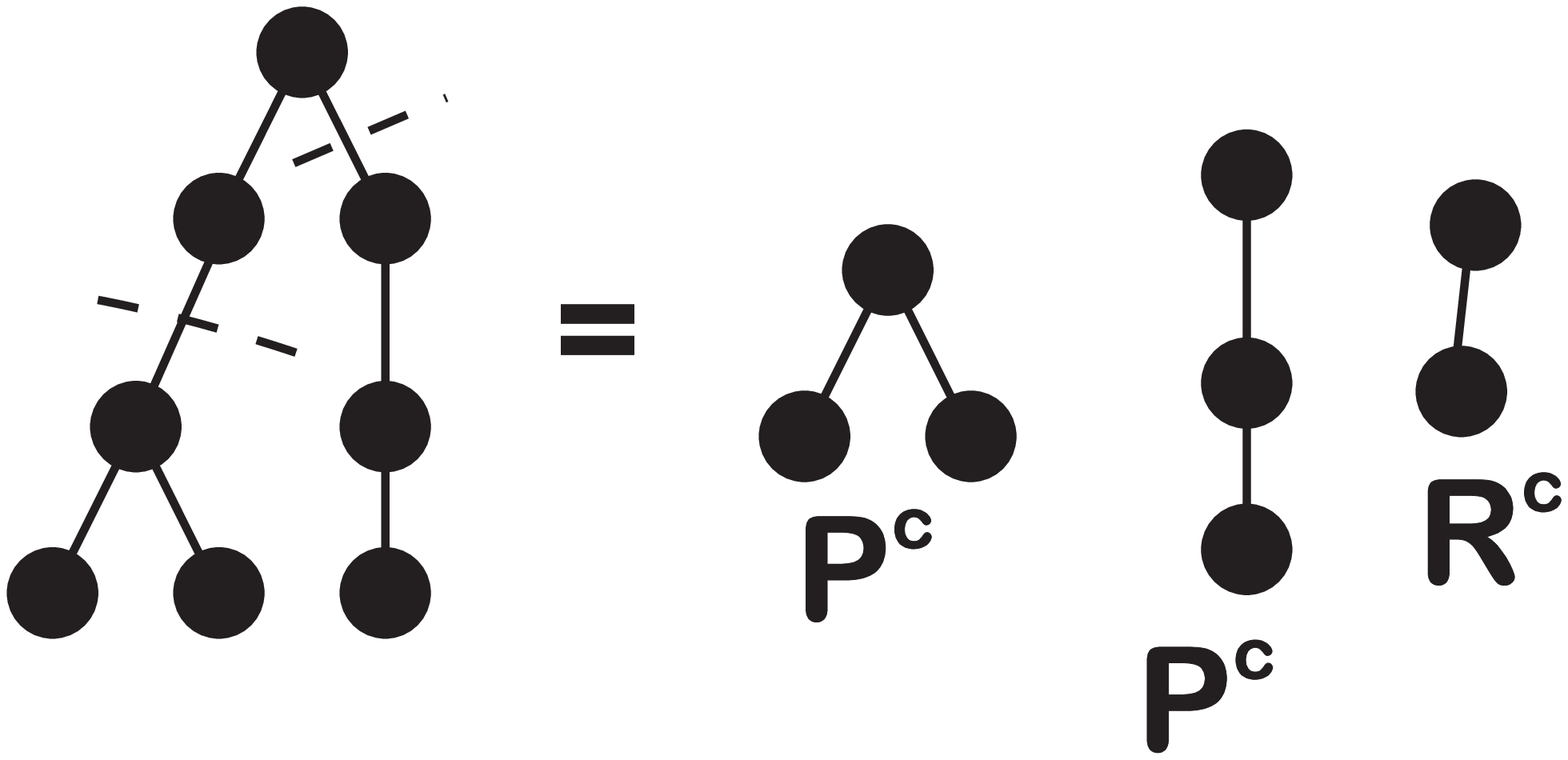}{An admissible cut.}{cut}{An
admissible cut $C$ acting
on a tree $t$. It produces a monomial of trees.
One of the factors, $R^C(t)$, contains the root of $t$.}{5}

%%XXX \special{src: 1155 CK1.TEX} %Inserted by TeXtelmExtel

An admissible cut $C$ maps a tree to a monomial in trees.
If the cut $C$ contains $n$ elementary cuts,
it induces a map
\begin{equation}
C: t\to C(t)=\prod_{i=1}^{n+1} t_{j_i}.
\end{equation}

%%XXX \special{src: 1164 CK1.TEX} %Inserted by TeXtelmExtel

Note that precisely one of these trees $t_{j_i}$
will contain the root of $t$.
Let us denote this distinguished tree
by $R^C(t)$. The monomial which is delivered by the
$n-1$ other factors is denoted by $P^C(t)$.

%%XXX \special{src: 1172 CK1.TEX} %Inserted by TeXtelmExtel

The definitions of $C,P,R$ can be extended to monomials of trees in the obvious 
manner, by choosing a cut $C^i$ for every tree $t_{j_i}$ in the monomial:
\begin{eqnarray*}
C(t_{j_1}\ldots t_{j_n}) & := & C^1(t_{j_1})\ldots C^n(t_{j_n}),\\
P^C(t_{j_1}\ldots t_{j_n}) & := & P^{C^1}(t_{j_1})\ldots
P^{C^n}(t_{j_n}),\\
R^C(t_{j_1}\ldots t_{j_n}) & := & R^{C^1}(t_{j_1})\ldots
R^{C^n}(t_{j_n}).
\end{eqnarray*}

%%XXX \special{src: 1184 CK1.TEX} %Inserted by TeXtelmExtel

We have now collected a sufficient amount of structure
to define a Hopf algebra on rooted trees.
Our aim is to see the correspondence between the Hopf algebra formulated
on rooted trees and the generation of a local counterterm for the functions
$x_t(c)$ introduced above, and finally to see the correspondence
between the Hopf algebra of rooted trees and the Hopf algebra of the previous section.

%%XXX \special{src: 1193 CK1.TEX} %Inserted by TeXtelmExtel

Before we define the Hopf algebra of rooted trees, 
we leave it as an exercise to the reader to convince himself
that any admissible cut in a rooted tree determines in the representation on
functions $x_t(c)$ a divergent subintegration, and that vice versa any 
divergent subintegration corresponds to an admissible cut.
For example, the single cut possible at $x_2(c)\equiv x_{t_2}(c)$
corresponds to the single divergent subintegration in this function.

%%XXX \special{src: 1203 CK1.TEX} %Inserted by TeXtelmExtel

Let us now establish the Hopf algebra structure. Following \cite{K} we define the 
counit and the coproduct. The {\em counit} $\epsilon$: ${\cal A} \to {\bf Q}$ is 
simple:
$$
\epsilon(X)=0
$$
for any $X\not= e$,
$$
\epsilon(e)=1.
$$

%%XXX \special{src: 1216 CK1.TEX} %Inserted by TeXtelmExtel

The {\em coproduct} $\Delta$
is defined by the equations
\begin{eqnarray}
\Delta(e) & = & e\otimes e\\
\Delta(t_1\ldots t_n) & = & \Delta(t_1)\ldots 
\Delta(t_n)\\
\Delta(t) & = & t \otimes e +(id\otimes B_+)[\Delta(B_-(t))],\label{cop2}
\end{eqnarray}
which defines the coproduct on trees with $n$ vertices
iteratively through the coproduct on trees with a lesser number of vertices.

%%XXX \special{src: 1229 CK1.TEX} %Inserted by TeXtelmExtel

The reader should work out the examples in Fig.(6) himself.
One checks coassociativity of $\Delta$ \cite{K}.
Also, we will give a formal proof in the next section.

%%XXX \special{src: 1235 CK1.TEX} %Inserted by TeXtelmExtel

%%XXX \special{src: 1238 CK1.TEX} %Inserted by TeXtelmExtel

\bookfig{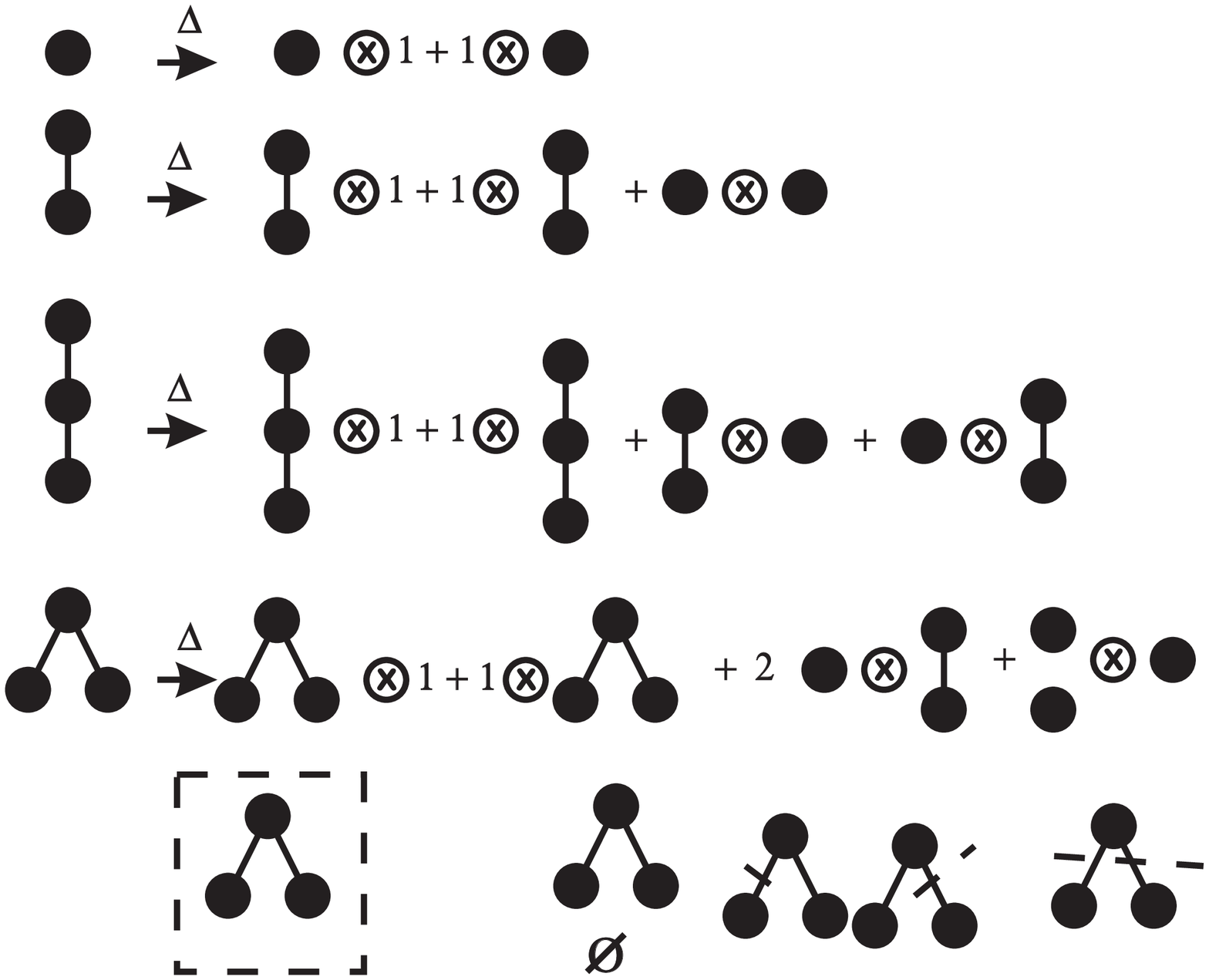}{The coproduct.}{cop}{The
coproduct. We work it out for the trees
$t_1,t_2,t_{3_1},t_{3_2}$.
For the latter, the last line gives 
explicitly the simple admissible cuts which were used
in the construction of the coproduct. The first two terms are generated
by the full admissible and the empty cut, while the last three
terms are generated by proper admissible cuts.}{8}

%%XXX \special{src: 1249 CK1.TEX} %Inserted by TeXtelmExtel

The following statement follows directly from the results in the next section,
but it is instructive to prove it here by elementary means to make contact with the 
previous section. We claim that the coproduct can be written as
\begin{equation}
\Delta(t)=e\otimes t+ t\otimes e+
\sum_{\mbox{\tiny adm.~cuts $C$ of $t$}}P^{C}(t)\otimes R^C(t).\label{cop1}
\end{equation}

%%XXX \special{src: 1259 CK1.TEX} %Inserted by TeXtelmExtel

\noindent {\bf Proof.} The result is true for the tree $t_1$ having only
one vertex. The induction is on the number of vertices.
We use that $B_-(t)$ has $n$ vertices if $t$ has $n+1$.
Thus,
\begin{eqnarray*}
\Delta(t) & = & t\otimes e + (id\otimes B_+)\Delta(B_-(t))\\
 & = & t\otimes id + (id\otimes B_+)\left(e\otimes B_-(t)
+B_-(t)\otimes e\right.\\
 & & \left. +
\sum_{\mbox{\tiny adm.~cuts $C$ of $B_-(t)$}}P^C(B_-(t))
\otimes R^C(B_-(t))\right)\\
 & = & t\otimes e + e \otimes B_+(B_-(t))
+B_-(t)\otimes \delta_1\\ 
 & & +
\sum_{\mbox{\tiny adm.~cuts $C$ of $B_-(t)$}}P^C(B_-(t))
\otimes B_+(R^C(B_-(t))))\\
 & = &
t\otimes e + e \otimes t\\ 
 & & +
\sum_{\mbox{\tiny adm.~cuts $C$ of $t$}}P^C(t)
\otimes R^C(t)).
\end{eqnarray*}
We used the fact that $B_+B_-=id$ and that
the only cut which distinguishes
$$
\sum_{\mbox{\tiny adm.~cuts $C$ of $B_-(t)$}}P^C(B_-(t))
\otimes B_+(R^C(B_-(t)))
$$
from
$$
\sum_{\mbox{\tiny adm.~cuts $C$ of $t$}}P^C(t)
\otimes R^C(t)
$$
is the cut which generates $B_-(t)\otimes \delta_1$. ~$\Box$

%%XXX \special{src: 1296 CK1.TEX} %Inserted by TeXtelmExtel

\smallskip

%%XXX \special{src: 1300 CK1.TEX} %Inserted by TeXtelmExtel

\noindent Note that the above formula can be streamlined.
$$
\Delta(t)=\sum{}^\prime_{\mbox{\tiny adm.~cuts $C$ of $t$}}P^C(t)\otimes 
R^C(t),
$$ 
where the primed sum indicates that we include the empty cut and the full admissible 
cut in the definition of admissible cuts in the manner indicated in Fig.(\ref{cop}).

%%XXX \special{src: 1310 CK1.TEX} %Inserted by TeXtelmExtel

Any cut corresponds to a choice of a subset of edges on the set 
$t^{(1)}$ of all edges of a given rooted tree $t$. The empty cut $C=\emptyset$ 
corresponds to the empty set in this sense. Thus, 
$$
P^\emptyset(t)=e,
$$ 
$$
R^\emptyset(t)=t.
$$

%%XXX \special{src: 1322 CK1.TEX} %Inserted by TeXtelmExtel

The full admissible cut $C_f$ is defined by  the complementary   result:
$$
P^{C_f}(t)=t,
$$ 
$$R^{C_f}(t)=e.
$$
It can be regarded as a cut $c$ on the one new edge
of $B_+(t)$, defined as the intersection $c=\{t^{(1)}\cap B_+(t)^{(1)}\}$.
Note that 
$$
P^{C_f}(t)\otimes R^{C_f}(t)=(id\otimes B_-)[P^c(B_+(t))\otimes 
R^c(B_+(t))],
$$
with the cut $c$ determined as above. In Fig.(\ref{cop}) we indicate the full cut 
$C_f(t_{3_2})$ by a dashed box around the rooted tree $P^{C_f}(t_{3_2})$.

%%XXX \special{src: 1340 CK1.TEX} %Inserted by TeXtelmExtel

The coproduct introduced here is linear in rooted trees in the right factor and 
polynomial in the left as one clearly
sees in Eq.(\ref{cop1}). This is a fundamental property shared with the coproduct of 
the previous section. We will explore this fact in some detail soon.

%%XXX \special{src: 1347 CK1.TEX} %Inserted by TeXtelmExtel

Up to now we have established a bialgebra structure. It is actually a Hopf algebra.
Following \cite{K} we find the antipode $S$ as
\begin{eqnarray}
S(e) & = & e\\
S(t) & = & -t-\sum_{\mbox{\tiny adm.~cuts $C$ of $t$}}S[P^C(t)]R^C(t),
\end{eqnarray}
and one immediately checks that
\begin{eqnarray}
m[(S\otimes id)\Delta(t)] & = &
t+S(t)+
\sum_{\mbox{\tiny adm.~cuts $C$ of $t$}}S[P^C(t)]R^C(t)\\
 & = & 0 = \epsilon(t) \, .
\end{eqnarray}
To show that $m[(id\otimes S)\Delta(t)]=0$ one uses induction
on the number of vertices \cite{K}.

%%XXX \special{src: 1365 CK1.TEX} %Inserted by TeXtelmExtel

We mentioned already that a cut on a tree $t$ is given by a subset of the set 
$t^{(1)}$ of the set of edges of $t$. So far, we allowed for a restricted class
of subsets, corresponding to admissible cuts. We actually enlarged the set
already and considered the set $B_+(t)^{(1)}$ of all edges of $B_+(t)$,
to construct the full admissible cut. We now consider all cuts corresponding to this 
set, that is all possible subsets of $B_+(t)^{(1)}$, including the empty set.
These subsets fall in two classes, one which contains the edge $c(t)=t^{(1)}\cap 
B_+(t)^{(1)}$, one which does not contain this edge.

%%XXX \special{src: 1376 CK1.TEX} %Inserted by TeXtelmExtel

Cuts corresponding to the first class we call full cuts, cuts not containing
this distinguished edge we call normal cuts. Thus, the empty cut is a normal cut. 
Non-empty normal cuts are also called proper cuts. Note that for a given normal cut 
$C\subset t^{(1)}$ and the corresponding full cut $\{C,c(t)\}=C\cap c(t)$ we have
$$
P^C(t)=P^{\{C,c(t)\}}(t),
$$
$$
R^C(t)=R^{\{C,c(t)\}}(t),
$$
while $n_{\{C,c(t)\}}=n_C+1$, where $n_C$ is the cardinality of the set $C$.

%%XXX \special{src: 1390 CK1.TEX} %Inserted by TeXtelmExtel

Let us give yet another formula to write the antipode, which one easily derives 
using 
induction on the number of vertices:
\begin{eqnarray*}
S(t) & = & \sum_{\mbox{\tiny all full cuts $C$ of $t$}}(-1)^{n_C}P^C(t)R^C(t).
\end{eqnarray*}

%%XXX \special{src: 1399 CK1.TEX} %Inserted by TeXtelmExtel

This time, we have a non-recursive expression, summing over all full cuts $C$, 
relaxing the restriction to admissible cuts. We introduced full cuts so that the 
overall sign agrees with the number of cuts employed.

%%XXX \special{src: 1405 CK1.TEX} %Inserted by TeXtelmExtel

Note that we have for all $t\not=e$
$$
m[(S\otimes id)\Delta(t)]
=\sum_{\mbox{\tiny all cuts $C$ of $t$}}(-1)^{n_C}P^C(t)R^C(t)=0=\epsilon(t),
$$
as each cut appears twice, either as a full cut or as a normal cut,
with opposite sign, due to the fact that the cardinality of $\{C,c(t)\}$
extends the cardinality of $C$ by one.

%%XXX \special{src: 1416 CK1.TEX} %Inserted by TeXtelmExtel

By now we have established a Hopf algebra on rooted trees.
It is instructive to calculate the coproduct and antipode
for some simple trees. Figs.(\ref{cop},\ref{ant}) give examples.
Note that in Fig.(\ref{ant}) we represented cuts as boxes. This is possible
in a unique way, as each cut on a simply connected rooted tree can be closed
in the plane to a box without further intersecting the tree and so that the root is 
in the exterior of the box.
%

%%XXX \special{src: 1426 CK1.TEX} %Inserted by TeXtelmExtel

At this time  we can make contact to the previous section.
We note that the sum of the two trees with three vertices
behaves under the coproduct as the element $\delta_3$ in
the first section. Defining $\delta_1:=t_1,\delta_2:=t_2,\delta_3:=
t_{3_1}+t_{3_2}$, we find
\begin{eqnarray*}
\Delta(\delta_1) & = & \delta_1\otimes e+e\otimes \delta_1\\
\Delta(\delta_2) & = & \delta_2\otimes e+e\otimes \delta_2+\delta_1\otimes\delta_1\\
\Delta(\delta_3) & = & \delta_3\otimes 
e+e\otimes\delta_3+3\delta_1\otimes\delta_2+\delta_2\otimes
\delta_1+\delta_1^2\otimes \delta_1,
\end{eqnarray*}
in accordance with Eq.(\ref{eq10}) in the previous section.
This is no accident, as we will soon see.

%%XXX \special{src: 1441 CK1.TEX} %Inserted by TeXtelmExtel

The reader should also check  the formulas 
for the coproduct and antipode on examples
himself.

%%XXX \special{src: 1447 CK1.TEX} %Inserted by TeXtelmExtel

\bookfig{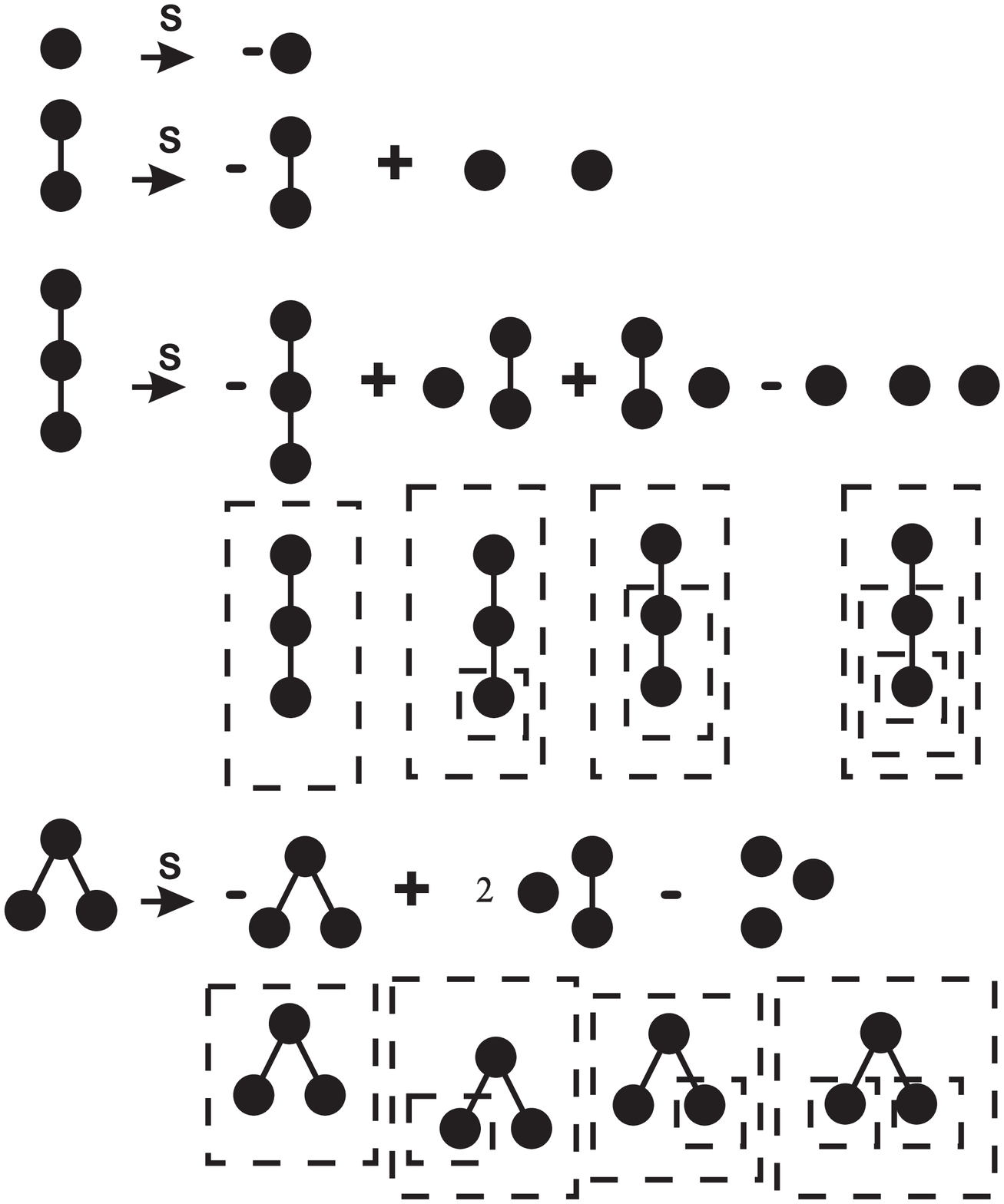}{The antipode.}{ant}{The antipode
for some simple trees.
Introducing full cuts, we find a very convenient way to express it
using boxes for cuts.
The sign for each term can be easily memorized as $(-1)^{n_c}$,
where $n_c$ is the number of full cuts.}{9}

%%XXX \special{src: 1456 CK1.TEX} %Inserted by TeXtelmExtel

It is now not difficult to employ this Hopf algebra to regain the local counterterms 
for the toy model. By construction, subdivergent sectors correspond to admissible 
cuts. Also, forests in the sense of renormalization theory are in one to one 
correspondence with arbitrary cuts, with full cuts corresponding to full forests, 
and 
normal cuts to normal forests. This allows to recover local counterterms from the 
formula for the antipode in our Hopf algebra. The recursive and the non-recursive 
manner to write the antipode give rise to two equivalent formulas for the local 
counterterm, as it is standard in renormalization theory \cite{Collins}. 

%%XXX \special{src: 1468 CK1.TEX} %Inserted by TeXtelmExtel

To see all this, note that any cut defines a natural bracket structure on $P^c(t)$. 
In a moment we will see how this fact allows to introduce various different 
renormalization schemes $R$.

%%XXX \special{src: 1474 CK1.TEX} %Inserted by TeXtelmExtel

It is instructive to come back to the toy model. Fig.(\ref{toyr}) summarizes how the 
standard notions of renormalization theory derive from the Hopf algebra of rooted 
trees. 

%%XXX \special{src: 1480 CK1.TEX} %Inserted by TeXtelmExtel

\bookfig{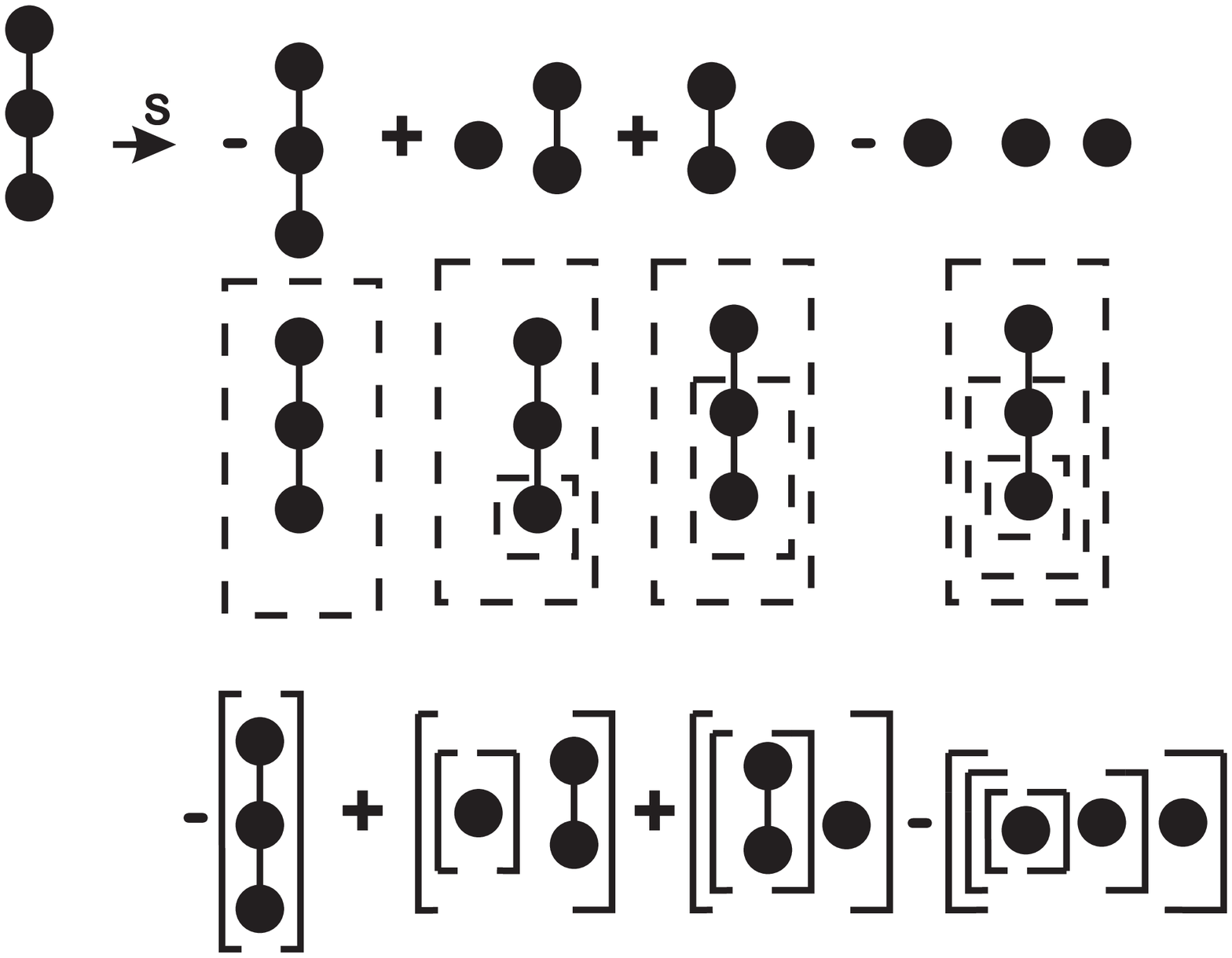}{Renormalization of the toy model.}{toyr}{Cuts induce
a bracket structure on trees.
Exploring this fact, the toy model can 
be easily renormalized
using the Hopf algebra structure of rooted trees.
}{7}

%%XXX \special{src: 1489 CK1.TEX} %Inserted by TeXtelmExtel

In Fig.(\ref{toyr}) we see the tree $t_{3_1}$, corresponding to the function
$x_{t_{3_1}}(c)$. The antipode
$$
S(t_{3_1})=-t_{3_1}+t_1 t_2+t_2 t_1-t_1 t_1 t_1
$$
derives from full cuts which induce the following bracket structure on toy model 
functions:
\[
-[x_{t_{3_1}}(c)]+[[x_{t_2}(c)]x_{t_1}(c)]+[[x_{t_1}(c)]x_{t_2}(c)]
-[[[x_{t_1}(c)] x_{t_1}(c)] x_{t_1}(c)].
\]
There is a certain freedom how to evaluate this bracket structure.
Such a freedom is always there in renormalization. It corresponds to a choice of 
renormalization map $R$ in the notation of \cite{K}, while in 
Collins textbook \cite{Collins} it corresponds to the choice of the map
$T$ which extracts the divergent part of a given expression.

%%XXX \special{src: 1508 CK1.TEX} %Inserted by TeXtelmExtel

As long as the evaluation of the bracket leaves the divergent part 
unchanged, it corresponds to a valid renormalization scheme,
and we obtaine the finite renormalized Green function from the consideration of
$
m[(S\otimes id)\Delta(t_{3_1})],
$
which gives rise to the following expression by summing over brackets
induced by full and normal cuts
\begin{eqnarray*}
x^R_{t_{3_1}}(c) & = & 
x_{t_{3_1}}(c)-[x_{t_2}(c)]x_{t_1}(c)-[x_{t_1}(c)]x_{t_2}(c)
+[[x_{t_1}(c)]x_{t_1}(c)]x_{t_1}(c)\\
 & - &[x_{t_{3_1}}(c)]+[[x_{t_2}(c)]x_{t_1}(c)]+[[x_{t_1}(c)]x_{t_2}(c)]
-[[[x_{t_1}(c)]x_{t_1}(c)]x_{t_1}(c)].
\end{eqnarray*}
It is not to difficult to check the finiteness of this expression
for the typical choices of renormalization schemes, on shell, 
$$
[x_t(c)]=x_t(1),
$$
minimal subtraction, 
$$
[x_t(c)]=Pole Part_\epsilon(x_t(1)),
$$
or BPHZ type schemes.

%%XXX \special{src: 1536 CK1.TEX} %Inserted by TeXtelmExtel

Note that the only reason that the renormalized function $x_t^R(c)$ does not vanish
identically is that full cuts involve one more bracket evaluation
than normal cuts. As the bracket evaluation respects the divergent part,
it is clear that functions $x_t^R(c)$ must be finite, due to the very fact
that $m[(S\otimes id)\Delta(t)]$ involves a sum over all cuts in pairs of 
normal cuts and associated full cuts. This gives rise to a sum of pairs of 
contributions, each pair being a difference $X-[X]$ between an analytic contribution
$X$ and its bracket evaluation. Thus, as long as the bracket evaluation
respects the divergent part of $X$, we will obtain a finite result for
$x^R_t(c)$.

%%XXX \special{src: 1549 CK1.TEX} %Inserted by TeXtelmExtel

Let us now turn away from toy models and address QFTs. By its very definition, pQFT 
deals with the calculation of Feynman diagrams. As such, it confronts the problem
of the presence of ultraviolet divergences in the diagrams. Hence, the diagrams 
refer 
to ill-defined analytic quantities. This is reflected by the fact that the analytic 
expressions provided by the diagrams become Laurent series in a regularization 
parameter. The presence of pole terms in this Laurent series then indicates the 
presence of UV-divergences in the first place. 

%%XXX \special{src: 1560 CK1.TEX} %Inserted by TeXtelmExtel

Equivalently, in the BPHZ spirit, we can Taylor expand the integrand in external 
momenta and would find that the first few terms are ill-defined quantities.

%%XXX \special{src: 1565 CK1.TEX} %Inserted by TeXtelmExtel

The art of obtaining meaningful physical quantities from these Laurent series is 
known as renormalization. It is in this process that we will find the Hopf algebra
structure realized.

%%XXX \special{src: 1571 CK1.TEX} %Inserted by TeXtelmExtel

Let us recall a few basic properties of renormalization.
Feynman diagrams consist of edges (propagators) and vertices
of different types. To each such type we can assign a weight,
and to a Feynman diagram we can assign an integral  weight
called {\em degree of divergence} which can be calculated
from the dimension of spacetime, the numbers of closed
cycles in the Feynman diagram, and the weights of its
propagators and vertices.

%%XXX \special{src: 1582 CK1.TEX} %Inserted by TeXtelmExtel

One finds that the analytic expression provided
by a Feynman diagram under consideration provides
UV-divergences if and only if its degree of divergence $\omega$
is $\geq 0$. One speaks of logarithmic, linear, quadratic,
$\ldots$ divergences for $\omega=0,1,2,\ldots$.

%%XXX \special{src: 1590 CK1.TEX} %Inserted by TeXtelmExtel

A Feynman diagram usually contains subdiagrams,
which have their own degree of divergence, and thus might provide
UV-divergent analytic expressions by themselves.
All these divergences are to be compensated by local
counterterms, which are to be calculated from a Feynman graph and its 
divergent subgraphs.

%%XXX \special{src: 1599 CK1.TEX} %Inserted by TeXtelmExtel

Fig.(\ref{exa2}) shows how the previous discussion extends
to a Feynman graph with one subdivergent graph.

%%XXX \special{src: 1604 CK1.TEX} %Inserted by TeXtelmExtel

\bookfig{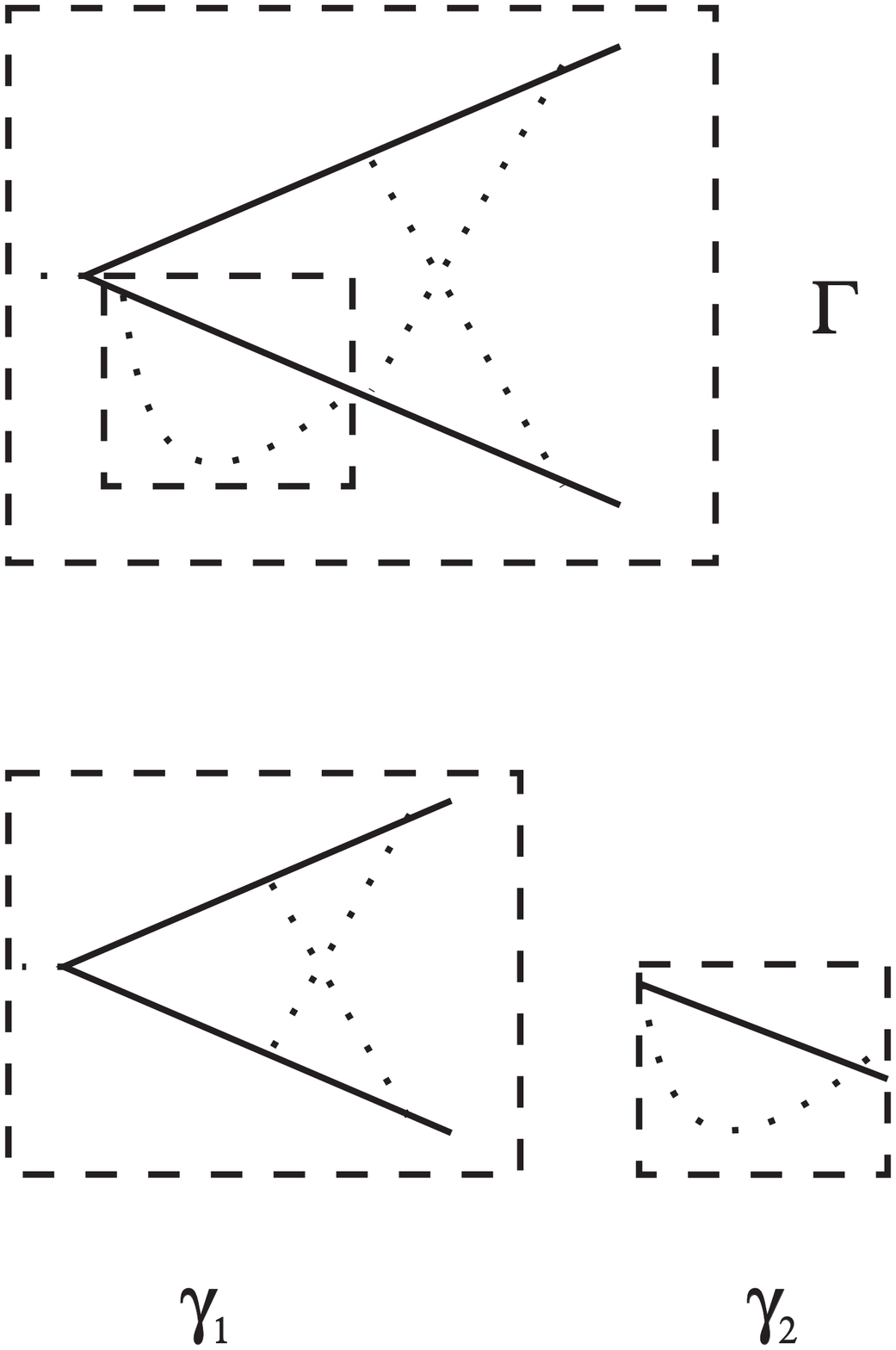}{A Feynman graph $\Gamma$ with a subdivergence.}{exa2}{This
Feynman graph $\Gamma$ behaves in the same manner as the toy function $x_2(c)$.
It contains a subdivergent graph $\gamma_2\subset \Gamma$, contained in the smaller 
box. This subdivergence sits in a graph $\gamma_1$ which we obtain if we shrink
$\gamma_2$ to a point in $\Gamma$, $\gamma_1=\Gamma/\gamma_2$.
To get a local counterterm, we follow the same steps as before:
we replace the divergent subgraph $\gamma_2$ by its subtracted 
renormalized form $\gamma_2^R=\gamma_2-[\gamma_2]$,
and can calculate the local counterterm 
for $\Gamma$ from the resulting expression.}{6}

%%XXX \special{src: 1617 CK1.TEX} %Inserted by TeXtelmExtel

It is the main result of \cite{K} that the transition from a bare Feynman 
diagram to its local counterterm, and to the renormalized Feynman graph, is 
described 
by a Hopf algebra structure. In this paper, we have changed 
the notation of \cite{K}
and formulate the Hopf algebra on rooted trees. A glance at Fig.(\ref{exa}) shows 
how 
to assign a rooted tree to a Feynman graph with subdivergences. This is possible
in a unique manner as long as all subdivergences of the
Feynman graph are either disjoint or nested
\cite{K}. In such circumstances, we can associate a unique rooted tree
to the graph. If the subdivergences are overlapping, the renormalization
will nevertheless follow the combinatorics dictated by the Hopf algebra of rooted trees.
In such circumstances, a Feynman graph corresponds to a sum of rooted trees \cite{habil,K}.
We will comment on this fact in an appendix.

%%XXX \special{src: 1635 CK1.TEX} %Inserted by TeXtelmExtel

%%XXX \special{src: 1638 CK1.TEX} %Inserted by TeXtelmExtel

Each Feynman diagram $\Gamma$ furnishes a tree whose vertices are decorated
by Feynman graphs $\gamma\subset\Gamma$ which are free of subdivergences themselves.
These decorations correspond to the letters in the parenthesized words of \cite{K}.

%%XXX \special{src: 1644 CK1.TEX} %Inserted by TeXtelmExtel

Note that the set of all admissible cuts is by construction in one to one
correspondence with the set of all superficially divergent subgraphs of $\Gamma$.
Further the set of all cuts is in one to one correspondence with the
set of all forests in the sense of renormalization theory.

%%XXX \special{src: 1651 CK1.TEX} %Inserted by TeXtelmExtel

This is the main result of \cite{K}. The antipode of the Hopf algebra ${\cal 
A}_{QFT}$ delivers the $Z$-factor of a Feynman diagram.

%%XXX \special{src: 1656 CK1.TEX} %Inserted by TeXtelmExtel

\bookfig{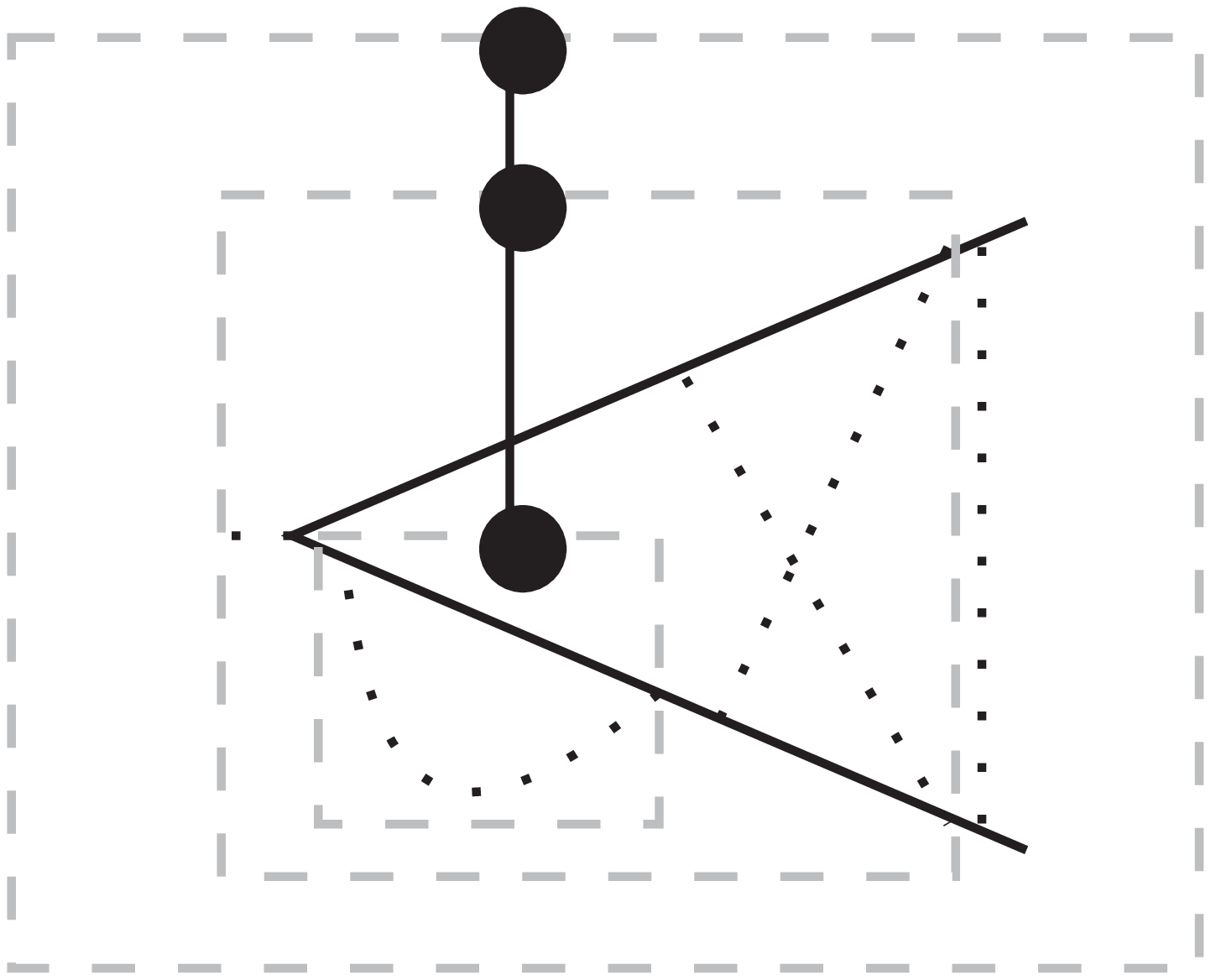}{Rooted trees from Feynman
diagrams.}{exa}{Rooted trees from Feynman diagrams.
They are in one-to-one correspondence
with the parenthesized words of \cite{K}.
In this example it is the rooted tree $t_{3_1}$ which is associated to the
configuration of subdivergences given by the graph.}{6}

%%XXX \special{src: 1665 CK1.TEX} %Inserted by TeXtelmExtel

At this stage, we can summarize the results in \cite{K} using the language of 
the Hopf algebra of rooted trees. This is done in Fig.(\ref{renormalization}).
Note that we even do not have to specify the renormalization scheme, but that the
cuts used in the definition of the antipode on rooted trees extend to forests,
so that we can apply any chosen renormalization prescription to evaluate
the content
of these forests. Thus, as before in the toy model, each cut corresponds to the 
operation $T$ in Collins book \cite{Collins}, some operation which extracts the 
divergence of the expression on which it acts. As we mentioned already, this 
operation was called $R$ in \cite{K}.

%%XXX \special{src: 1678 CK1.TEX} %Inserted by TeXtelmExtel

\bookfig{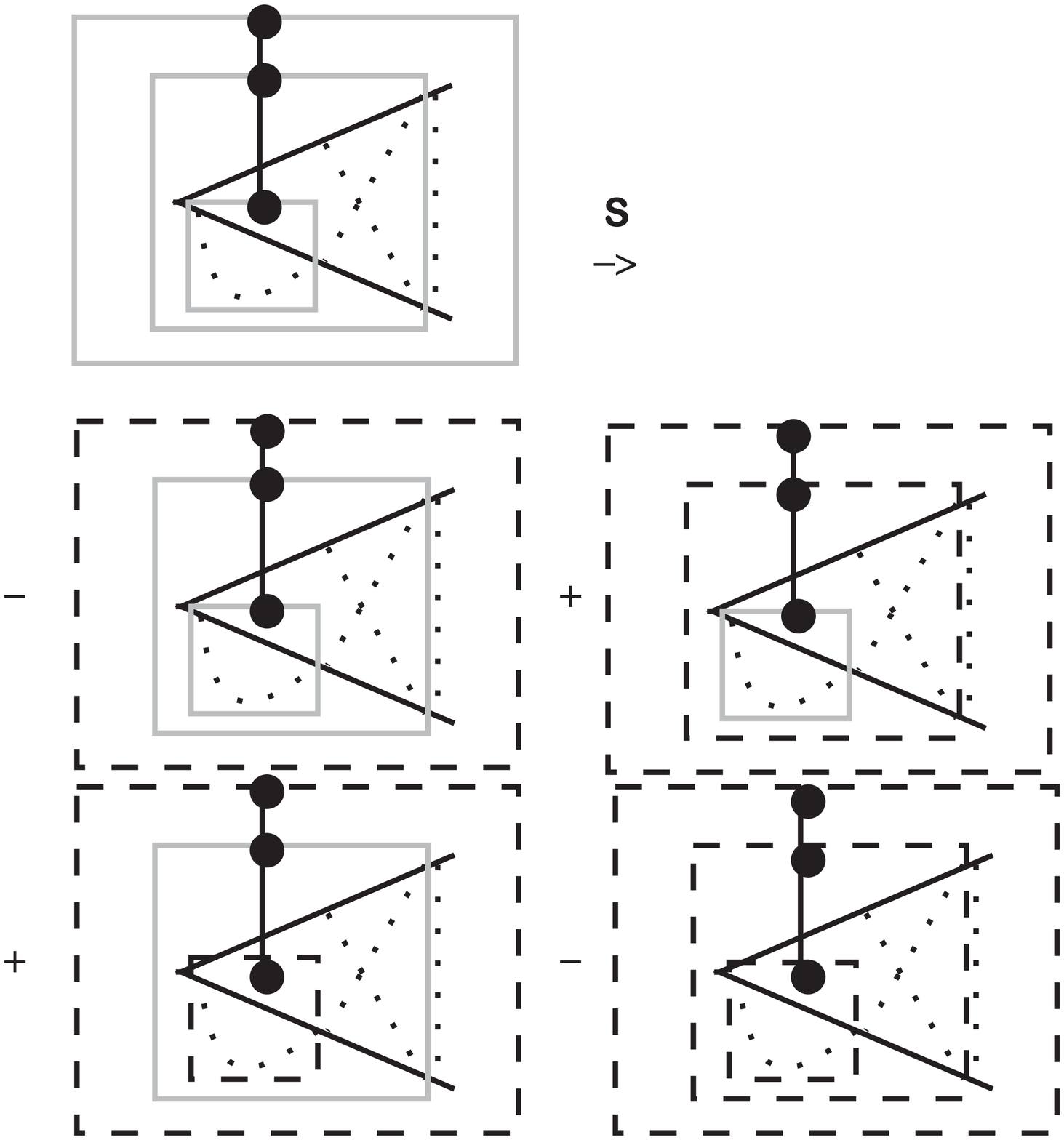}{Renormalization using the Hopf algebra of
rooted trees.}{renormalization}{The steps involved in the
process of renormalization are governed by the Hopf algebra on rooted
trees, as this figure clearly exhibits. We indicate how the Hopf algebra of rooted 
trees acts on the Feynman diagrams. Subgraphs are indicated by grey rectangles and 
determine the tree structure, forests corresponding to cuts generated by the 
antipode 
of the associated rooted tree are given as dashed black rectangles.}{10}

%%XXX \special{src: 1689 CK1.TEX} %Inserted by TeXtelmExtel

At this point, we succeeded in deriving the renormalization procedure from the Hopf 
algebra of rooted trees. The attentive reader will have noticed that we ignored 
overlapping divergences in our discussion. As promised,
we will discuss them in an appendix, 
where 
it is shown how to assign a unique sum of rooted trees to any graph containing 
overlapping divergences, to which then our previous considerations apply.

%%XXX \special{src: 1699 CK1.TEX} %Inserted by TeXtelmExtel

For us, these considerations of renormalization and the underlying Hopf algebra of 
rooted trees are sufficient motivation to get interested in this Hopf algebra. We 
will continue our exploration of this subject by showing how it relates to the Hopf 
algebra of the first section. We saw some of these relations already,
and now continue to make this more precise.
\clearpage

%%XXX \special{src: 1708 CK1.TEX} %Inserted by TeXtelmExtel

\clearpage
\section{The relation between ${\cal H}_{R}$ and ${\cal H}_{T}$}

%%XXX \special{src: 1712 CK1.TEX} %Inserted by TeXtelmExtel

Recall the relations
\begin{eqnarray*}
\Delta(\delta_1) & = & \delta_1\otimes e+e\otimes \delta_1\\
\Delta(\delta_2) & = & \delta_2\otimes e+e\otimes \delta_2+\delta_1\otimes\delta_1\\
\Delta(\delta_3) & = & \delta_3\otimes 
e+e\otimes\delta_3+3\delta_1\otimes\delta_2+\delta_2\otimes
\delta_1+\delta_1^2\otimes \delta_1,
\end{eqnarray*}
which indicate an intimate connection to the Hopf algebra
${\cal H}_T$ introduced in the first section.
To find the general relation between the two Hopf algebras under consideration we
first introduce naturally grown forests $\delta_k$.

%%XXX \special{src: 1717 CK1.TEX} %Inserted by TeXtelmExtel

To this end, we consider an operator $N$ which maps a tree $t$ with
$n$ vertices to a sum $N(t)$ of $n$ trees $t_i$, each having $n+1$
vertices, by attaching one more outgoing edge and vertex to each vertex of $t$, as 
in 
Fig.(\ref{ng}). The root remains the same in this operation.

%%XXX \special{src: 1725 CK1.TEX} %Inserted by TeXtelmExtel

\bookfig{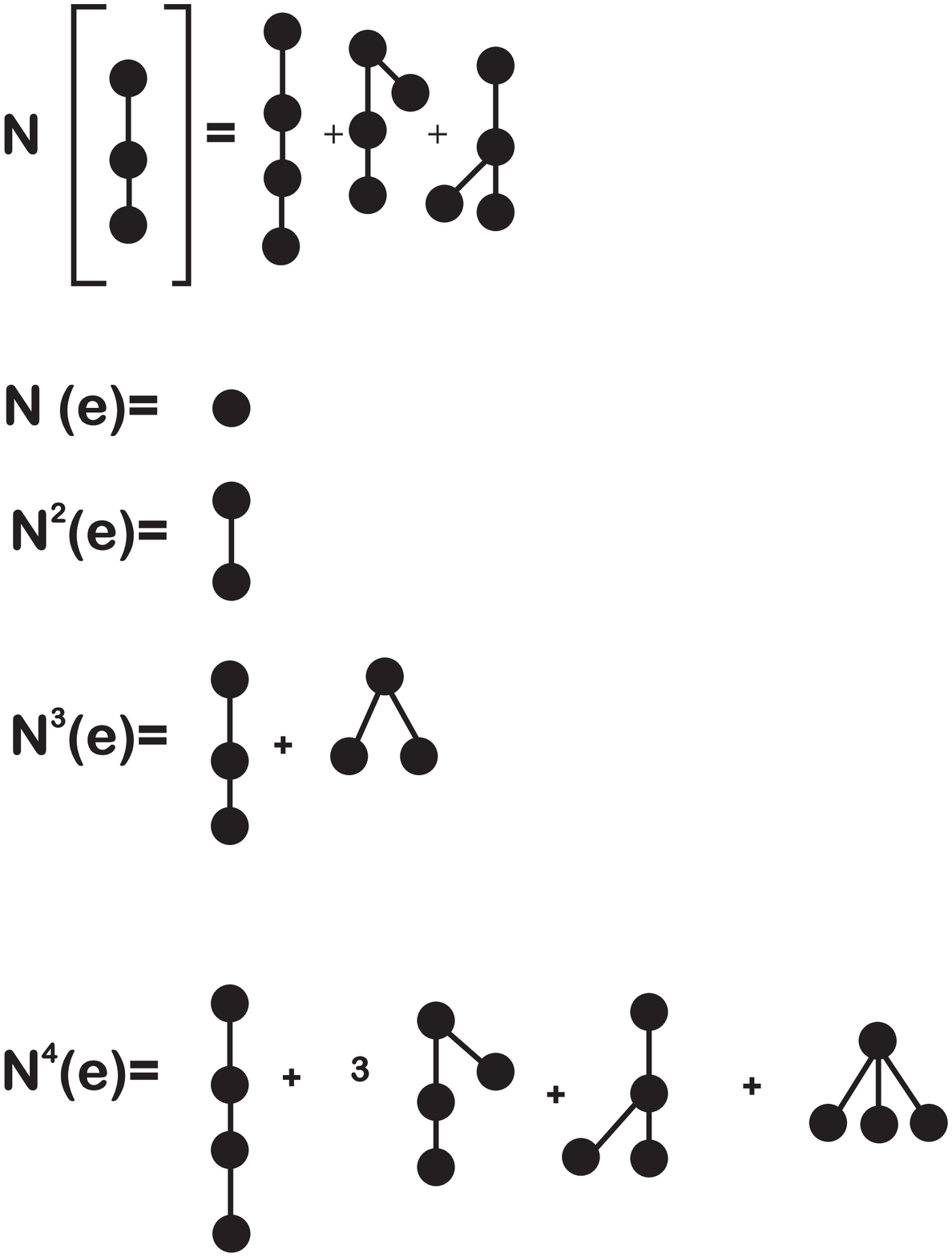}{the operator $NG$.}{ng}{The operator $N$
and the elements $\delta_k$.}{7}

%%XXX \special{src: 1730 CK1.TEX} %Inserted by TeXtelmExtel

Now we define
\begin{equation}
\delta_k:=N^k(e)
\end{equation}
so that $\delta_{k+1}=N(\delta_k)$. On products of trees $N$ will act as a derivation,
comparable to the derivation $D$ introduced in Eq.(\ref{eq29}).

%%XXX \special{src: 1739 CK1.TEX} %Inserted by TeXtelmExtel

In Fig.(\ref{ng}) we see the first few elements $\delta_k$.
Note that there are non-trivial multiplicities as in $\delta_4$.

%%XXX \special{src: 1744 CK1.TEX} %Inserted by TeXtelmExtel

Let 
$$[X,\delta_n]=\delta_{n+1},$$
$$[Y,\delta_n]=n\delta_n.$$
The following result, which is a trivial consequence of the results in the next 
section, initiated this paper:
\begin{itemize}
\item[i)] With the coproduct of ${\cal H}_{R}$,
the $\delta_k$ span a closed Hopf subalgebra of ${\cal H}_{R}$.
\item[ii)]
\begin{eqnarray}
\Delta(\delta_n) & = & e\otimes \delta_n+\delta_n \otimes e
+R_{n-1}\\
R_0 & = & 0\\
R_1 & = & \delta_1\otimes \delta_1\\
R_k & = & [X\otimes e+ e\otimes X,R_{k-1}]+
k\delta_1 \otimes \delta_k+[\delta_1\otimes Y,R_{k-1}].
\end{eqnarray}
\end{itemize}
The proof follows from the results in the next section, but it is instructive
to investigate directly  the compatibility of the operation of natural growth
and the notion of an admissible cut: We note that $\delta_n$ is a sum of trees: 
$\delta_n=\sum_i t_i$, say. Thus,
\begin{equation}
\Delta(\delta_n)=e\otimes\delta_n+\delta_n\otimes e
+\sum_i \sum_{\mbox{\tiny all cuts $C^i$ of $t_i$}}
P^{C^i}(t_i)\otimes R^{C^i}(t_i).
\end{equation}

%%XXX \special{src: 1774 CK1.TEX} %Inserted by TeXtelmExtel

Hence we can write, with the same $t_i$ as before,
\begin{eqnarray}
\Delta(\delta_{n+1}) & = & e\otimes \delta_{n+1}+
\delta_{n+1} \otimes e \nonumber\\
 & & +\sum_i \sum_{\mbox{\tiny all cuts $C^i$ of $t_i$}}\left\{
N[P^{C^i}(t_i)]\otimes R^{C^i}(t_i)\right.\nonumber\\
 & & \left. + (P^{C^i}(t_i)\otimes N[R^{C^i}(t_i)]\right\}\nonumber\\
 & & +n\delta_1\otimes \delta_n\nonumber\\
 & & 
+\sum_i \sum_{\mbox{\tiny all cuts $C^i$ of $t_i$}}
l[R^{C^i}(t_i)]\delta_1 P^{C^i}(t_i)\otimes R^{C i}(t_i),
\end{eqnarray}
where l(t) gives the number of vertices of a tree $t$.
Thus, we decomposed the cuts at the components of $\delta_{n+1}$
in four classes: either the edge to the new grown vertex
is not cut, then we will have natural growth on either
the former $P^C$ or $R^C$ part. Thus, the first two contributions
deliver the operator $N$ on either side of the tensorproduct.
Or, for the remaining two cases, the edge to the new grown vertex is cut.
These cases will always have a factor $\delta_1$ on the lhs of the tensorproduct.

%%XXX \special{src: 1797 CK1.TEX} %Inserted by TeXtelmExtel

In these cases, it either was grown from  the former $R^C$ part (admissibility of cuts
forbid that it was  grown from the $P^C$ part), or it was grown from the whole uncut former
$\delta_n$, which gives the term $n\delta_1\otimes \delta_n$.

%%XXX \special{src: 1803 CK1.TEX} %Inserted by TeXtelmExtel

Hence we have decomposed the cuts possible at the trees 
of $\delta_{n+1}$ in terms of the cuts at the
trees of $\delta_n$.

%%XXX \special{src: 1809 CK1.TEX} %Inserted by TeXtelmExtel

Fig.(\ref{instrr}) gives an instructive example.
To finally prove the result, we note the following identities
\begin{eqnarray}
N(\delta_{i_1}\ldots\delta_{i_k}) & = & [X,\delta_{i_1}\ldots\delta_{i_k}]\\
   l(\delta_k)\delta_k & = & k \delta_k=[Y,\delta_k],
\end{eqnarray}
where we note that $l[\delta_k]=k$ is well-defined,
as $\delta_k$ is a homogenous combination of trees
with $k$ vertices.~$\Box$

%%XXX \special{src: 1821 CK1.TEX} %Inserted by TeXtelmExtel

\smallskip

%%XXX \special{src: 1825 CK1.TEX} %Inserted by TeXtelmExtel

\bookfig{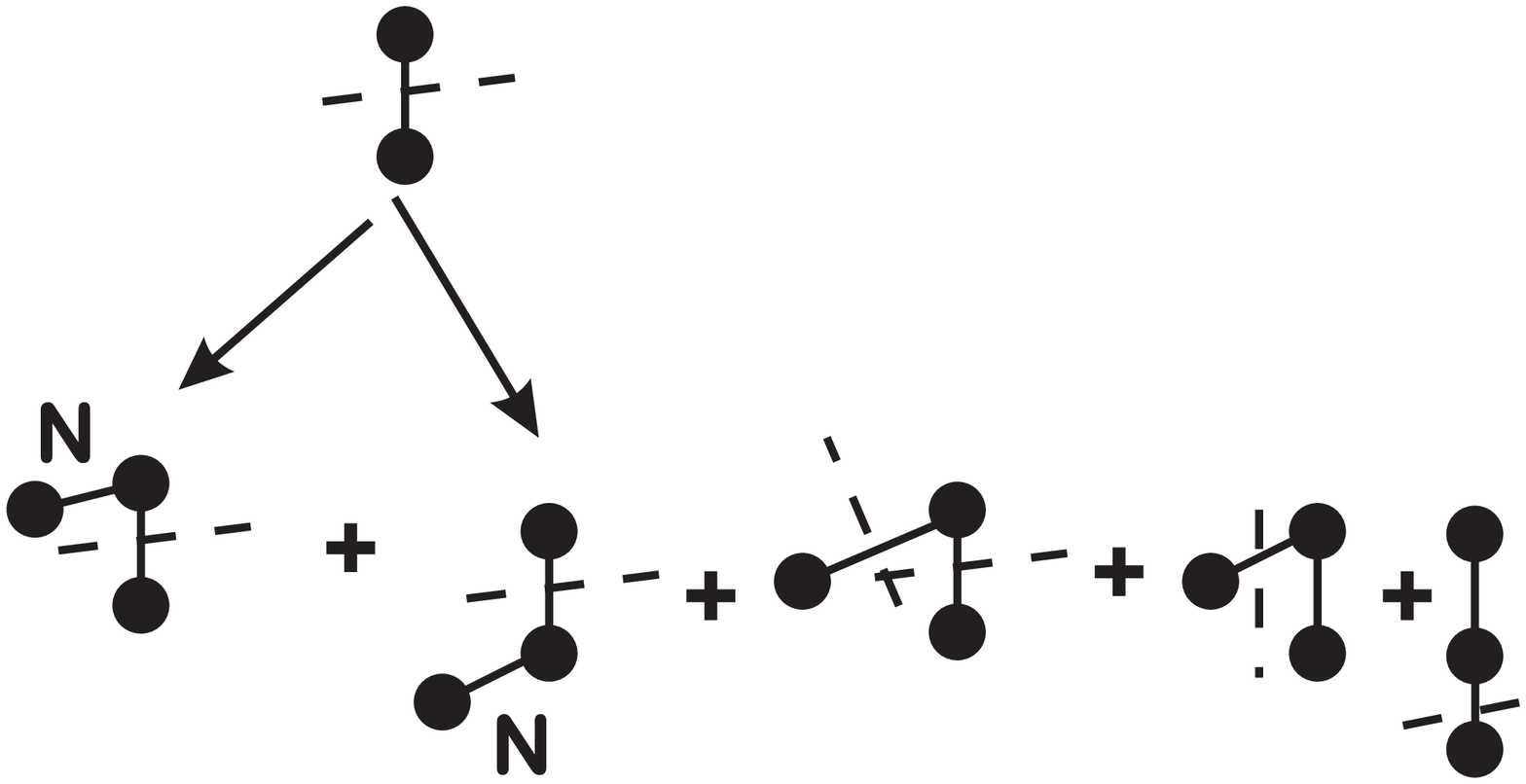}{The decomposition of $\delta_{n+1}$.}{instrr}{The 
decomposition of the cuts at $\delta_{n+1}$ in terms
of the cuts at $\delta_n$ and the operator $N$.
The first two terms of the bottom line indicate
natural growth on the $P^C$ or the $R^C$ part.
The third term gives the contribution for the case
that the natural growth carries a cut itself.
This can only happen at the $R^C$ part,
due to admissibility of cuts.
The last two terms are generated by the remaining possibility
that the natural growth carries the sole cut.}{6}

%%XXX \special{src: 1839 CK1.TEX} %Inserted by TeXtelmExtel

At this stage, we begin to see a fundamental connection between the
process of renormalization and the results of \cite{CM}.
Thus, we will now set out to define the Hopf algebra of rooted 
trees more formally and repeat the analysis of \cite{CM}
for it.

%%XXX \special{src: 1847 CK1.TEX} %Inserted by TeXtelmExtel

We shall formalize the simplest example from the last section as the 
Hopf algebra of rooted trees, and extend many of the results of the first section to 
this more involved case.

%%XXX \special{src: 1855 CK1.TEX} %Inserted by TeXtelmExtel

By a rooted tree $T$ we mean a finite, connected, simply connected, one 
dimensional simplicial complex with a base point $* \in T^{(0)} = \{ \hbox{set 
of vertices of}$ $T \}$. This base point is called the root. By the degreee of the 
tree we mean
\begin{equation}
\deg (T) = {\rm Card} T^{(0)} = \# \ \hbox{of vertices of} \ T \, . \label{eq3.1}
\end{equation}
For each $n$ we have a finite set of rooted trees $T$ with $\deg (T) = n$ where 
we only consider isomorphism classes of trees and choose a representative in each 
isomorphism class. Thus for $n=1$ we have one element $t_1\equiv*$, 
for $n=2$ we also have 
only one, $t_2$, and for $n = 3$ we have two, $t_{3_1}$ and $t_{3_2}$,
all defined in Fig.(\ref{f1}).

%%XXX \special{src: 1871 CK1.TEX} %Inserted by TeXtelmExtel

\noindent By a {\it simple} cut of a tree $T$ we mean a subset $c \subset T^{(1)}$ of the set 
of edges of $T$ such that,
\begin{equation}
\hbox{for any} \ x \in T^{(0)} \, \hbox{the path} \ (*,x) \ \hbox{only contains at 
most one element of} \ c \, . \label{eq3.3}
\end{equation}
Thus what is excluded is to have two cuts of the same path or branch. Given a 
cut $c$ the new simplicial complex $T_c$ with $T_c^{(0)} = T^{(0)}$ and
\begin{equation}
T_c^{(1)} = T^{(1)} \backslash c \, , \label{eq3.4}
\end{equation}
is no longer connected, unless $c = \emptyset$. We let $R_c (T)$ be the connected 
component of $*$ with the same base point and call it the trunk, while for each 
other connected component, called a cut branch, we endow it with the base point 
which is the edge of the cut. We obtain in this way a set, with multiplicity, of 
finite rooted trees.

%%XXX \special{src: 1890 CK1.TEX} %Inserted by TeXtelmExtel

For each $n$ we let $\Sigma_n$ be the set of trees of degree $\leq n$, up to 
isomorphism, and let ${\cal H}_n$ be the polynomial commutative algebra generated by 
the symbols,
\begin{equation}
\delta_T \ , \ T \in \Sigma_n \, . \label{eq3.5}
\end{equation}
One defines a coproduct on ${\cal H}_n$ by,
\begin{equation}
\Delta \, \delta_T = \delta_T \otimes 1 + 1 \otimes \delta_T + \sum_c \left( 
\prod_{P_c (T)} \delta_{T_i} \right) \otimes \delta_{R_c (T)} \, , \label{eq3.6}
\end{equation}
where the last sum is over all non trivial simple cuts ($c \not= \emptyset$) of $T$, 
while the product ${\displaystyle \prod_{P_c (T)}}$ is over the cut branches,
in accordance with Eq.(\ref{cop1}).

%%XXX \special{src: 1907 CK1.TEX} %Inserted by TeXtelmExtel

Equivalently, one can write (\ref{eq3.6}) as,
\begin{equation}
\Delta \, \delta_T = \delta_T \otimes 1 +  \sum_c \left( \prod_{P_c (T)} 
\delta_{T_i} \right) \otimes \delta_{R_c (T)} \, , \label{eq3.7})
\end{equation}
where the last sum is over all simple cuts. 

%%XXX \special{src: 1916 CK1.TEX} %Inserted by TeXtelmExtel

This defines $\Delta$ on generators and it extends uniquely as an algebra 
homomorphism,
\begin{equation}
\Delta : {\cal H}_n \rightarrow {\cal H}_n \otimes {\cal H}_n \, . \label{eq3.8}
\end{equation}

%%XXX \special{src: 1924 CK1.TEX} %Inserted by TeXtelmExtel

\smallskip

%%XXX \special{src: 1928 CK1.TEX} %Inserted by TeXtelmExtel

\noindent {\bf Lemma 1.} {\it The coproduct $\Delta$ is coassociative.}

%%XXX \special{src: 1932 CK1.TEX} %Inserted by TeXtelmExtel

\smallskip

%%XXX \special{src: 1936 CK1.TEX} %Inserted by TeXtelmExtel

\noindent {\bf Proof.} It is enough to check the equality
\begin{equation}
(1 \otimes \Delta) \, \Delta \, \delta_T = (\Delta \otimes 1) \, \Delta \, \delta_T 
\qquad \forall \, T \in \Sigma_n \, , \label{eq3.9}
\end{equation}
one can do it directly by introducing the notion of a double cut of $T$, but we 
shall use instead the following map from ${\cal H}_R = \cup \, {\cal H}_n$ to ${\cal 
H}_R$,
\begin{equation}
L (\delta_{T_1} \ldots \delta_{T_m}) = \delta_T \ , \ \forall \, T_j \in \Sigma = 
\cup 
\Sigma_n \, , \label{eq3.10}
\end{equation}
where $T$ is the pointed tree obtained by connecting a new base point $*$ to the 
base points of the pointed trees $T_j$. The map $L$ is the unique linear map from 
${\cal H}_R$ to ${\cal H}_R$ satisfying (\ref{eq3.10}). It agrees with the map
$B_+$ introduced in the previous section. Let us show that,
\begin{equation}
\Delta \circ L = L \otimes 1 + ({\rm id} \otimes L) \circ \Delta \, . \label{eq3.11}
\end{equation}
Let $a = \delta_{T_1} \ldots \delta_{T_m}$ and $T$ be as in (\ref{eq3.10}) so that 
$L(a) = \delta_T$. From (\ref{eq3.7}), one gets,
\begin{equation}
\Delta (L(a)) - L(a) \otimes 1 = {\displaystyle \sum_c \prod_{P_c}} \ 
\delta_{T'_i} \otimes \delta_{R_c}, \label{eq3.12}
\end{equation}
where all simple cuts of $T$, (including $c = \emptyset$) are allowed. Moreover,
\begin{equation}
\Delta (a) = \prod_{i=1}^n \left( \delta_{T_i} \otimes 1 + \sum_{c_i} 
\prod_{P_{c_i}} \, \delta_{T''_{i_j}} \otimes \delta_{R_{c_i}} \right) \, , 
\label{eq3.13}
\end{equation}
where again all simple cuts $c_i$ of $T_i$ are allowed.

%%XXX \special{src: 1972 CK1.TEX} %Inserted by TeXtelmExtel

Let $t_n$ be the tree with base point $*$ and $n$ other vertices $v_i$ labelled from 
$i=1$ to $i=n$, all directly connected to the base point $*$. We view $t_n$ in an 
obvious way as a subgraph of the tree $T$, where the base points are the same and 
the 
vertex $v_i$ is the base point of $T_i$. Given a simple cut $c$ of $T$ one gets by 
restriction to the subgraph $t_n \subset T$ a cut of $t_n$, it is characterized by 
the subset $I \subset \{1,...,n \}$, $I= \{i\,;\, (*,v_i) \in c \}$. The simple cut 
$c$ is uniquely determined by the subset $I$ and for each $i \in I^c$, i.e. each 
branch $(*,v_i)$ of $t_n$ which is not cut, by the simple cut $c_i$ of $T_i$ given 
by 
the restriction of $c$ to this subgraph. Thus the simple cuts $c$ of $T$ are in one 
to one correspondence with the various terms of the expression (\ref{eq3.13}), 
namely 
the $\prod_{k \in I}  \delta_{T_k} \otimes 1 \prod_{i \in I^c} \, \prod_{P_{c_i}} \, 
\delta_{T''_{i_j}} \otimes \delta_{R_{c_i}}$.

%%XXX \special{src: 1990 CK1.TEX} %Inserted by TeXtelmExtel

The two sums match termwise and, applying ${\rm id} \otimes L$ to (\ref{eq3.13}) one 
gets,
\begin{equation}
\Delta (L(a)) = L(a) \otimes 1 + ({\rm id} \otimes L) \, \Delta (a) \, . 
\label{eq3.14}
\end{equation}
This is Eq.(\ref{cop2}) of the previous section.
(Note that $L(1) = \delta_*$ by definition.)

%%XXX \special{src: 2001 CK1.TEX} %Inserted by TeXtelmExtel

One has,
\begin{equation}
\Delta \, \delta_* = \delta_* \otimes 1 + 1 \otimes \delta_* \label{eq3.15}
\end{equation}
so that ${\cal H}_1$ is coassociative. Let us assume that ${\cal H}_n$ is 
coassociative and prove it for ${\cal H}_{n+1}$. It is enough to check (\ref{eq3.9}) 
for the generators $\delta_T $, with $ deg(T) \leq n+1$ one has $\delta_T= L 
(\delta_{T_1} \ldots \delta_{T_m})= L(a)$ where the degree of all $T_j$ is $\leq n$, 
i.e. $a \in {\cal H}_n$. Using (\ref{eq3.14}) we can replace $\Delta \, \delta_T$ by
\begin{equation}
L(a) \otimes 1 + ({\rm id} \otimes L) \, \Delta (a) \, , \label{eq3.16}
\end{equation}
where $\Delta$ is the coassociative coproduct in ${\cal H}_n$. Thus we can use the 
notation which encodes the coassociativity of ${\cal H}_n$,
\begin{equation}
\Delta a = a_{(1)} \otimes a_{(2)} \, , \, ({\rm id} \otimes \Delta) \, \Delta (a) = 
(\Delta \otimes {\rm id}) \, \Delta (a) = a_{(1)} \otimes a_{(2)} \otimes a_{(3)} \, 
. \label{eq3.17}
\end{equation}
The first term of (\ref{eq3.9}) is then: $L(a) \otimes 1 \otimes 1 + a_{(1)} \otimes 
\Delta \circ L \, a_{(2)}$, which by (\ref{eq3.14}) gives
\begin{equation}
L(a) \otimes 1 \otimes 1 + a_{(1)} \otimes L (a_{(2)}) \otimes 1 + a_{(1)} \otimes 
a_{(2)} \otimes L \, a_{(3)} \, . \label{eq3.18}
\end{equation}
The second term of (\ref{eq3.9}) is $\Delta \circ L (a) \otimes 1 + \Delta a_{(1)} 
\otimes L a_{(2)}$, which by (\ref{eq3.14}) gives,
\begin{equation}
L(a) \otimes 1 \otimes 1 + a_{(1)} \otimes L \, a_{(2)} \otimes 1 + a_{(1)} \otimes 
a_{(2)} \otimes L \, a_{(3)} \, . \label{eq3.19}
\end{equation}
Thus we conclude that $\Delta$ is coassociative.~$\Box$

%%XXX \special{src: 2036 CK1.TEX} %Inserted by TeXtelmExtel

\smallskip

%%XXX \special{src: 2040 CK1.TEX} %Inserted by TeXtelmExtel

We shall now characterize the Hopf algebra ${\cal H}_R = \cup \, {\cal H}_n$ as the 
solution of a universal problem in Hochschild cohomology. First, given an algebra 
$\cal A$ with augmentation $\varepsilon$, let us consider the Hochschild cohomology 
of $\cal A$ with coefficients in the following bimodule ${\cal M}$. As a vector 
space 
${\cal M} = {\cal A}$, the {\it left} action of ${\cal A}$ on ${\cal M}$ is $(a,\xi) 
\rightarrow a\xi$, for all $a \in {\cal A}$, $\xi \in {\cal M}$. The {\it 
right} action of $\cal A$ on ${\cal M}$ is by $(\xi , a) \rightarrow \xi \, 
\varepsilon (a)$, $\xi \in {\cal M}$, $a \in \cal A$. Thus the right module 
structure 
is through the augmentation. Let us denote the corresponding cocycles by 
$Z_{\varepsilon}^n ({\cal A})$, the coboundaries by $B_{\varepsilon}^n ({\cal A})$ 
and the cohomology as $H_{\varepsilon}^n ({\cal A})$.

%%XXX \special{src: 2056 CK1.TEX} %Inserted by TeXtelmExtel

Thus for instance a 1-cocycle $D \in Z_{\varepsilon}^n ({\cal A})$ is a linear map 
${\cal A} \stackrel{D}{\rightarrow} {\cal A}$ such that $D(ab) = D(a) \, \varepsilon 
(b) + a \, D(b) \quad \forall \, a,b \in {\cal A}$. Next, given a Hopf algebra $\cal 
H$ we use the unit of $\cal H$ and its coalgebra structure to transpose (as in the 
Harrison cohomology), the above complex.

%%XXX \special{src: 2064 CK1.TEX} %Inserted by TeXtelmExtel

More precisely an $n$-cochain $L$ is a linear map,
\begin{equation}
L : {\cal H} \rightarrow \underbrace{{\cal H} \otimes \ldots \otimes {\cal H}}_{n \, 
{\rm times}} \label{eq3.20}
\end{equation}
and the coboundary $b$ is given by,
\begin{equation}
(bL)(a) = ({\rm id} \otimes L) \, \Delta (a) \, - \Delta_{(1)} \, L(a) + 
\Delta_{(2)} 
\, L(a) + \ldots + (-1)^j \, \Delta_{(j)} \, L(a) \label{eq3.21}
\end{equation}
\[
+ \ldots + (-1)^n \, \Delta_{(n)} \, L(a) + (-1)^{n+1} L(a) \otimes 1 \, , 
\]
where the lower index $(j)$ in $\Delta_{(j)}$ indicates where the coproduct is 
applied. For $n=0$, $L$ is just a linear form on $\cal H$ and one has
\begin{equation}
(bL) (a) = ({\rm id} \otimes L) \, \Delta (a) \, -  L(a) \, 1 \, . \label{eq3.22}
\end{equation}
For $n=1$, $L$ is a linear map from $\cal H$ to $\cal H$ and
\begin{equation}
(bL) (a) =  ({\rm id} \otimes L) \, \Delta (a) - \Delta \, L (a) + L(a) \otimes 1 
\in 
{\cal H} \otimes {\cal H} \, . \label{eq3.23}
\end{equation}
We shall use the notation $Z_{\varepsilon}^n ({\cal H}^*)$, $H_{\varepsilon}^n 
({\cal 
H}^*) \ldots$ for the corresponding cocycles, cohomology classes, etc $\ldots$

%%XXX \special{src: 2095 CK1.TEX} %Inserted by TeXtelmExtel

\smallskip

%%XXX \special{src: 2099 CK1.TEX} %Inserted by TeXtelmExtel

\noindent {\bf Theorem 2.} {\it There exists a pair $({\cal H} , L)$, unique up to 
isomorphism, where $\cal H$ is a {\rm commutative} Hopf algebra and $L \in 
Z_{\varepsilon}^1 ({\cal H}^*)$ which is {\rm universal} among all such 
pairs. In other words for any pair $({\cal H}_1 , L_1)$ where ${\cal H}_1$ is a 
commutative Hopf algebra and $L \in Z_{\varepsilon}^1 ({\cal H}_1^*)$, there exists 
a unique Hopf algebra morphism ${\cal H} \stackrel{\rho}{\rightarrow} {\cal H}_1$ 
such that $L_1 \circ \rho = \rho \circ L$.}

%%XXX \special{src: 2109 CK1.TEX} %Inserted by TeXtelmExtel

\smallskip

%%XXX \special{src: 2113 CK1.TEX} %Inserted by TeXtelmExtel

\noindent {\bf Proof.} Let ${\cal H}_R$ be the Hopf algebra of rooted 
trees and $L$ be the linear map defined by (\ref{eq3.10}).

%%XXX \special{src: 2118 CK1.TEX} %Inserted by TeXtelmExtel

The equality (\ref{eq3.11}) shows that $bL = 0$. This shows that $L$ is a 1-cocycle. 
It is clear that it is not a coboundary, indeed one has
\begin{equation}
L(1) = \delta * \not= 0 \label{eq3.24}
\end{equation}
where $*$ is the tree with only one vertex.

%%XXX \special{src: 2127 CK1.TEX} %Inserted by TeXtelmExtel

Moreover, for any coboundary $T = bZ$ one has
\begin{equation}
T(1) = 0 \, , \label{eq3.25}
\end{equation}
since $T(1) = Z(1) \, 1 - ({\rm id} \otimes Z) \, \Delta (1) = 0$.

%%XXX \special{src: 2135 CK1.TEX} %Inserted by TeXtelmExtel

Next consider a pair $({\cal H}_1 , L_1)$ where ${\cal H}_1$ is a commutative Hopf 
algebra and $L_1 \in Z_{\varepsilon}^1 ({\cal H}_1^*)$ is a 1-cocycle. The equality 
$L_1 \circ \rho = \rho \circ L$ uniquely determines an algebra homomorphism $\rho : 
{\cal H}_R \rightarrow {\cal H}_1$. Indeed on the linear basis $\Pi \, \delta_{T_i}$ 
of ${\cal H}_R$ one must have,
\begin{equation}
\rho \, (\Pi \, \delta_{T_i}) = \Pi \, \rho \, (\delta_{T_i}) \, , \label{eq3.26}
\end{equation}
by multiplicativity of $\rho$, while $\rho \, (\delta_T)$ is determined by 
induction by $\rho \, (\delta_*) = L_1 \, (1)$, and,
\begin{equation}
\rho \, (L \, (\Pi \, \delta_{T_i})) = L_1 \, \rho \, (\Pi \, \delta_{T_i}) \, . 
\label{eq3.27}
\end{equation}

%%XXX \special{src: 2152 CK1.TEX} %Inserted by TeXtelmExtel

We need to check that it is a morphism of Hopf algebras, i.e. that it is compatible 
with the coproduct,
\begin{equation}
(\rho \otimes \rho) \, (\Delta (a)) = \Delta_1 \, \rho (a) \qquad \forall \, a \in 
\cal H_R \, . \label{eq3.28}
\end{equation}
It is enough to check (\ref{eq3.28}) on generators of the form $\delta_T=  \, L (\Pi 
\, \delta_{T_i})$. To do this one uses the cocycle property of $L_1$ which allows to 
write,
\begin{equation}
\Delta_1 \, L_1 \, (\rho \, (\Pi \, \delta_{T_i})) = L_1 \, (\rho \, (\Pi \, 
\delta_{T_i})) \otimes 1 + ({\rm id} \otimes L_1) \, \Delta_1 \, \rho \, (\Pi \, 
\delta_{T_i}) \, . \label{eq3.29}
\end{equation}
One uses an induction hypothesis on the validity of (\ref{eq3.28}), to write,
\begin{equation}
({\rm id} \otimes L_1) \, \Delta_1 \, \rho \, (\Pi \, \delta_{T_i}) \, = (\rho 
\otimes (\rho \circ L)) \, \Delta (\Pi \, \delta_{T_i}) \label{eq3.30}
\end{equation}
making use of the identity $\rho \otimes \rho \circ L = ({\rm id} \otimes L_1) \, 
(\rho \otimes \rho)$. Thus one has,
\begin{equation}
\Delta_1 \, L_1 \, (\rho \, (\Pi \, \delta_{T_i})) = L_1 \, (\rho \, (\Pi \, 
\delta_{T_i})) \otimes 1 + (\rho \otimes (\rho \circ L)) \, \Delta (\Pi \, 
\delta_{T_i}) \, , \label{eq3.31}
\end{equation}
and the validity of (\ref{eq3.28}) for $a=\delta_T$ follows from the equality $(\rho 
\otimes \rho) \, \Delta (L (\Pi \, \delta_{T_i})) = (\rho \otimes \rho) \, ( L \, ( 
\Pi \, \delta_{T_i})) \otimes 1 + ({\rm id} \otimes L) \, \Delta \,  (\Pi \, 
\delta_{T_i}))$.

%%XXX \special{src: 2185 CK1.TEX} %Inserted by TeXtelmExtel

We have thus shown the existence and uniqueness of the Hopf algebra 
morphism $\rho$.~$\Box$

%%XXX \special{src: 2190 CK1.TEX} %Inserted by TeXtelmExtel

\smallskip

%%XXX \special{src: 2194 CK1.TEX} %Inserted by TeXtelmExtel

As the simplest example, let ${\cal H}_1$ be the Hopf algebra of polynomials $P( 
\delta_1 )$, as above, with,
\begin{equation}
\Delta \, \delta_1 = \delta_1 \otimes 1 + 1 \otimes \delta_1 \, . \label{eq3.32}
\end{equation}
The cohomology group $H_{\varepsilon}^1 ({\cal H}_1^*)$ is one dimensional, and the 
natural generator is the cocycle, 
\begin{equation}
L_1 (P)(x) = \, \int_0^x P(a)da \qquad  \forall P= P( \delta_1 ) \in {\cal H}_1 \, . 
\label{eq3.33}
\end{equation}
The cocycle identity follows from the equality,
\begin{equation}
\int_0^{x+y} P(a)da= \,\int_0^x P(a)da + \, \int_0^y P(x+a)da \, . \label{eq3.34}
\end{equation}
The coboudaries are of the form,
\begin{equation}
L_0 (P)= \, \int ( P(x+a)- P(a)) \phi (a)da \qquad  \forall P= P( \delta_1 ) \in 
\cal H_1 \, , \label{eq3.35}
\end{equation}
where $\phi$ is a distribution with support the origin, and possibly infinite 
order.

%%XXX \special{src: 2219 CK1.TEX} %Inserted by TeXtelmExtel

The tranpose $\rho^t$ of the morphism of Hopf algebras given by Theorem 2 
determines a Lie algebra homomorphism from the one dimensional Lie algebra
(${\cal A}_1^1 $ with the notations of section I), to the Lie algebra ${\cal L}^1$ 
which corresponds, by the Milnor-Moore theorem to the commutative Hopf algebra 
${\cal H}_R$.

%%XXX \special{src: 2227 CK1.TEX} %Inserted by TeXtelmExtel

We proceed as in section I to determine ${\cal L}^1$ .

%%XXX \special{src: 2231 CK1.TEX} %Inserted by TeXtelmExtel

Let $\cal L^1$ be the linear span of the elements $Z_T$, indexed by rooted trees. We 
introduce an operation on ${\cal L}^1$ by the equality,
\begin{equation}
Z_{T_1} * Z_{T_2} = \sum_T \, n(T_1,T_2;T) \,  Z_{T} \, , \label{eq3.36}
\end{equation}
where the integer $n(T_1,T_2;T)$ is determined as the number of simple cuts $c$ of 
cardinality 1 such that the cut branch is $T_1$ while the remaining trunk is $T_2$.\footnote{The reader shall not confuse the operation
which relates $T_1$ and $T_2$ to $T$ with the transplantation
used in the theory of operads. Indeed, in the latter, the root of
the tree $T_1$ is restricted to be the end of a branch of
$T_2$.}

%%XXX \special{src: 2241 CK1.TEX} %Inserted by TeXtelmExtel

\smallskip

%%XXX \special{src: 2245 CK1.TEX} %Inserted by TeXtelmExtel

\noindent {\bf Theorem 3.} a) {\it Let ${\cal L}^1$ be the linear span of the 
elements $Z_T$, indexed by rooted trees. The following equality defines a 
structure of Lie algebra on ${\cal L}^1$. The Lie bracket $[Z_{T_1} , Z_{T_2}]$ is 
$Z_{T_1} * Z_{T_2} \, - \, Z_{T_2} * Z_{T_1}$.} 

%%XXX \special{src: 2252 CK1.TEX} %Inserted by TeXtelmExtel

b) {\it The Hopf algebra ${\cal H}_R$ is the dual of the envelopping algebra of 
the Lie algebra ${\cal L}^1$.}

%%XXX \special{src: 2257 CK1.TEX} %Inserted by TeXtelmExtel

\smallskip

%%XXX \special{src: 2261 CK1.TEX} %Inserted by TeXtelmExtel

\noindent Define
\begin{equation}
A \, (T_1 , T_2 , T_3) = Z_{T_1} * (Z_{T_2} * Z_{T_3}) - (Z_{T_1} * Z_{T_2}) * 
Z_{T_3} \, . \label{eq3.37}
\end{equation}
We shall need the following lemma,

%%XXX \special{src: 2270 CK1.TEX} %Inserted by TeXtelmExtel

\smallskip

%%XXX \special{src: 2274 CK1.TEX} %Inserted by TeXtelmExtel

\noindent {\bf Lemma 4.} {\it One has $A (T_1 , T_2 , T_3) = \Sigma \, n (T_1 , T_2 
, T_3 ; T) \, Z_T$, where the integer $n$ is the number of simple cuts $c$ of 
$T$, $\vert c \vert = 2$ such that the two branches are $T_1 , T_2$ while $R_c 
(T) = T_3$.}

%%XXX \special{src: 2281 CK1.TEX} %Inserted by TeXtelmExtel

\smallskip

%%XXX \special{src: 2285 CK1.TEX} %Inserted by TeXtelmExtel

\noindent {\bf Proof.} When one evaluates (\ref{eq3.37}) against $Z_T$ one gets the 
coefficient,
\begin{equation}
\sum_{T'} n (T_1 , T' ; T) \, n (T_2 , T_3 ; T') - \sum_{T''} n (T_1 , T_2 ; 
T'') \, n (T'' , T_3 ; T) \, , \label{eq3.38}
\end{equation}
the first sum corresponds to pairs of cuts, $c$, $c'$ of $T$ with $\vert c \vert 
= \vert c' \vert = 1$ and where $c'$ is a cut of $R_c (T)$. These pairs of cuts fall 
in two classes, either $c \cup c'$ is a simple cut or it is not. The second sum 
corresponds to pairs of cuts $c_1$, $c'_1$ of $T$ such that $\vert c_1 \vert 
= \vert c_1' \vert = 1$, $R_{c_1} (T) = T_3$ and $c'_1$ is a cut of $P_{c_1} 
(T)$. In such a case $c_1 \cup c'_1$ is never a simple cut so the difference 
(\ref{eq3.38}) amounts to substract from the first sum the pairs $c$, $c'$ such that 
$c \cup c'$ is not a simple cut. This gives,
\begin{equation}
A \, (T_1 , T_2 , T_3) = \, \sum_T \, n(T_1,T_2,T_3;T) \,  Z_T \, , \label{eq3.39}
\end{equation}
where $n(T_1,T_2,T_3;T)$ is the number of simple cuts $c$ of $T$ of cardinality $2$ 
such that the two cut branches are $T_1$ and $T_2$.~$\Box$

%%XXX \special{src: 2307 CK1.TEX} %Inserted by TeXtelmExtel

\smallskip

%%XXX \special{src: 2311 CK1.TEX} %Inserted by TeXtelmExtel

It is thus clear that,
\begin{equation}
A \, (T_1 , T_2 , T_3) = A \, (T_2 , T_1 , T_3) \, . \label{eq3.40}
\end{equation}
One then computes $[[Z_{T_1} , Z_{T_2}], Z_{T_3}] + [[Z_{T_2} , Z_{T_3}], 
Z_{T_1}] + [[Z_{T_3} , Z_{T_1}], Z_{T_2}]$.

%%XXX \special{src: 2320 CK1.TEX} %Inserted by TeXtelmExtel

One can write it, for short, as a sum of 24 terms,
\begin{eqnarray*}
& & (T_1 * T_2) * T_3 - (T_2 * T_1) * T_3 - T_3 * (T_1 * T_2) + T_3 * (T_2 
* T_1) \\
&+ & (T_2 * T_3) * T_1 - (T_3 * T_2) * T_1 - T_1 * (T_2 * T_3) + T_1 * 
(T_3 * T_2) \\
&+ & (T_3 * T_1) * T_2 - (T_1 * T_3) * T_2 - T_2 * (T_3 * T_1) + T_2 * 
(T_1 * T_3) \\
&= & - A \, (T_1 , T_2 , T_3) + A \, (T_2 , T_1 , T_3) - A \, (T_3 , T_1 , 
T_2) + A \, (T_3 , T_2 , T_1) \\
& & - A \, (T_2 , T_3 , T_1) + A \, (T_1 , T_3 , T_2) = 0 \, .
\end{eqnarray*}

%%XXX \special{src: 2335 CK1.TEX} %Inserted by TeXtelmExtel

b) For each rooted tree $T$ let us define a linear form $Z_T$ on ${\cal H}_R$ by 
the equality,
\begin{equation}
\langle Z_T , P (\delta_{T_i}) \rangle = (\partial / \partial \delta_T \, P) (0) \, 
. 
\label{eq3.41}
\end{equation}
Thus $Z_T$ vanishes when paired with any monomial $\delta_{T_1}^{n_1} \ldots 
\delta_{T_k}^{n_k}$ except when this monomial is $\delta_T$ while,
\begin{equation}
\langle Z_T , \delta_T \rangle = 1 \, . \label{eq3.42}
\end{equation}
Since $P \rightarrow P (0)$ is the counit of ${\cal H}_R$ and since $Z_T$ satisfies
\begin{equation}
\langle Z_T , PQ \rangle = \langle Z_T , P \rangle \, \varepsilon (Q) + \varepsilon 
(P) \, \langle Z_T , Q \rangle \, , \label{eq3.43}
\end{equation}
it follows that the coproduct of $Z_T$ is,
\begin{equation}
\Delta \, Z_T = Z_T \otimes 1 + 1 \otimes Z_T \, , \label{eq3.44}
\end{equation}
where the coproduct on ${\cal H}_R^*$ is defined, when it makes sense, by 
dualizing the product of ${\cal H}_R$.

%%XXX \special{src: 2361 CK1.TEX} %Inserted by TeXtelmExtel

Similarly the product of two elements of ${\cal H}_R^*$ is defined by
\begin{equation}
\langle Z_1 \, Z_2 , P \rangle = \langle Z_1 \otimes Z_2 , \Delta \, P \rangle \, . 
\label{eq3.45}
\end{equation}
Since the bracket of two derivations is still a derivation, the subspace of 
${\cal H}_R^*$ of elements satisfying (\ref{eq3.39}) is stable under bracket. What 
remains is to show that,
\begin{equation}
Z_{T_1} \, Z_{T_2} - Z_{T_2} \, Z_{T_1} = [Z_{T_1} , Z_{T_2}] \, , \label{eq3.46}
\end{equation}
where the r.h.s. is defined by the Lie algebra structure of theorem 3.

%%XXX \special{src: 2376 CK1.TEX} %Inserted by TeXtelmExtel

Let ${\cal H}_0$ be the augmentation ideal of ${\cal H}_R$, ${\cal H}_0 = {\rm Ker} 
\, \varepsilon$. The formula defining the coproduct in ${\cal H}_R$ shows that,
\begin{equation}
\Delta \, \delta_T = \delta_T \otimes 1 + 1 \otimes \delta_T + R_T \label{eq3.47}
\end{equation}
where $R_T \in {\cal H}_0 \otimes {\cal H}_0$. In fact one can compute $R_T$ modulo 
higher powers of ${\cal H}_0$, i.e. modulo ${\cal H}_0^2 \otimes {\cal H}_0$, it 
gives,
\begin{equation}
R_T^{(0)} = \sum_c \, \delta_{T'_c} \otimes \delta_{T_c} \label{eq3.48}
\end{equation}
where $c$ varies among {\it single cuts} of the tree $T$, where $T_c$ is 
the part of $T$ that contains the base point, while $T'_c$ is the tree 
which remains. When one computes $\langle Z_{T_1} \, Z_{T_2} , \Pi \, \delta_{T_i} 
\rangle = \langle Z_{T_1} \otimes Z_{T_2} , \Pi \, \Delta \, \delta_{T_i} \rangle$ 
the part which is not symmetric in $T_1 , T_2$ is zero unless $\Pi \, \delta_{T_i}$ 
is equal to a single $\delta_T$. When one computes
\begin{equation}
\langle Z_{T_1} \, Z_{T_2} , \delta_T \rangle = \langle Z_{T_1} \otimes Z_{T_2} , 
\Delta \, \delta_T \rangle \, , \label{eq3.49}
\end{equation}
the only part which contributes comes from $R_T^{(0)}$ and it counts the 
number of ways of obtaining $T$ from $T_1$ and $T_2$, which gives 
(\ref{eq3.46}).~$\Box$

%%XXX \special{src: 2403 CK1.TEX} %Inserted by TeXtelmExtel

\smallskip

%%XXX \special{src: 2407 CK1.TEX} %Inserted by TeXtelmExtel

\noindent {\bf Proposition 5.}  {\it The equality degree $Z_T = \#$ of 
vertices of $T$ defines a grading of the Lie algebra ${\cal L}^1$.}

%%XXX \special{src: 2412 CK1.TEX} %Inserted by TeXtelmExtel

\smallskip

%%XXX \special{src: 2416 CK1.TEX} %Inserted by TeXtelmExtel

\noindent {\bf Proof.} The number of vertices of any tree obtained by 
gluing $T_1$ to $T_2$ is the sum of the number of vertices of $T_1$ and 
$T_2$.~$\Box$

%%XXX \special{src: 2422 CK1.TEX} %Inserted by TeXtelmExtel

\smallskip

%%XXX \special{src: 2426 CK1.TEX} %Inserted by TeXtelmExtel

We shall now show how to extend the Hopf algebra ${\cal H}_R$ to include
the generators $X, Y$ of the Lie algebra of the affine group as in section I.
The commutator of $Y$ with $\delta_T$ will simply be given by,
\begin{equation}
[Y,\,\delta_T]=\, deg(T)\delta_T \label{eq3.50}
\end{equation}
i.e. by the above grading.

%%XXX \special{src: 2436 CK1.TEX} %Inserted by TeXtelmExtel

The commutator with $X$ will generate a derivation $N$ of ${\cal H}_R$,
uniquely determined by its value on the generators $\delta_T$, by
\begin{equation}
N \, \delta_T = \sum \, \delta_{T'} \label{eq3.51}
\end{equation}
where the trees $T'$ are obtained by adding one vertex and one edge to $T$ 
in all possible ways without changing the base point. It is clear that the 
sum (\ref{eq3.51}) contains $\deg (T)$ terms.

%%XXX \special{src: 2447 CK1.TEX} %Inserted by TeXtelmExtel

Using the derivation property of $N$, one has,
\begin{equation}
N \, \left( \prod_1^n \, \delta_{T_i} \right) = \sum_1^n \, \delta_{T_1} \ldots N 
(\delta_{T_i}) \ldots \delta_{T_n} \, . \label{eq3.52}
\end{equation}
Our first task will be to get a formula for $\Delta \, N (\delta_T)$.

%%XXX \special{src: 2456 CK1.TEX} %Inserted by TeXtelmExtel

\smallskip

%%XXX \special{src: 2460 CK1.TEX} %Inserted by TeXtelmExtel

\noindent {\bf Proposition 6.}  {\it For any $ a \in \cal H_R $ one has} 
\[
\Delta \, N \, a = (N \otimes {\rm id}) \, \Delta \, a + ({\rm id} \otimes N) \, 
\Delta \, a + [\delta_1 \otimes Y , \Delta \, a] \, . 
\]

%%XXX \special{src: 2468 CK1.TEX} %Inserted by TeXtelmExtel

\smallskip

%%XXX \special{src: 2472 CK1.TEX} %Inserted by TeXtelmExtel

\noindent {\bf Proof.} First, it is enough to check the equality when $a= \, 
\delta_T$. Indeed, both $\Delta \circ N$ and $(N \otimes {\rm id} + {\rm id} \otimes 
N + ad \, (\delta_1 \otimes Y)) \circ \Delta$ are derivations from ${\cal H}_R$ to 
the ${\cal H}_R$-bimodule ${\cal H}_R \otimes {\cal H}_R$ (using $\Delta$ to define 
the bimodule structure). Thus so is their difference $\varepsilon_0$ which vanishes 
provided it does on the generators $\delta_T$. Let thus $T$ be a pointed tree and 
$T'$ be obtained from $T$ by adjoining an edge at $v_0 \in \Delta^0(T)$. One has
\begin{equation}
\Delta \, \delta_{T'} = \delta_{T'} \otimes 1 + 1 \otimes \delta_{T'} + \sum_{c'} \, 
(\Pi \, \delta_{T'_j} \otimes \delta_{R'_{c'}}) \label{eq3.53}
\end{equation}
where $c' \in \Delta^1(T') $ varies among the simple cuts of $T'$. One has 
$\Delta^1(T')=\, \Delta^1(T) \cup \{ \varepsilon \}$ where $\varepsilon$ the new 
edge. Now the cuts $c'$ for $T'$ are of two kinds,

%%XXX \special{src: 2489 CK1.TEX} %Inserted by TeXtelmExtel

(A) The new edge is not cut, (A') It is cut.

%%XXX \special{src: 2493 CK1.TEX} %Inserted by TeXtelmExtel

\noindent There is also another dichotomy,

%%XXX \special{src: 2497 CK1.TEX} %Inserted by TeXtelmExtel

(B) The vertex $v_0$ belongs to the trunk, (B') It belongs to one of the cut 
branches.

%%XXX \special{src: 2502 CK1.TEX} %Inserted by TeXtelmExtel

\noindent If we sum (\ref{eq3.53}) over all possible $T'$ we get,
\begin{equation}
\Delta \, N \, \delta_T = N \, \delta_T \otimes 1 + 1  \otimes N \, \delta_T + 
\sum_v \, \sum_{c'} \, \Pi \, \delta_{T'_i} \otimes \delta_{R_{c'}} \, . 
\label{eq3.54}
\end{equation}
Let us concentrate on the last term and consider first only the cuts $c'$ 
which satisfy (A) We also consider the term,
\begin{equation}
\sum_c \, (\Pi \, \delta_{T_c}) \otimes \delta_{R_c} \label{eq3.55}
\end{equation}
over all the cuts $c$ of the tree $T$. If we apply $({\rm id} \otimes N)$ to 
(\ref{eq3.55}), 
we obtain all possible cuts of a $T'$ such that (A) (B) holds so that,
\begin{equation}
\sum_{(A) \, (B)} = ({\rm id} \otimes N) \, \sum_c \, (\Pi \, \delta_{T_c}) \otimes 
\delta_{R_c} \, . \label{eq3.56}
\end{equation}

%%XXX \special{src: 2523 CK1.TEX} %Inserted by TeXtelmExtel

It follows that,
\begin{equation}
\sum_{(A) \, (B')} = (N \otimes {\rm id}) \, \sum_c \, (\Pi \, \delta_{T_c}) \otimes 
\delta_{R_c} \, . \label{eq3.57}
\end{equation}
We can thus summarize what we obtained so far by,
\begin{equation}
\Delta \, N \, \delta_T = (N \otimes {\rm id}) \, \Delta \, \delta_T + ({\rm id} 
\otimes N) \, \Delta \, \delta_T + \sum_{(A')} \, . \label{eq3.58}
\end{equation}
Now consider the sum ${\displaystyle \sum_{(A')}^{}}$, the first case is when the 
{\it only} cut is the cut of the new edge. The only cut branch gives us a 
$\delta_1$ and the number of ways of doing it is $n = \deg T$, thus we get
\begin{equation}
[\delta_1 \otimes Y , \delta_T \otimes 1 + 1 \otimes \delta_T ] \ , \ [Y , \delta_T 
] = n \, \delta_T \, . \label{eq3.59}
\end{equation}
The next case is when a non trivial cut $c$ remains after we remove the new 
edge. For that cut $c$ the new vertex necessarily belongs to the trunk (so 
that (A) (B) is excluded) as follows from the very definition of a cut. For 
such cuts, the result is to get an additional $\delta_1$ among the $\delta_{T_i}$, 
which comes from the cut new edge. The number of ways of doing it is 
exactly the degree of the trunk. Thus we get
\begin{equation}
[\delta_1 \otimes Y , \, \, \sum_c \, (\Pi \, \delta_{T_c}) \otimes \delta_{R_c}] \, 
. \label{eq3.60}
\end{equation}
Combining (\ref{eq3.59}) and (\ref{eq3.60}) we get,
\begin{equation}
\sum_{(A')} = [\delta_1 \otimes Y , \, \, \Delta \, \delta_T ] \, . \label{eq3.61}
\end{equation}
This is enough to assert that for any tree $T$ one has,
\begin{equation}
\Delta \, N \, \delta_T = (N \otimes {\rm id}) \, \Delta \, \delta_T + ({\rm id} 
\otimes N) \,\, \Delta \, \delta_T + [\delta_1 \otimes Y , \Delta \, \delta_T] 
\label{eq3.62}
\end{equation}
which ends the proof of Proposition 6.~$\Box$

%%XXX \special{src: 2564 CK1.TEX} %Inserted by TeXtelmExtel

\smallskip

%%XXX \special{src: 2568 CK1.TEX} %Inserted by TeXtelmExtel

In other words we can enlarge ${\cal H}_R$ to $\widetilde{\cal H}_R$ by adjoining 
the 
elements $X,Y$ with
\begin{equation}
[X,a] = N(a) \, , \ [Y ,a ] = (\deg a) \, a \qquad \forall \, a \in {\cal H}_R 
\label{eq3.63}
\end{equation}
\[
[Y,X] = X \, , \ \Delta Y = Y \otimes 1 + 1 \otimes Y \, , \ \Delta X = X \otimes 1 
+ 1 \otimes X + \delta_1 \otimes Y \, . 
\]
Let us translate Proposition 6 in terms of the transposed map $N^t$ acting on ${\cal 
H}_R^*$.

%%XXX \special{src: 2584 CK1.TEX} %Inserted by TeXtelmExtel

One has $\langle N^t (AB) , a \rangle = \langle AB , N(a) \rangle =$
\[
\langle A \otimes B , \Delta \, N (a) \rangle = \langle A \otimes B , (N \otimes 
{\rm 
id} + {\rm id} \otimes N + \delta_1 \otimes \deg) \, \Delta a \rangle
\]
\[
= \langle N^t (A) \otimes B + A \otimes N^t (B) + (\delta_1)^t \otimes \deg^t (A 
\otimes B) , \Delta a \rangle 
\]
\[
= \langle N^t (A) \, B +  A \, N^t (B) + \delta_1^t (A) \, \deg^t (B) , a \rangle \, 
;
\]
thus,
\begin{equation}
N^t (AB) = N^t (A) \, B + A \, N^t (B) + \delta_1^t (A) \, \deg^t (B) \, , 
\label{eq3.64}
\end{equation}
where $\delta_1^t$ (resp. $\deg^t$) is the transposed of the multiplication by 
$\delta_1$ (resp. $\deg$)
\begin{equation}
\langle \delta_1^t A ,a \rangle = \langle A , \delta_1 \, a \rangle \, . 
\label{eq3.65}
\end{equation}
One has $\langle \delta_1^t (AB) , a \rangle = \langle AB , \delta_1 \, a \rangle = 
\langle A \otimes B , \Delta \, \delta_1 \, \Delta \, a \rangle = \langle A \otimes 
B 
, ( \delta_1 \otimes 1+1 \otimes \delta_1) \, \Delta \, a \rangle$. Thus,
\begin{equation}
\delta_1^t (AB) = \delta_1^t (A) \, B + A \, \delta_1^t (B) \, , \label{eq3.66}
\end{equation}
i.e. $\delta_1^t$ is a derivation. Moreover on the generator $Z_T$,
\begin{equation}
\delta_1^t (Z_T) = 0 \quad \hbox{unless} \quad T = \{ * \} \, , \ \delta_1^t (Z_1) 
= 1 \, . \label{eq3.67}
\end{equation}
Indeed, $\langle Z_T , \delta_1 \, a \rangle = 0$ unless $T = \{ * \}$, while for $T 
= \{ * \}$ one gets that $\langle Z_1 , \delta_1 \, a \rangle = \varepsilon (a)$. 
Thus,
\begin{equation}
\delta_1^t = {\partial \over \partial \, Z_1} \label{eq3.68}
\end{equation}
where we use the Poincar\'e-Birkhoff-Witt theorem to write elements of ${\cal U} 
({\cal L}^1)$ in the form $\sum \, \Pi \, Z_{T_i} \, Z_1^a$.

%%XXX \special{src: 2632 CK1.TEX} %Inserted by TeXtelmExtel

Let us compute $N^t (Z_T)$ where $T$ is a tree with more than one vertex. 
One has $\langle N^t \, Z_T , \delta_{T_1} \, \delta_{T_2} \ldots \delta_{T_n} 
\rangle = \langle Z_T , N (\delta_{T_1} \ldots \delta_{T_n})\rangle$, and this 
vanishes unless $n=1$. Moreover for $n=1$,
\begin{equation}
\langle Z_T , N \, (\delta_{T_1}) \rangle = n( T; T_1) \label{eq3.69}
\end{equation}
where $n( T; T_1)$ is the number of times the tree $T$ is obtained by adjoining an 
edge and vertex to $T_1$.

%%XXX \special{src: 2644 CK1.TEX} %Inserted by TeXtelmExtel

Thus one has,
\begin{equation}
N^t \, Z_T = \sum n( T; T_1) Z_{T_1} \quad \ , \ N^t \, Z_1 = 0 \, . \label{eq3.70}
\end{equation}
We can now state the analogue of Lemma 4 of Section I as follows, where we let 
${\cal L}^k$ be the Lie subalgebra of ${\cal L}^1$ generated by the $Z_T$ with 
$deg(T)\geq k$.

%%XXX \special{src: 2654 CK1.TEX} %Inserted by TeXtelmExtel

\smallskip

%%XXX \special{src: 2658 CK1.TEX} %Inserted by TeXtelmExtel

\noindent {\bf Lemma 7.} {\it When restricted to ${\cal U} ({\cal L}^2)$, 
$N^t$ is the unique derivation, with values in ${\cal U} ({\cal L}^1)$ 
satisfying (\ref{eq3.70}), moreover, for $\deg (T_i)>1$ and $A= \Pi Z_{T_i}$ one has
\[
N^t (A \, Z_1^{a_1}) = N^t (A) \, Z_1^{a_1} + A \, {a_1 (a_1 - 1) 
\over 2} \, Z_1^{a_1 - 1} \, .
\]
}

%%XXX \special{src: 2669 CK1.TEX} %Inserted by TeXtelmExtel

\smallskip

%%XXX \special{src: 2673 CK1.TEX} %Inserted by TeXtelmExtel

\noindent {\bf Proof.} The first statement follows from (\ref{eq3.64}) and 
(\ref{eq3.66}). The second statement follows from,
\begin{equation}
N^t (Z_1^m) = {m (m-1) \over 2} \, Z_1^{m-1} \label{eq3.71}
\end{equation}
which one proves by induction on $m$.~$\Box$

%%XXX \special{src: 2682 CK1.TEX} %Inserted by TeXtelmExtel

\smallskip

%%XXX \special{src: 2686 CK1.TEX} %Inserted by TeXtelmExtel

\noindent Motivated by Section I and the first part of the lemma, we enlarge the Lie 
algebra ${\cal L}^1$ by adjoining two elements $Z_{0}$ and $Z_{-1}$ such that, 
\[
[Z_{-1} , Z_1] = Z_0 \ , \ [Z_0 , Z_T] = deg(T) \, Z_T \,
\]
\[
[Z_{-1} , Z_T] = \sum n( T; T_1) Z_{T_1} \qquad \forall \, T, \, \deg (T)>1 \, .
\]

%%XXX \special{src: 2697 CK1.TEX} %Inserted by TeXtelmExtel

The obtained Lie algebra $\cal L$, is an extension of the Lie algebra of formal 
vector fields with $Z_0 = x \, {\partial \over \partial \, x}$, $Z_{-1} = {\partial 
\over \partial \, x}$ and as above $Z_n = {x^{n+1} \over (n+1)!} \, {\partial \over 
\partial \, x}$, as follows from,

%%XXX \special{src: 2704 CK1.TEX} %Inserted by TeXtelmExtel

\smallskip

%%XXX \special{src: 2708 CK1.TEX} %Inserted by TeXtelmExtel

\noindent {\bf Theorem 8.} {\it The following equality defines a surjective Lie 
algebra homomorphism from $\cal L$ to $\cal A$, 
\[
\Theta (Z_T)= \,  n(T) Z_n , \quad \Theta (Z_i)= \,  Z_i , \, i=0,1 
\]
where $n(T)$ is the number of times $\delta_T$ occurs in $N^{deg(T)-1}(\delta_1)$.}

%%XXX \special{src: 2717 CK1.TEX} %Inserted by TeXtelmExtel

\smallskip

%%XXX \special{src: 2721 CK1.TEX} %Inserted by TeXtelmExtel

\noindent {\bf Proof.} The elements $X$, $Y$, and $\delta_*$ of the Hopf algebra 
$\widetilde{\cal H}_R $ fulfill the presentation of section I for the 
Hopf algebra $\widetilde{\cal H}_T $, thus there exists a unique homorphism of Hopf 
algebras $h$ from $\widetilde{\cal H}_R $ to $\widetilde{\cal H}_T $ such that,
\[
h(X)=X,\qquad h(Y)=Y , \qquad h(\delta_1)= \delta_* \, .
\]
By construction, $h$ restricts to the subalgebra ${\cal H}_R $ and defines a 
homomorphism to the Hopf algebra ${\cal H}_T $. Transposing this homomorphism to the 
Lie algebras, one obtains the restriction of $\Theta$ to the subalgebra ${\cal 
L}^1$.~$\Box$

\smallskip

At this stage we completed our understanding of
the relation between the two Hopf algebras.
It is best expressed by the Lie algebra homomorphism
$\Theta$ from ${\cal L}^1$ to ${\cal A}^1$.
Its extension to the full ${\cal L}$ justifies the construction
of the latter Lie algebra.

By Theorem 3 the Hopf algebra ${\cal H}_R$ should be thought of
as the algebra of coordinates on a nilpotent formal group ${\cal G}$
whose Lie algebra is the graded Lie algebra ${\cal L}^1$.
Given a field $K$, elements of the group ${\cal G}_K$
are obtained precisely as the characters of the algebra 
${\cal H}_R\otimes_{\bf Q}K$. Indeed, such characters
correspond to group-like elements $u$
(i.e.~elements $u$ satisfying $\Delta(u)=u\otimes u$)
of a suitable completion of the envelopping algebra of ${\cal L}^1$.
Viewing $u$ as a linear form on ${\cal H}_R$ gives us the desired
character.
If we let $K$ be the field of formal power series
in a variable $\epsilon$ we thus obtain as points of ${\cal G}_K$
the homomorphisms from ${\cal H}_R$ to $K$.

It is not difficult to check that the map
which to every bare Feynman diagram $\Gamma$ associates the corresponding
Laurent expansion (in Dimensional Regularization, say,
with regularized dimension $D=4-2\epsilon$, in four dimensions, say)
is precisely such a character.

This allows to reduce by the above conceptual
mathematical structure of inversion in ${\cal G}$
the computation of renormalization in QFT to the primitive
elements of the Hopf algebra,i.e.~to Feynman diagrams without
subdivergences.

In order to better understand the extension of the group
of diffeomorphisms provided by the group ${\cal G}$, it would
be desirable to find a non-commutative manifold $X$,
whose diffeomorphism group
is ${\cal G}$.

The coordinates $\delta_n=-(\log(\psi^\prime(x))^{(n)}$
of a diffeomorphism $\psi$ allow to reconstruct the latter by the formula
\[
\psi(x)=\int_0^x \exp(-\sum\frac{\delta_n}{n!}u^n)du.
\] 
This formula provides the clear meaning both for composition and inversion
of diffeomorphisms.

Of course, we would love to have a similar formula for the group ${\cal G}$
and it is tantalizing to consider the Feynman integral
\[
\int \exp(-L_0+\sum_\Gamma L_\Gamma)
\]
as a direct analogue of the above expression.

\subsection*{Acknowledgements}
D.K.~thanks the I.H.E.S., Bures-sur-Yvette, for hospitality during a stay
Jan.-Feb.~1998
and the theory group at the CPT (Marseille) for interest and discussions.
Also, support by a Heisenberg Fellowship of the DFG
for D.K.~is gratefully acknowledged.
D.K.~thanks R.~Stora for motivating the investigation which will be 
reported in \cite{overl}.

\section{Appendix}
\subsection{$\phi^3$ Theory and Overlapping Divergences}
A prominent problem in renormalization theory is the presence of overlapping
divergences. We will soon see that to Green functions which suffer 
from such overlapping divergences we will have to associate
a sum of trees, while so far our experience only lead to the identification
of single trees with a given Green function.

%%XXX \special{src: 2778 CK1.TEX} %Inserted by TeXtelmExtel

We will proceed by studying the example of $\phi^3$ theory in six dimensions.
A full study will be given elsewhere \cite{overl}, but we also mention
that solutions to the problem of overlapping divergences were already found in
\cite{habil}, using combinatorical considerations concerning  divergent sectors,
in \cite{K} and \cite{bdk} using Schwinger Dyson equations,
and were also known to others. In \cite{overl} we will show how
overlapping divergences give rise to a slightly modified Hopf algebra, which eventually
turns out to be identical to the Hopf algebra of rooted trees considered here.
We sketch this more formal argument after the consideration of $\phi^3$ theory
as an example.
 
%%XXX \special{src: 2791 CK1.TEX} %Inserted by TeXtelmExtel
 
In whatever approach one takes, the final message is the same:
Overlapping divergent functions can be resolved in sums of functions
having only nested and disjoint divergences.
To see how this comes about, we will here employ yet another approach,
using differential equations on bare Green functions.

%%XXX \special{src: 2799 CK1.TEX} %Inserted by TeXtelmExtel

Green functions in $\phi^3_6$ theory which are overall divergent are provided
by two- and three-point functions, to which we refer as
$G^{[2]}_n(q;m)$ and $G^{[3]}_n(p,q;m)$. Here the subscript $n$
refers to the number of loops in the Green-function, and $m$ is the
mass of the propagator, while $p,q$ are external momenta.

%%XXX \special{src: 2807 CK1.TEX} %Inserted by TeXtelmExtel

We first consider $G^{[3]}_n(p,q;m)$:
\[
G^{[3]}_n(p,q;m)=\int d^6l_1\ldots d^6l_n
\prod_{i:=1}^{3n}\frac{1}{P_i}.
\]
For $n\geq 1$, it is a product of $3n$ propagators
$P_i=1/(k_i^2-m^2+i\eta)$, where the $k_i$ are momentum vectors
which are linear combinations of external momenta $p,q$ and $n$
internal momenta $l_1,\ldots, l_n$ such that momentum conservation
holds at each vertex.

%%XXX \special{src: 2820 CK1.TEX} %Inserted by TeXtelmExtel

%%XXX \special{src: 2823 CK1.TEX} %Inserted by TeXtelmExtel

As each propagator $1/P_i$ contributes with weight two
to the powercounting, we find that $G^{[3]}$ is
overall logarithmic divergent, $3\times n\times 2-6\times n=0$.

%%XXX \special{src: 2829 CK1.TEX} %Inserted by TeXtelmExtel

For each $P_i$, let $\overline{P_i}:=k_i^2+i\eta$,
so that $P_i=\overline{P_i}-m^2$.

%%XXX \special{src: 2834 CK1.TEX} %Inserted by TeXtelmExtel

%%XXX \special{src: 2837 CK1.TEX} %Inserted by TeXtelmExtel

Then, 
one immediately sees that $G^{[3]}_n(p,q;m)-
G^{[3]}_n(p,q;0)$ is overall convergent.
This follows directly from powercounting in the expression
\[
\prod_{i:=1}^{3n}\frac{\prod_j \overline{P_j}-\prod_j
(\overline{P_j}-m^2)}{P_i\overline{P_i}}
\]
Thus, to determine the counterterm for a vertex function,
it suffices to consider the massless case.
\footnote{Even better, again using powercounting, one immediately shows
that it is sufficient to consider $G^{[3]}_n(0,q;0)$.}

%%XXX \special{src: 2852 CK1.TEX} %Inserted by TeXtelmExtel

%%XXX \special{src: 2855 CK1.TEX} %Inserted by TeXtelmExtel

Hence all possible subdivergences of $G^{[3]}_n(p,q;0)$ are given by
functions of the type $G^{[3]}_r(k_i,k_j;0)$ and $G^{[2]}_s(k_i;0)$,
with $s<n$ and $r<n$. 

%%XXX \special{src: 2861 CK1.TEX} %Inserted by TeXtelmExtel

%%XXX \special{src: 2864 CK1.TEX} %Inserted by TeXtelmExtel

In the context of $\phi^3$ theory in six dimensions, 
overlapping divergences can only be 
provided by two-point functions. The only circumstance which stops
us to assign a unique tree to $G^{[3]}_n(p,q;m)$ is the fact
that there might be overlapping subdivergences provided by 
massless two-point functions 
$G^{[2]}_s(k_i;0)$, $s<n$.

%%XXX \special{src: 2874 CK1.TEX} %Inserted by TeXtelmExtel

%%XXX \special{src: 2877 CK1.TEX} %Inserted by TeXtelmExtel

Before we handle these subdivergences, we turn to
$G^{[2]}_n(q;m)$ itself.
At $n$ loops, it consists of $3n-1$ propagators
\[
G^{[2]}_n(q;m)=\int d^6l_1\ldots d^6 l_n
\prod_{i:=1}^{3n-1}\frac{1}{P_i}
\] 
Consider the difference
\begin{eqnarray*}
G^{[2]}_n(q;m)-G^{[2]}_n(q;0) & = & m^2 
\int d^6l_1\ldots d^6l_n \prod_{i:=1}^{3n-1}\frac{1}{P_i}
\sum_{j:=1}^{3n-1}\frac{1}{\overline{P_j}}\\
 & & +\mbox{overall finite terms}
\end{eqnarray*}
which is of overall logarithmic degree of divergence.
As far as the overall counterterm is concerned, we can even
nullify masses in this difference
and thus find that the divergences of 
$G^{[2]}_n(q;m)$ can be separated as
\begin{eqnarray*}
G^{[2]}_n(q;m) & = & 
m^2\int d^6l_1\ldots d^6l_n \prod_{i:=1}^{3n-1}\frac{1}{\overline{P_i}}
\sum_{j:=1}^{3n-1}\frac{1}{\overline{P_j}}\\
 & & +G^{[2]}_n(q;0)+U(q,m)
\end{eqnarray*}
where $U(q;m)$ collects all the overall finite terms.

%%XXX \special{src: 2906 CK1.TEX} %Inserted by TeXtelmExtel

%%XXX \special{src: 2909 CK1.TEX} %Inserted by TeXtelmExtel

The first term on the rhs is overall logarithmic divergent. It only
can provide overlapping divergences through massless
functions 
$G^{[2]}_s(q;0)$ appearing as subgraphs in it, quite similar to the
analysis of the vertex function, as the sum over $j$
squares one propagator in turn.

%%XXX \special{src: 2918 CK1.TEX} %Inserted by TeXtelmExtel

We have thus reduced all appearances of overlapping divergences to
the presence of functions $G^{[2]}_i(q;0)$, $i\leq n$.
It remains to show how the overlapping divergences in 
$G^{[2]}_n(q;0)$ can be handled for all $n$.

%%XXX \special{src: 2925 CK1.TEX} %Inserted by TeXtelmExtel

%%XXX \special{src: 2928 CK1.TEX} %Inserted by TeXtelmExtel

This is actually not that difficult.
Necessarily, $G^{[2]}_n(q;0)$ has the form
\[
G^{[2]}_n(q;0)=(q^2)^{1-n\epsilon}F_{G_n}(\epsilon)
\]
where $F_{G_n}(\epsilon)$ is a Laurent series in $\epsilon$.
Hence $G^{[2]}_n(q;0)$ fulfils the differential equation
\[
\frac{1}{2D(1-n\epsilon)}
q^2\frac{\partial}{\partial q_\mu}\frac{\partial}{\partial q^\mu}
G^{[2]}_n(q;0)=G^{[2]}_n(q;0).
\]
This solves the problem. The remaining source of overlapping divergences,
$G^{[2]}_n(q;0)$, is expressed in terms of the overall logarithmic
divergent function
$q^2\frac{\partial}{\partial q_\mu}\frac{\partial}{\partial q^\mu}
G^{[2]}_n(q;0)$ which is free of overlapping divergences.
Such an approach is also very useful in practice \cite{bdk}.

%%XXX \special{src: 2949 CK1.TEX} %Inserted by TeXtelmExtel

Fig.(\ref{f2}) gives two  examples for the resolution of overlapping divergences.
Crosses in the figure indicate where the derivatives with respect to $q$ act for a chosen
momentum flow through the graph.
\bookfig{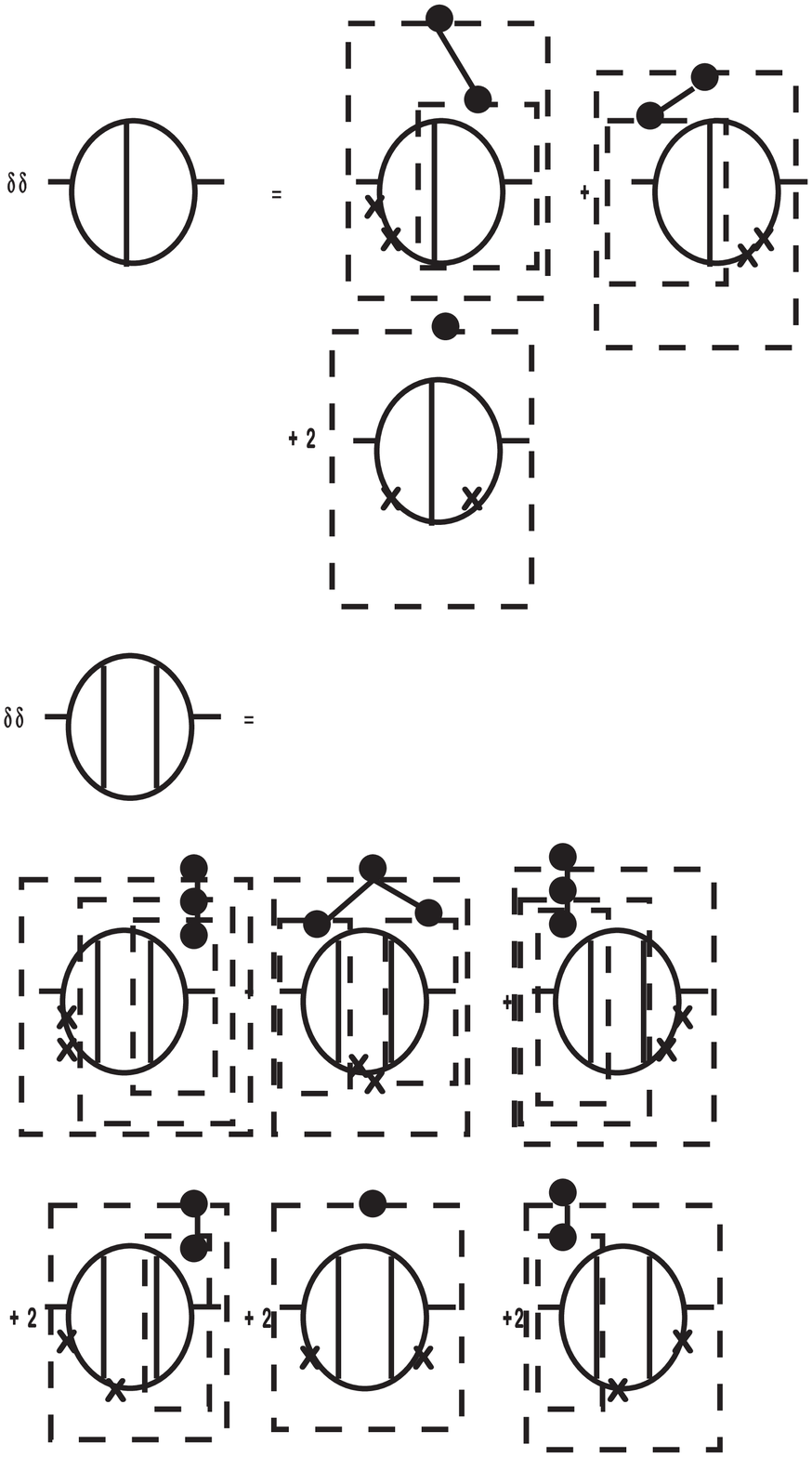}{Example.}{f2}{The resolution of overlapping divergences
and the resulting sum of trees. A double derivative
with respect to the external momentum resolves the graph in
contributions
each of which is free of overlapping divergences.
We indicate by crosses on propagators the places where the derivative
acts,
for a chosen momentum flow.
}{11}
\subsection*{A general Argument}
So far, we decomposed graphs which have overlapping divergences into a sum of 
contributions each of which delivers a rooted tree. Thus, overlapping divergences
correspond to a linear combination of rooted trees, while any Feynman diagram without
overlapping divergences corresponds to a single rooted tree.

%%XXX \special{src: 2969 CK1.TEX} %Inserted by TeXtelmExtel

%%XXX \special{src: 2972 CK1.TEX} %Inserted by TeXtelmExtel

One might suggest to enlargen the Hopf algebra $H_R$ of rooted trees to
another Hopf algebra, $H_O$ say, so that $H_O$
directly contains elements which correspond to graphs with overlapping
divergences \cite{KW}. 

%%XXX \special{src: 2979 CK1.TEX} %Inserted by TeXtelmExtel

Let us at this stage mention a general fact which shows
that any such Hopf algebra $H_O$ is nothing else than 
the Hopf algebra of rooted trees.
If we take into account 
the decorations of vertices by Feynman diagrams without subdivergences, any 
such Hopf algebra $H_O$ is a Hopf algebra $H_R$ for an appropriate set of decorations.

%%XXX \special{src: 2988 CK1.TEX} %Inserted by TeXtelmExtel

Consider a Feynman graph $\Gamma$ which has overlapping subdivergences, but in a way that any
of its divergent subgraphs $\gamma\subset \Gamma$ and any of the complementary graphs
$\Gamma/\gamma$ is free of overlapping subdivergences.
The first example in Fig.(\ref{f2}) is of this type.
The cases we have excluded here will be handled later by a recursive argument.

%%XXX \special{src: 2996 CK1.TEX} %Inserted by TeXtelmExtel

We want to construct a Hopf algebra $H_O$ which contains a single element
$t_\Gamma$ such that the antipode $S(t_\Gamma)$ delivers the counterterm
without making recourse to the methods of the previous paragraph to disentangle
$t_\Gamma$ first as a sum of trees $t_i$ in some decorated algebra $H_R$.
The question is: Could such an algebra have a structure different from $H_R$?

%%XXX \special{src: 3004 CK1.TEX} %Inserted by TeXtelmExtel

Now, as $H_O$ shall also be able to treat Feynman diagrams which only have nested
or disjoint subdivergences, it will contain the Hopf algebra of rooted trees as a subalgebra.

%%XXX \special{src: 3009 CK1.TEX} %Inserted by TeXtelmExtel

Let us actually construct $H_O$ by fairly general arguments.
Let $H_R\subset H_O$ be given, and let in particular all Feynman graphs
without subdivergences be identified. Hence all possible
decorations, and thus all primitive elements of $H_R$ are assumed to be determined.
Note that the primitive elements of $H_O$ are identical with the primitive elements
of $H_R$ as graphs with overlapping divergences necessarily contain subdivergences,
and thus do not provide primitive elements per se.

%%XXX \special{src: 3019 CK1.TEX} %Inserted by TeXtelmExtel

A Feynman graph $\Gamma$ chosen as above has only subgraphs which can be
described by proper rooted trees.
Thus, its coproduct in $H_O$ will have the general form
$$
\Delta(t_\Gamma)=t_\Gamma\otimes e+e\otimes t_\Gamma+
\sum_{\gamma}t(\gamma)\otimes t(\Gamma/\gamma),
$$
where the sum is over all subgraphs of $\Gamma$, while $t(\gamma)$ and
$t(\Gamma/\gamma)$ are the rooted trees assigned to the corresponding graphs.
By the constraints which we imposed on  
$\Gamma$ this is always possible.
In $H_O$ we consider the above  equation as the definition for the coproduct
on elements $t_\Gamma\not\in H_R\subset H_O$.

%%XXX \special{src: 3035 CK1.TEX} %Inserted by TeXtelmExtel

On the rhs of the above coproduct, the only part which is not
in $H_R\otimes H_R$  is 
$$
t_\Gamma\otimes e+e\otimes t_\Gamma,
$$
and we write 
$$
\Delta(t_\Gamma)=t_\Gamma\otimes e+e\otimes t_\Gamma+R_\Gamma,
$$
with $R_\Gamma \in H_R \otimes H_R$.

%%XXX \special{src: 3048 CK1.TEX} %Inserted by TeXtelmExtel

Now, we know that there exists an element $T_\Gamma \in H_R$
such that  
$$
\Delta(T_\Gamma)=T_\Gamma\otimes e+e\otimes T_\Gamma + R_\Gamma.
$$
This element $T$ is just the linear combination of rooted trees
constructed in the previous section, but ist existences can be established on general grounds
from the consideration of maximal forests \cite{habil,overl}.

%%XXX \special{src: 3059 CK1.TEX} %Inserted by TeXtelmExtel

Finally we set $U:=t_\Gamma-T_\Gamma$ and calculate
$$
\Delta(U)=U\otimes e+e\otimes U.
$$
Now, if $U$ is superficially divergent at all it is a primitive element.
It thus can be described by the rooted
tree $t_1$. To be able to do so we only have to enlarge the algebra
$H_R$ to contain the decoration $U$.
An easy recursion argument finally allows to drop the constraint on $\Gamma$
\cite{overl}.

%%XXX \special{src: 3072 CK1.TEX} %Inserted by TeXtelmExtel

One concludes that any Hopf algebra which contains $H_R$ but also elements
$t_\Gamma\not\in H_R$ is isomorphic to the algebra of rooted trees $H_R$ with 
an enlarged  set of primitive elements.
In Fig.(\ref{f2}) we see some contributions which only generate the tree
$t_1$. They correspond to such new primitive elements.
A detailed version of this argument will be given elsewhere \cite{overl}.
\clearpage
%%%%%%%%%%%%%%%%%%%%%%%%%%%%%%%%%%%%%%%%%%%%%%%%%%%%%%

%%XXX \special{src: 3083 CK1.TEX} %Inserted by TeXtelmExtel

\end{document}